\documentclass[aps,prd,nofootinbib,superscriptaddress,eqsecnum,twocolumn,reprint,amsmath,amssymb,floatfix]{revtex4-2}

\usepackage{graphicx}
\usepackage{mathrsfs}
\usepackage{graphicx}
\usepackage{xcolor}
\usepackage[colorlinks=true,allcolors=teal]{hyperref}
\usepackage{orcidlink}
\usepackage{verbatim} 
\usepackage[normalem]{ulem}

\newcommand{\m}{\textrm{mem}}

\newcommand{\nm}{\textrm{osc}}
\newcommand{\s}{\textrm{sum}}
\newcommand{\ud}{\textrm{d}}
\newcommand{\ef}{\textrm {eff}}
\newcommand{\E}[1]{#1^{(e)}}
\newcommand{\B}[1]{#1^{(b)}}
\newcommand{\p}[1]{#1^{(p)}}
\newcommand{\mass}{\textrm m}
\newcommand{\ord}[2]{\underset{^{(#1)}}{#2}{}}
\newcommand{\seq}{\stackrel{\mathrm s}{=}}

\newcommand{\tl}{\mathbf{l}}
\newcommand{\tn}{\mathbf{n}}
\newcommand{\tm}{\mathbf{m}}

\DeclareMathOperator{\STF}{STF}

\newcommand{\UVA}{Department of Physics, University of Virginia, P.O.~Box 400714, Charlottesville, Virginia 22904-7414, USA}
\newcommand{\Southampton}{School of Mathematical Sciences and STAG Research Centre, University of Southampton,
  Southampton, SO17 1BJ, United Kingdom}

\begin{document}

\title{Defining higher memory signals and forecasting their observation prospects for binary-black-hole mergers with next-generation gravitational-wave detectors}

\author{S.\ Siddhant\,\orcidlink{0000-0001-5344-0008}}
\email{sbp4ab@virginia.edu}
\affiliation{\UVA}%

\author{Alexander M.\ Grant}
\email{amg425@cornell.edu}
\affiliation{\Southampton}%

\author{David A.\ Nichols\,\orcidlink{0000-0002-4758-9460}}
\email{david.nichols@virginia.edu}
\affiliation{\UVA}%

\date{\today}

\begin{abstract}
In asymptotically flat spacetimes far from an isolated source, gravitational waves (GWs) undergo nonlinear interactions with themselves and with the parts of the spacetime curvature related to the time-dependent four-momentum and angular momentum of the spacetime.
These spacetime nonlinearities produce distinctive offsets in the GW strain and its time integrals, which have been referred to as the displacement memory effect for the strain and ``higher'' GW memory effects for the integrals of the strain.
There are existing data analysis pipelines that search for evidence for the displacement memory effect in the population of binary-black-hole mergers observed by current GW detectors (though conclusive evidence for the effect has not yet been found).
The first set of higher memory effects include the spin and center-of-mass effects (collectively, ``drift'' memory), and the second is the ``ballistic'' memory effect.
Prior work has shown that next-generation, ground-based GW detectors could find evidence for the spin memory effect in the large population of binary black holes that these detectors will be capable of observing.
In this paper, we investigate how well a detector network of two Cosmic Explorer detectors can measure the displacement through ballistic memory signals.
We first formulate what are appropriate notions of time-dependent GW signals associated with these memory effects.
With these definitions of displacement and higher memory signals, we find that the Cosmic Explorer network is capable of detecting the displacement memory from tens of individual mergers per year and capable of finding evidence for the spin and center-of-mass memory effects in a population of mergers after a roughly one-year observation run at design sensitivity.
We compute a lower limit on the detection prospects of the ballistic memory, which shows that Cosmic Explorer will not be able to detect a limited portion of the signal that can be evaluated in the approximation we use to compute the higher memory effects in this paper.
\end{abstract}

\maketitle 

\tableofcontents

\listoffigures

\section{Introduction} \label{sec:intro}

In 1962, Bondi, van der Burg, and Metzner~\cite{Bondi:1962px} and Sachs~\cite{Sachs:1962zza,Sachs:1962wk} published analyses of the Einstein field equations of general relativity using coordinates that are well adapted for studying outgoing gravitational waves (GWs).
After imposing asymptotically flat boundary conditions on the spacetime metric and expanding in a power series in inverse luminosity distance, they determined that the metric can be described by a shear tensor which contains information about the outgoing gravitational waves, and a (countably infinite) number of ``Bondi aspects'' which characterize the gravitational properties of the source of the waves (for a review, see also, e.g.,~\cite{Madler:2016xju,Grant:2021hga}).

In these coordinates, the Einstein field equations reduce to a set of nonlinear partial differential equations that determine the evolution of the Bondi aspects in response to the outgoing gravitational waves.
Two of these, the mass and angular momentum aspects, contain information about the energy-momentum and relativistic angular momentum of the spacetime in their lowest multipole moments, and the energy-momentum and angular momentum evolve because of nonlinear GW terms in their evolution equations (see, e.g.,~\cite{Madler:2016xju}).
The evolution of the higher multipole moments of the mass and angular momentum aspects and the dynamics of the higher Bondi aspects contain additional information about nonlinear GW effects in the wave zone.
These nonlinear effects have persistent features, which have been referred to as the displacement and ``higher'' GW memory effects~\cite{Grant:2023jhd,Grant:2023ged,Siddhant:2024nft}.
Analyzing these phenomena, their GW signatures from binary-black-hole mergers, and the detection prospects of their associated GW signals will be the subject of this paper.

\subsection{Overview of gravitational-wave memory and its higher generalizations} \label{subsec:memoryOverview}

Although we will analyze (higher) memory effects using the Bondi-Sachs framework, it is not necessary to do so, and, in fact, the first calculations of the GW memory effect took place using other coordinate choices.\footnote{In fact, it has been argued that it can be beneficial to understand the memory effect with calculations that directly use the Riemann tensor in linearized gravity, which is gauge invariant in this context~\cite{Bieri:2013ada,Garfinkle:2022dnm}.}
Specifically, Zeldovich and Polnarev~\cite{Zeldovich:1974gvh} computed the memory effect (which was coined as such in~\cite{Braginsky:1985vlg}) in linearized gravity for gravitationally unbound binaries.
A nonlinear memory effect was later predicted by Christodoulou~\cite{Christodoulou:1991cr} and in the context of
the multipolar post-Minkowski formalism by Blanchet and Damour~\cite{Blanchet:1992br}.\footnote{See also~\cite{Wiseman:1991ss} for a calculation of the nonlinear memory from compact binaries, and~\cite{Frauendiener:1992dmu} for a translation of Christodoulou's results into the Newman-Penrose formalism~\cite{Newman:1961qr}.}
It has been interpreted as being of the same form as the linear memory effect of Zeldovich and Polnarev when the stress-energy tensor of the matter fields is replaced by the effective stress-energy of gravitational waves~\cite{Thorne:1992sdb}.
The linear and nonlinear memory effects are collectively termed the ``displacement memory''~\cite{Strominger:2014pwa} because they produce a change in the relative separation of co-moving observers which depends on the initial separation of the observers~\cite{Flanagan:2018yzh}. 

Within the Bondi-Sachs formalism, the linear and nonlinear contributions to the displacement memory arise from the quadrupolar and higher multipole moments of one of the asymptotic Einstein equations (specifically, the evolution equation for the mass aspect, the lowest multipoles of which describe the evolution of the Bondi mass and linear momentum).
The memory effect also can be interpreted in terms of the supermomentum charges~\cite{Geroch:1977jn,Wald:1999wa} conjugate to the supertranslation asymptotic symmetries of the Bondi-Metzner-Sachs (BMS) algebra~\cite{Bondi:1962px,Sachs:1962zza} and the action of supertranslation symmetries on the gravitational-wave strain.\footnote{A related interpretation exists in transverse-traceless gauge in terms of ``large'' local residual gauge transformations~\cite{DeLuca:2024cjl,DeLuca:2024asq}.}
Specifically, this part of the asymptotic Einstein's equation can be interpreted as a conservation (or continuity) equation for the supermomentum charges (see, e.g.,~\cite{Strominger:2014pwa,Ashtekar:2014zsa}).
In addition, the lasting strain associated with the memory effect can also be understood as resulting from a difference in the initial and final ``canonical'' rest frames (in which the shear vanishes) before and after the waves, which are related by a supertranslation~\cite{Strominger:2014pwa,Ashtekar:2014zsa,Flanagan:2015pxa}. 

Analyzing the quadrupole and higher multipole moments of the evolution equation for the Bondi angular momentum aspect led to the identification of two additional memory effects.
The first was associated with the ``magnetic-parity'' part of the equation and was called ``spin memory''~\cite{Pasterski:2015tva}  
(see also~\cite{Flanagan:2015pxa,Nichols:2017rqr}).
The second was related to the ``electric-parity'' part of angular momentum [called the ``center-of-mass (CM),'' ``mass dipole,'' or ``boost charge'' part], and the corresponding memory effect was referred to as ``center-of-mass'' memory~\cite{Nichols:2018qac}.
The connection between these memory effects and extensions of the BMS algebra~\cite{Barnich:2009se, Barnich:2010eb, Campiglia:2014yka,Campiglia:2015yka} is more involved (see, e.g.,~\cite{Compere:2018ylh}).
The spin and CM memories are related to lasting changes in the time integral of the shear.
For nearby freely falling observers, the effects produce a lasting displacement that depends on the initial relative velocity of the observers~\cite{Flanagan:2018yzh} (which led to it also being called ``drift'' memory in~\cite{Grant:2023ged}).
Specifically, it is the part of the velocity that is proportional to the time integral of the GW strain, as opposed to the kinematic drift that grows linearly with time arising from the relative velocity.

The evolution equations for the higher Bondi aspects were also shown to contribute to a class of higher memory effects called the (temporal) moments of the news (where in the BMS framework, the news tensor is related to the time derivative of the shear)~\cite{Grant:2021hga}.
The moments of the news, in fact, encompass the displacement and drift GW memory effects, because the zeroth moment of the news corresponds to the displacement memory and the first moment contains the spin and CM (collectively, drift) memories.
The higher moments of the news are related to the higher-order memory effects.

The evolution equations from which the higher memory effects arise have a similar form to the equation which determines the displacement memory (which is equivalent to the continuity equation satisfied by the supermomentum charge and its flux).
For this reason, the nonlinear terms in the higher memory equations were called ``flux'' or ``pseudoflux'' terms (which, respectively, vanish or do not vanish in the absence of radiation), and the remaining terms were called the ``charge'' terms. 
We will use this ``charge and flux'' language henceforth.\footnote{The charge and flux nomenclature is less ambiguous than the ``linear and nonlinear'' terminology for several reasons.
Specifically, the definitions of the charge terms are often nonlinear in the Bondi metric functions, and even when they are not, the Bondi metric functions themselves satisfy nonlinear evolution equations which makes them implicitly nonlinear, even when an expression is linear in a given Bondi metric function.
Furthermore, in the post-Newtonian and multipolar post-Minkowskian approach, the radiative multipole moments are nonlinear functions of the source multipoles (see, e.g.,~\cite{Blanchet:2013haa}).
The ``ordinary and null'' terminology used in~\cite{Bieri:2013ada} is a reasonable alternative for the displacement and drift memory effects, where the nonlinear GW terms can be written in the form of components of an effective null stress-energy of GWs. However, it is less apparent that the flux and pseudoflux terms in the higher memory effects can be cast in the form of an effective null stress-energy tensor; thus, we will not use the ``ordinary-null'' nomenclature either. \label{fn:charge_flux}}
The precise relationship between these higher memories, larger asymptotic symmetry algebras, and their corresponding charges is under investigation (see, e.g.,~\cite{Compere:2022zdz,Freidel:2021ytz}).

While we have already described how freely falling observers can measure the zeroth and first moments of the news, to measure the higher moments, the observers would need to have a nonzero relative acceleration.
In this case, the relative separation of the observers will change because of both the initial relative velocity and the relative acceleration, but once these kinematical effects are removed, there is a residual change in the final relative separation that depends on the GW strain and the time derivatives of the initial acceleration. 
This residual change is the ``curve deviation'' observable (see~\cite{Flanagan:2018yzh,Grant:2021hga}).
It follows that there are an infinite hierarchy of higher memory effects that can be computed from the asymptotic Einstein equations in Bondi gauge, and which can be measured by observers who are accelerating with respect to each other.

\subsection{Overview of efforts to detect gravitational-wave memory signals} \label{subsec:memoryDetect}

In Sec.~\ref{subsec:memoryOverview}, we summarized how the Bondi-Sachs equations can be used to compute different memory effects, and we also described ``proof-of-principle'' measurement procedures of these effects using geodesic or curve deviation.
Here we review the definition and the status of the computation of higher memory signals from astrophysical sources---specifically, binary black holes (BBHs)---which produce all of the higher memory effects that have been described thus far~\cite{Grant:2023jhd,Siddhant:2024nft}.

Depending on the specific GW observatory, the nature of the GW source, and the type of search being performed, different levels of fidelity are required in defining and modeling the part of the GW signal corresponding to the memory, which we call the ``memory signal''~\cite{Grant:2022bla,Zosso:2026czc}.
For example, pulsar timing arrays (specifically Parkes and NANOGrav) have performed searches for GW bursts with memory using a step-function approximation to the memory signal~\cite{Wang:2014zls,Agazie:2025oug}.
Such a signal model makes an implicit assumption that the memory signal grows on a timescale that is short compared to the cadence of the pulsar observations.\footnote{Note, however, that it is possible to search for memory signals from BBH mergers in pulsar timing arrays without this assumption~\cite{Tomson:2025oox}, and the method of~\cite{Tomson:2025oox} has been applied to data from the Parkes and European Pulsar Timing Arrays~\cite{Tomson:2025ixn}.}
In this paper, however, we will focus on ground-based gravitational-wave detectors, where modeling the precise time dependence of the memory signals is typically necessary to make accurate forecasts and perform searches in GW strain data.\footnote{Note that there have been searches for the memory effect from subsolar-mass black holes~\cite{Ebersold:2020zah}, which use the coherent WaveBurst search pipeline~\cite{Klimenko:2008fu,Klimenko:2015ypf}.
This pipeline uses a wavelet basis in the time-frequency domain to recover coherent excess power present in the GW data in multiple detectors and distinguish it from instrumental noise in individual detectors.
While memory signals would be useful for restricting the region of time-frequency space where the memory signal is likely to reside, the searches themselves do not make use of GW templates of the memory signal, because they use a superposition of wavelets for their signal model instead.}

Modeling the displacement GW memory signal has been most systematically studied for the nonlinear displacement memory effect from BBH mergers, given that its signal is larger than the higher memory effects and that the LIGO-Virgo-KAGRA (LVK) Collaboration has detected hundreds of such mergers since the first detection in 2015~\cite{LIGOScientific:2016aoc,LIGOScientific:2018mvr,LIGOScientific:2020ibl,KAGRA:2021vkt,LIGOScientific:2025slb,LIGOScientific:2026wfs}.
An important subtlety in computing the memory effect is that numerical-relativity (NR) simulations that extract waveforms from finite radii are generally missing the memory effect and thus are inconsistent with the predictions of general relativity in asymptotically flat spacetimes.
They can be made more consistent by adding the nonlinear memory signal defined through the conservation equation for the supermomentum (see, e.g.,~\cite{Mitman:2020bjf}).
The accuracy of this ``correction'' procedure was confirmed by comparisons with simulations that were performed with Cauchy-characteristic evolution (CCE)~\cite{Bishop:1996gt,Bishop:1997ik} (which solves Einstein's equations out to null infinity and which includes the memory portion of the GW strain~\cite{Pollney:2010hs,Mitman:2020pbt}).

To search for or forecast the detection prospects of observing the GW memory signal in ground- and space-based interferometers (which do not have sensitivity to the zero-frequency lasting offset associated with the memory effect), one needs to have a prescription for defining a time-domain signal associated with the memory offset.
For BBH mergers, because the flux (nonlinear) contribution to the displacement memory has been shown to be larger than the charge contribution~\cite{Mitman:2020pbt},
the time-dependent signal obtained from using the flux term in the equation of supermomentum conservation has been used as a common definition of the memory signal (e.g.,~\cite{Nichols:2017rqr,Talbot:2018sgr,Mitman:2020bjf,Grant:2022bla}).
However, including only nonoscillatory or both oscillatory and nonoscillatory effects in the nonlinear memory signal have both been considered.
While there are arguments to support both proposals (e.g.,~\cite{Grant:2023jhd,Siddhant:2024nft} for both and~\cite{Inchauspe:2024ibs,Zosso:2026czc,Cogez:2026frh} for only nonoscillatory), for reasons that we describe in more detail in Sec.~\ref{subsec:nonosc}, we will focus on nonoscillatory, nonlinear memory signals in this paper.\footnote{As will be discussed further in Sec.~\ref{subsec:nonosc}, oscillatory and nonoscillatory here refers primarily to spherical-harmonic modes of the flux term (that is, those with $m \neq 0$ and $m = 0$, respectively).
Even within ``nonoscillatory'' spherical-harmonic modes, there are oscillatory phenomena, such as quasinormal mode ringing.
The CCE waveforms contain both memory and quasinormal ringing (as do some waveform models of the nonoscillatory spherical harmonic modes~\cite{Yoo:2023spi,Rossello-Sastre:2024zlr,Rossello-Sastre:2025dep}).
However, it is understood that the quasinormal mode contributions should not be included in searches for the memory effect~\cite{Cheung:2024zow,Mitman:2026zfg,Rossello-Sastre:2026gah}, so stand-alone waveform models for the GW memory are needed (as in~\cite{Elhashash:2024thm,Elhashash:2025hqi}).}

For the current LVK detectors, earlier studies focused on the detection prospects of the memory effect for individual compact-object mergers (e.g.,~\cite{Braginsky:1987kwo,Kennefick:1994nw,Pollney:2010hs}).
After the first detections by the LVK, it was determined that detecting the memory effect from individual BBH mergers was unlikely with the LVK detectors, but it would be possible to detect the effect in the population of all BBH mergers, even though each individual event might be substantially below a threshold for detection~\cite{Lasky:2016knh}.
Current forecasts estimate that it will require thousands of events (several years of the LVK detectors running at O5 sensitivity~\cite{KAGRA:2013rdx}) to detect the memory effect~\cite{Hubner:2019sly,Boersma:2020gxx,Hubner:2021amk,Grant:2022bla,Mitman:2026zfg} (see also~\cite{Rossello-Sastre:2026gah}).
Next-generation gravitational-wave detectors including LISA~\cite{Amaro-Seoane:2017ADS} (in space) or Einstein Telescope~\cite{Punturo:2010zz} and Cosmic Explorer (CE)~\cite{Reitze:2019iox} (on the ground) are expected to detect memory signals from individual BBH systems (see~\cite{Favata:2009ii,Islo:2019qht,Goncharov:2023woe,Inchauspe:2024ibs,Zosso:2026czc,Cogez:2026frh} for forecasts with LISA, \cite{Goncharov:2023woe} for Einstein Telescope, and \cite{Grant:2022bla} for CE\footnote{\label{fn:forecastsCE}There was an error in the forecasts for CE in~\cite{Grant:2022bla} related to the use of the amplitude spectral density for the CE detector noise, which caused the signal-to-noise ratios to be underestimated by a factor of a few on average. 
This was identified by one of the authors (S.\ Siddhant); corrected results will be presented in Sec.~\ref{sec:results}.}).

Similar considerations to those described above for the displacement memory arise for computing the measurement prospects for the higher memory effects.
Because the higher memory effects have been predicted in theoretical calculations more recently and because they have smaller amplitudes in BBH mergers, their signals and detection prospects have been less well studied.
For the spin memory, the flux term has been computed in~\cite{Nichols:2017rqr,Mitman:2020pbt}, and the forecasts for detecting the effect in a population on BBH mergers were investigated in CE~\cite{Grant:2022bla} (see Footnote~\ref{fn:forecastsCE}) and Einstein Telescope~\cite{Goncharov:2023woe}.
For the CM memory effect and the next higher memory (the ballistic memory), the memory signals have been computed in the post-Newtonian approximation~\cite{Nichols:2018qac,Siddhant:2024nft} and in a few NR simulations~\cite{Grant:2023jhd}.
The detection prospects for the CM or the ballistic memory effects have not been assessed.

\subsection{Aims of this paper}

Given the discussion in Sec.~\ref{subsec:memoryDetect}, two aims are apparent.
The first is to update the displacement and spin-memory calculations for CE.
This will show the differences between using oscillatory and nonoscillatory portions of the memory signal and correct for the error in~\cite{Grant:2022bla}, as mentioned in Footnote~\ref{fn:forecastsCE}.
The second is to perform new calculations of the detection prospects for the CM and ballistic memory effects from BBH mergers with CE.
There are several subtle aspects of the calculations of the CM and ballistic memories, which we will summarize here, but the detailed exposition of them will be deferred to the main sections of this paper.

Unlike the displacement and spin memory effects, where the flux terms are larger than the charge terms in the nonoscillatory modes of the memory signals for BBH systems~\cite{Mitman:2020pbt}, in the CM and ballistic memory effects, the charge contributions can be larger~\cite{Grant:2023jhd}.
Thus, we will need a procedure to compute both charge and flux contributions to the CM and ballistic memory signals.
However, the $l\geq 2$ and $m=0$ spherical-harmonic modes of the charge terms contain oscillatory features (related to the ringdown of the remnant black hole) which do not produce lasting offsets in the moments of the news. 
It is therefore necessary to separate the higher-memory contribution from an oscillatory contribution even in these $m=0$ modes.
We give a prescription for computing these charge contributions based on the expected value of the moment of the news that arises in asymptotically flat spacetimes that undergo stationary-to-stationary transitions.
We show that this gives a good approximation to the offset arising in the corresponding moment of the news when the BBH systems do not have significant GW kicks.

With this prescription for defining GW signals associated with the CM and ballistic memory effects, we compute forecasts for how well CE can find evidence in favor of these effects in simulations of populations of merging BBH systems.
We find that there are sufficiently many GW mergers for two CE detectors (at design sensitivity) to accumulate evidence for the CM memory effect after an observing run of about one year.
For the ballistic memory, we compute lower limits that suggest that the signals are sufficiently small that it is unlikely to find evidence in the population of BBH mergers for the component of the signal that we model.
However, these estimates are neglecting portions of the signal that we are unable to compute in the stationary-to-stationary approximation, but which are likely larger.

\subsection{Organization of this paper and conventions} 

In Sec.~\ref{sec:review}, we review the Bondi-Sachs formalism in Sec.~\ref{subsec:BMSformalism} and describe how the conservation and evolution equations can be expressed in a charge-flux form in Sec.~\ref{subsec:chargeFlux}.
Section~\ref{subsec:higherMemory} describes an integral form of these evolution equations involving the temporal moments of the news, and expressions for the corresponding strain associated with the offsets in the higher moments of the news.
Next, Sec.~\ref{sec:multipole} gives our conventions for multipolar expansions (Sec.~\ref{subsec:multipoles}) and derives multipolar expressions for the flux and pseudoflux terms (Sec.~\ref{subsec:fluxes}).
Section~\ref{sec:stationary} reviews the computation of Bondi metric functions in stationary regions.
It uses the expressions in these regions to define stationary-to-stationary parts of the metric functions that give rise to the displacement and higher memory signals in this context.
We will use these expressions to compute the displacement and higher memory signals from BBH mergers in the subsequent parts of the paper.

The remainder of the paper specializes to BBH systems.
Section~\ref{sec:computingMemory} contains four parts which describe why we focus on nonoscillatory moments of the strain for our forecasts (Sec.~\ref{subsec:nonosc}), what simplifications arise for the nonoscillatory modes of precessing BBH systems (Sec.~\ref{subsec:signals_nonprecessing}), how accurately the stationary-to-stationary approximation represents the offsets in the moments of the news (Sec.~\ref{subsec:accuracy}), and examples of the oscillatory and memory signals for a binary like the GW150914 event~\cite{LIGOScientific:2016aoc} (Sec.~\ref{subsec:higherMemoryBBH}).
The methods for our forecasts are presented in Sec.~\ref{sec:forecastMethods}, which include details about how we model the BBH populations and compute the detection forecasts.
The results of these forecasts are given in Sec.~\ref{sec:results}, 
and our conclusions are presented in Sec.~\ref{sec:conclusions}.
Three technical appendices contain more results related to the stationary-to-stationary approximation, properties of spin-weighted and tensor spherical harmonics, and relationships between Bondi metric functions and Newman-Penrose scalars.

We use geometric units with $G = c = 1$ throughout this paper.
We use the metric and curvature-tensor conventions in~\cite{Wald:1984rg}.

\section{Review of the Bondi-Sachs metric and methods for computing memory effects} \label{sec:review}

In this section, we will review some aspects of the Bondi-Sachs formalism.
We focus on the conservation and evolution equations in these coordinates and on how these equations can be written in a ``charge-flux'' form.

\subsection{Bondi-Sachs formalism} \label{subsec:BMSformalism}

Our notation and conventions for the Bondi-Sachs metric and the corresponding Einstein equations closely follows that used in~\cite{Siddhant:2024nft} (though see also, e.g.,~\cite{Bondi:1962px,Sachs:1962wk,Barnich:2010eb,Flanagan:2015pxa,Grant:2021hga} upon which~\cite{Siddhant:2024nft} built).

Bondi coordinates, $\{u,r,\theta^A\}$, are adapted to spacetimes containing outgoing null radiation. 
The variable $u$ is a retarded time coordinate, $\theta^A$ are the angular coordinates which are held fixed along the outgoing null rays, and $r$ is an areal radius (namely, surfaces of constant $u$ and $r$ have area $4\pi r^2$).
These descriptive conditions defining Bondi coordinates can be summarized in four quantitative gauge conditions,
\begin{equation}
    g_{rr} = g_{rA} = 0, \qquad \det[g_{AB}]=r^4 q(\theta^A) ,
\end{equation}
where $q(\theta^A)$ is the determinant of the round 2-sphere metric $q_{AB}$.
In these coordinates, the Bondi-Sachs metric can be written as
\begin{align} \label{eqn:BMS metric}
    \ud s^2 =& -\left(1 - \frac{2V}{r}\right) e^{2 \beta/r} \ud u^2 - 2 e^{2 \beta/r} \ud u \ud r 
    \nonumber\\
    & + r^2 \mathcal{H}_{AB} \left(\ud \theta^A - \frac{\mathcal{U}^A}{r^2} \ud u\right) \left(\ud\theta^B - \frac{\mathcal{U}^B}{r^2} \ud u\right).
\end{align}
Neither asymptotically flat boundary conditions nor Einstein's equation have yet been imposed upon the metric in Eq.~\eqref{eqn:BMS metric}.
We next discuss the consequences of applying these conditions.

For Bondi coordinates to be asymptotically flat and inertial, the Bondi metric functions $V$, $\beta$, $\mathcal U^A$, and $\mathcal H_{AB}$ must fall off with appropriate powers of $1/r$.
Furthermore, it is frequently assumed that the expansion of these functions can be written as a series in powers of $1/r$: namely, there are no terms in the series involving powers of the natural logarithm of $r$ (such an assumption is not necessary, but it does seem to encompass the class of solutions that include BBH mergers\footnote{There also exists literature on the extension of this analysis to the case where logarithms are allowed, yielding a so-called ``polyhomogeneous'' expansion, see for example~\cite{winicour1985logarithmic, Chrusciel:1993hx, Geiller:2024ryw}.
Moreover, there appears to be some evidence that, at least to properly capture the behavior of BBH mergers in the infinite past, such logarithmic terms may be necessary~\cite{Kehrberger:2021uvf, Kehrberger:2021vhp, Kehrberger:2021azo}.}).
The factors of $r$ and its inverse in the metric functions in Eq.~\eqref{eqn:BMS metric} were chosen so that the leading-order terms approach $r$-independent functions of $u$ and $\theta^A$ in this expansion when asymptotically flat boundary conditions are applied (for $\beta$, imposing Einstein's equation will cause the leading-order term to vanish).
For the symmetric tensor $\mathcal H_{AB}$, its expansion will be given by
\begin{equation} \label{eqn: mexp H}
\mathcal{H}_{AB} = \sqrt{1+\frac{\mathcal{C}_{CD}\mathcal{C}^{CD}}{2r^2}} q_{AB} + \frac{1}{r} \mathcal{C}_{AB}(u,r,\theta^C). 
\end{equation}
The function is grouped into the sum of two terms: the first is proportional to the round metric on the 2-sphere, $q_{AB}$, and the second term has vanishing trace with respect to $q_{AB}$.
The expression multiplying $q_{AB}$ in the first term ensures that the determinant condition $\det(\mathcal H_{AB}) = q(\theta^C)$ is satisfied to all orders in $1/r$.
The tensor $\mathcal{C}_{AB}$ admits an expansion of the form (see, e.g.,~\cite{Grant:2021hga}):
\begin{equation} \label{eqn: mexp shear}
    \mathcal{C}_{AB} = C_{AB}(u,\theta^C) + \frac{1}{r^2}\sum_{n = 0}^\infty \frac{1}{r^n} \ord{n}{\mathcal E}_{AB}(u,\theta^C) \, .
\end{equation}
The 2-sphere tensor $C_{AB}$ is called the ``shear'' tensor, and it contains information about the two polarizations of the transverse-traceless gravitational waves.
No $1/r$ term was included in Eq.~\eqref{eqn: mexp shear}, because Einstein's equations imply that if such a term were present, then terms involving natural logarithms of $r$ (divided by powers of $r$) would be required in the expansion of other Bondi metric functions.

The different components of Einstein's equations in Bondi coordinates take the form of ``hypersurface,'' ``conservation,'' and ``evolution'' equations, respectively~\cite{Madler:2016xju}.
The vacuum hypersurface equations (specifically, those with $R_{r\mu}=0$) do not involve time derivatives, and they take the form of first-order partial differential equations in $r$.
They can be integrated on each surface of constant $u$, and, given the asymptotically flat and power-series in $1/r$ ans\"atze, the solutions can be written in terms of two ``functions of integration'' $\mass(u,\theta^A)$ and $N_A(u,\theta^B)$ and the Bondi metric functions in $\mathcal H_{AB}$:
\begin{subequations}
\label{eqn: mexp}
\begin{align}
    \label{eqn:mexp beta}
    \beta = {} & \frac 1r \tilde{\beta}(u,r,\theta^A) ,  \\
    \label{eqn: mexp V}
    V = {} & \mass(u,\theta^A) + \frac 1r \mathcal{M}(u,r,\theta^A) ,  \\
    \mathcal{U}^A = {} & -\frac 12 \mathscr{D}_B C^{AB} - \frac 1r \biggr[N^A(u,\theta^B)- \frac1{16} \mathscr{D}^A(C_{BC}C^{BC})\nonumber\\
    \label{eqn: mexp U}
    & - \frac1{2} C^{AB} \mathscr{D}^C C_{BC}\biggr]+ \frac 1{r^2} \Upsilon^A(u,r,\theta^B) .
\end{align}
\end{subequations}
The function, $\mass$, is the Bondi mass aspect and $N_A$ is the Bondi angular-momentum aspect.
The operator $\mathscr{D}_A$ is the covariant derivative compatible with the metric $q_{AB}$.
The remaining functions ($\tilde{\beta}$, $\mathcal{M}$, and $\Upsilon^A$) are (nonlinear) functions of the shear, mass aspect, angular-momentum aspect, higher Bondi aspects ($\ord{n}{\mathcal E}_{AB}$) and their derivatives.
Their forms are known up to a few orders in the expansion in $1/r$ (see, e.g.,~\cite{Flanagan:2015pxa,Madler:2016xju,Grant:2021hga}), but we do not give their full expressions in this paper.

Our focus will be on the conservation and evolution equations in the Bondi formalism.
The evolution equation for the symmetric, trace-free (STF) part of $R_{AB}=0$ (in vacuum) gives no constraints on the shear at leading order in $1/r$ [it reduces to a ``trivially satisfied'' equation $\partial_r (\partial_u C_{AB}) = 0$].
The interpretation of this result is that the $u$ derivative of the shear $C_{AB}$ is ``free data'' which gets called the ``news tensor'' and will be denoted by $N_{AB} = \partial_u C_{AB}$.
It vanishes in the absence of gravitational waves, and its presence indicates the existence of gravitational radiation in the spacetime.
The higher-order in $1/r$ parts of the evolution equations in the STF part of $R_{AB}=0$ give rise to the evolution equations for the higher Bondi aspects $\ord{n}{\mathcal E}_{AB}$.
For this paper, we will need the evolution equation for the first (quadrupole) Bondi aspect, which is given by
\begin{align}
\label{eqn:dot_E0}
    \partial_u \ord{0}{\mathcal E}_{AB} = {} & \frac{1}{4} N_{CD} C^{CD} C_{AB} + \frac{1}{3} \STF \mathscr{D}_A N_B \nonumber\\
    &+ \frac{1}{4} C_B{}^C \mathscr{D}_{[A} \mathscr{D}^D C_{C]D} + \frac{1}{2} \mass C_{AB} .
\end{align}
Note that if we introduce the left dual of a tensor on the 2-sphere,
\begin{equation}
    (^* T)_{A_1 \cdots A_s} \equiv \epsilon^B{}_{A_1} T_{BA_2 \cdots A_s} ,
\end{equation}
the term $2 C_B{}^C \mathscr{D}_{[A} \mathscr{D}^D C_{C]D}$ can be rewritten as $[\mathscr D_C \mathscr D_D (^*C)^{CD}](^*C)_{AB}$.
We also will use the second (octopole) Bondi aspect's evolution:\footnote{Note that the equivalent expression in~\cite{Grant:2021hga} had the $\STF$ acting on the two indices outside of the angular derivative on the second line.
However, these two expressions are equal due to an identity that holds for two-dimensional manifolds.}
\begin{subequations}
\begin{equation} \label{eqn:dot_E1}
    \begin{split}
        \partial_u \ord{1}{\mathcal E}_{AB} = &- \frac{1}{4}(\mathscr{D}^2 +2) \ord{0}{\mathcal E}_{AB} \\
        &+ \mathscr D^C \STF \left[L_A C_{BC} + \frac{5}{32} C_{DE} C^{DE} \mathscr D_A C_{BC}\right],
    \end{split}
\end{equation}
where $L_A$ is given by
\begin{equation} \label{eq:LA}
    \begin{split}
        L_A \equiv -\frac{1}{3} N_A + \frac{1}{4} \left[C_{AD} \mathscr{D}_E C^{DE} - \frac{3}{8} \mathscr{D}_A (C_{DE} C^{DE})\right].
    \end{split}
\end{equation}
\end{subequations}

The evolution of the mass and angular-momentum aspects can be obtained from the vacuum Einstein-equation components $R_{uu} = 0$ and $R_{uA}=0$, which are equivalent to certain components of the contracted Bianchi identities (specifically the $\beta=u$ and $\beta=A$ components of $\nabla_\mu {R^\mu}_\beta =0$, in vacuum).
Their evolution equations are given by
\begin{subequations}
\label{eqn:Evol}
\begin{align}
    \label{eqn:dot_m}
    \partial_u \mass &= \frac{1}{4} \mathscr{D}_A \mathscr{D}_B N^{AB} - \frac{1}{8} N_{AB} N^{AB} ,  \\
    \label{eqn:dot_N}
    \partial_u N_A &= \mathscr{D}_A \mass - \frac{1}{4} \epsilon_{BA} \mathscr{D}^B (\epsilon^{CD} \mathscr{D}_C \mathscr{D}^E C_{DE})\nonumber\\
    &+ \frac{1}{4} \left(N^{BC} \mathscr{D}_B C_{CA} + 3 C_{AB}\mathscr{D}_C N^{BC}\right) .
\end{align}
\end{subequations}

\subsection{Charge-flux form of the Bondi-Sachs equations} \label{subsec:chargeFlux}

As was described in~\cite{Grant:2021hga}, the Bondi-Sachs conservation and evolution equations in Sec.~\ref{subsec:BMSformalism} can be cast in a form that resembles the evolution of charges due to fluxes and pseudofluxes.
The evolution of the mass aspect is the canonical case, as it is a scalar equation which precisely describes the evolution of the supermomentum charge arising from the flux of gravitational waves.
The evolution equations for the angular momentum aspect and higher Bondi aspects can also be written in a charge-flux form.
There is not a unique prescription to split these equations into charge and flux contributions; we give our conventions for this split below.
The charge and flux terms are naturally expressed as 2-sphere scalars, vectors, and tensors; however, the vector and tensor terms can also be written in terms of scalars defined from divergences of the corresponding vectors and tensors (and their duals).
We review these notions of charges and fluxes below.

\subsubsection{2-sphere scalar, vector, and tensor charges and fluxes}

For the mass aspect, we define a flux
\begin{subequations} \label{eq:mdotDef}
\begin{equation} \label{eq:F0}
    \mathcal{F}_0 = -\frac{1}{8}N_{AB}N^{AB} ,
\end{equation}
so that its evolution equation is given by
\begin{equation} \label{eq:mdotF}
  \partial_u \mass  = \frac{1}{4} \mathscr D_A \mathscr D_B N^{AB} + \mathcal F_0 .
\end{equation}
\end{subequations}
To write the evolution equation for the angular-momentum aspect, we first define the modified angular-momentum aspect
\begin{subequations} \label{eq:hatNAdotDef}
\begin{align}
    \hat{N}^A = {} & N^A - \mathcal W^A , \\ 
    \label{eq:calWa}
    \mathcal W^A = {} & \frac{1}{8}(C_B{}^C \mathscr{D}_C C^{AB} + 3 C^{AB}\mathscr{D}^C C_{BC}) .
\end{align}
It has the form of the integrand in the Dray-Streubel~\cite{Dray:1984rfa} (or Wald-Zoupas~\cite{Wald:1999wa}) charge.
The evolution equation for $\hat N^A$ takes the form
\begin{equation} \label{eq:hatNAdotF}
    \partial_u \hat N^A = \mathscr{D}^A \mass - \frac{1}{4} \epsilon^{BA} \mathscr{D}_B (\epsilon^{CD} \mathscr{D}_C \mathscr{D}^E C_{DE}) + \mathcal F_1^A ,
\end{equation}
where there is a contribution from a charge (specifically $\mathscr D^A \mass$) and a flux term,
\begin{align} \label{eq:F1A}
    \mathcal F_1^A = {} & \frac{1}{8} \left[(N_{BC}\mathscr{D}^B C^{CA}-C_{BC}\mathscr{D}^B N^{CA}) \right. \nonumber \\
    & \left. + 3 (C^{AB}\mathscr{D}^C N_{BC}- N^{AB}\mathscr{D}^C C_{BC})\right] .
\end{align}
\end{subequations}

For the higher Bondi aspects, we define the aspects $\ord{n}{\mathcal E}_{AB}$ to be the charges.
We write the evolution equations for the first two higher Bondi aspects in terms of fluxes (which vanish in the absence of radiation) and pseudofluxes (which contain the shear, but do not necessarily vanish in the absence of radiation).
The evolution of the first (quadrupole) higher Bondi aspect, when expressed in terms of $\hat N^A$, can be written as
\begin{subequations} \label{eq:EABdotDef}
\begin{equation} \label{eq:EABdotF}
    \partial_u \ord{0}{\mathcal E}^{AB} = \frac 13 \STF \mathscr D^A \hat N^B + \mathcal F_2^{AB} + \mathcal G_2^{AB} , 
\end{equation}
where there is a charge contribution proportional to $\STF \mathscr D^A \hat N^B$ and flux and pseudoflux contributions given by
\begin{align} \label{eq:F2AB}
    \mathcal F_2^{AB} = {} & \frac 14 (C_{CD} N^{CD}) C^{AB} , \\
    \label{eq:G2AB}
    \mathcal G_2^{AB} = {} & \frac 12 \mass C^{AB} + \frac 18 [\mathscr D_C \mathscr D_D (^*C)^{CD}](^*C)^{AB} \nonumber \\
    & + \frac 13 \STF \mathscr D^A \mathcal W^B .
\end{align}
\end{subequations}

The evolution of the second (octopole) higher Bondi aspect has a vanishing flux, leaving just charge and pseudoflux terms.
These can be written as
\begin{subequations} \label{eq:EAB1dotDef}
    \begin{equation} \label{eq:EAB1dotG}
        \partial_u \ord{1}{\mathcal E}^{AB} = - \frac 14 (\mathscr D^2 + 2) \ord{0}{\mathcal E}^{AB} + \mathcal G_3^{AB} ,
    \end{equation}
    where the pseudoflux is given by 
    \begin{equation} \label{eq:G3AB}
        \mathcal G_3^{AB} = \mathscr{D}_C \STF \left[L^A C^{BC} + \frac{5}{32} C_{DE} C^{DE} \mathscr D^A C^{BC}\right].
    \end{equation}
\end{subequations}
We will not need the $O(C^3)$ pieces of the quantity $L^A$ defined in Eq.~\eqref{eq:LA} in Secs.~\ref{sec:computingMemory}--\ref{sec:results}, but we keep this expression in terms of $L^A$ (instead of $N^A$) for completeness.

\subsubsection{Electric- and magnetic-parity scalar charges}

There are two degrees of freedom in the angular-momentum aspect, and each higher Bondi aspect contains two degrees of freedom as well.
These Bondi aspects can be decomposed into electric- and magnetic-parity parts using a Helmholtz-Hodge decomposition.
Their evolution equations admit such a decomposition, which allows for all the evolution and conservation Bondi-Sachs equations to be written in terms of the evolution of scalar charges and corresponding fluxes (and, for the higher Bondi aspects, pseudofluxes as well).
This approach was used in~\cite{Siddhant:2024nft}, and we review it below.

For the electric charges, we make the definitions
\begin{subequations}
\begin{equation} \label{eqn:Def Echarges}
    \E{Q_0} \equiv \mass, \quad \E{Q_1}\equiv \mathscr{D}_A \hat N^A, \quad \E Q_{n+2} \equiv \mathscr{D}_A \mathscr{D}_B \ord{n}{\mathcal E}^{AB}. 
\end{equation}
For the fluxes and pseudofluxes, we define
\begin{align} \label{eqn:FGscalars}
    & \E{\mathcal F_0} \equiv \mathcal F_0, \qquad \E{\mathcal F_1}\equiv \mathscr{D}_A \mathcal F_1^A, \nonumber\\
    & \E{\mathcal F}_{n+2} \equiv \mathscr{D}_A \mathscr{D}_B {\mathcal F}_{n+2}^{AB} \qquad \E{\mathcal G}_{n+2} \equiv \mathscr{D}_A \mathscr{D}_B {\mathcal G}_{n+2}^{AB}. 
\end{align}
\end{subequations}
The corresponding magnetic charges are defined to be
\begin{subequations}
\begin{align} \label{eqn:Def Bcharges}
    & \B Q_0 \equiv - \frac{1}{4} \mathscr{D}_A \mathscr{D}_B (^*C)^{AB}, \qquad 
    \B Q_1 \equiv  \mathscr{D}_A (^*\hat{N})^A, \nonumber\\
    & \B Q_{n+2} \equiv \mathscr{D}_A \mathscr{D}_B (^*\ord{n}{\mathcal E})^{AB},
\end{align}
For the magnetic case, the zeroth flux $\mathcal{F}_0$ is trivial and the nonzero fluxes and pseudofluxes are given as:
\begin{align} \label{eqn:FGscalarsB}
    &  \B{\mathcal F_1}\equiv \mathscr{D}_A (^*\mathcal F_1)^A, \qquad\B{\mathcal F}_{n+2} \equiv \mathscr{D}_A \mathscr{D}_B (^*\mathcal F_{n+2})^{AB}, \nonumber\\
    & \E{\mathcal G}_{n+2} \equiv \mathscr{D}_A \mathscr{D}_B (^*\mathcal G_{n+2})^{AB}. 
\end{align}
\end{subequations}
The evolution equation for the mass aspect is given by
\begin{subequations} \label{eqn:Q_n_PDEs}
\begin{equation} \label{eqn: evol Q_0}
  \E {\dot Q_0}  = \frac{1}{4} \mathscr D_A \mathscr D_B N^{AB} + \E{\mathcal F}_0 .
\end{equation}
For the magnetic charge, $\B{\dot Q_0}$ is the trivial equation
\begin{equation} \label{eqn: evol Q_0 mag}
  \B{\dot Q_0} = - \frac{1}{4} \mathscr{D}_A \mathscr{D}_B (^*N)^{AB} . 
\end{equation}
The evolution of the electric and magnetic  charges with with $n \geq 1$ have the same form and are given as
\begin{align}
    \p{\dot Q_1} &= \mathscr{D}^2 \p Q_0 + \p{\mathcal F_1} , \label{eqn: evol Q_1} \\
    \p{\dot Q_2}&= \frac{1}{6}(\mathscr D^2+2) \p Q_1 + \p{\mathcal F_2} + \p{\mathcal G_2}, \label{eqn: evol Q_2} \\
    \p{\dot Q_3}&= -\frac{1}{4}(\mathscr D^2+6) \p Q_2 + \p{\mathcal G_3}. \label{eqn: evol Q_3}
\end{align}
\end{subequations}
Here, we introduce the parity label $(p)$, which can take either $(e)$ or $(b)$ as its value.

\subsection{Higher memory signals and moments of the news tensor} \label{subsec:higherMemory}

In this part, we review the integral form of the Bondi-Sachs evolution equations written in terms of the temporal moments of the news tensor.
We then discuss the corresponding parts of the GW signals that are responsible for a given offset in a moment of the news.

\subsubsection{Bondi-Sachs equations and moments of the news}

The higher memory signals are associated with the temporal moments of the news tensor.
As in~\cite{Siddhant:2024nft}, we use the following definition of the $n^\mathrm{th}$ moment of the news:\footnote{These moments are functions of angles on the 2-sphere. For brevity, we will not write the angular dependence explicitly.}
\begin{equation}
    \ord{n}{\mathcal{N}}_{AB}(u_1, u_0)= \int_{u_0} ^{u_1}  \! \ud u_2 \, \cdots \int_{u_0} ^{u_n} \ud u_{n+1} \;N_{AB}(u_{n+1}) .
\end{equation}
These moments can be related to integrals of the fluxes and changes in charges by repeatedly integrating the Bondi-Sachs conservation and evolution equations.
To obtain these equations, we first split the the two degrees of freedom in the news tensor into its electric- and magnetic-parity part as:\footnote{The electric and magnetic piece of the news only have $l \geq 2$ modes because the news is a rank-2 STF tensor on 2-spheres of constant $u$.}
\begin{equation}
    \E{\ord{n}{\mathcal{N}}} = \mathscr{D}^A \mathscr D^B {\ord{n}{\mathcal{N}}_{AB}} , \qquad 
    \B{\ord{n}{\mathcal{N}}} = \mathscr{D}^A \mathscr D^B (^*{\ord{n}{\mathcal{N}})_{AB}} .
\end{equation}
The zeroth moment can be obtained from integrating the evolution of the mass aspect in Eq.~\eqref{eqn: evol Q_0},
\begin{align} \label{eqn: zeroth moment}
    \frac{1}{4} \E{\ord{0}{\mathcal{N}}}(u_1, u_0) =  \Delta \E Q_0(u_1, u_0) - \int_{u_0} ^{u_1} \ud u\;\mathcal F_0 .
\end{align}
The notation $\Delta \E Q_0(u_1, u_0) \equiv \E Q_0(u_1) - \E Q_0(u_0)$ denotes the change between retarded times $u_1$ and $u_0$ (and this ``$\Delta$'' notation will be used to denote such differences for other quantities below).
The first term on the right is the charge part and the second term is the flux part.

By algebraically manipulating and integrating the conservation equations~\eqref{eqn:dot_m} and~\eqref{eqn:dot_N}, one can obtain the first moment of the news in terms of its charge and flux pieces (see~\cite{Siddhant:2024nft} for further details). 
The electric- and magnetic-parity pieces of the first moments, respectively, are given by
\begin{subequations} \label{eqn:first moment}
\begin{align} \label{eqn:first moment elec}
    \frac{1}{4}\mathscr{D}^2 \E{\ord{1}{\mathcal{N}}} (u_1, u_0) = {} & \Delta \E{Q}_1(u_1, u_0)- (u_1-u_0)\mathscr{D}^2 \E Q_0(u_0)  \nonumber\\
    & - \mathscr{D}^2 \int_{u_0} ^{u_1} \ud u_2 \int_{u_0} ^{u_2} \ud u_3 \mathcal{F}_0(u_3) \nonumber \\
    & -\int_{u_0} ^{u_1} \ud u_2 \mathcal{F}_1(u_2) , \\
    \label{eqn:first moment mag}
    \frac{1}{4} \mathscr{D}^2 \B{\ord{1}{\mathcal{N}}}(u_1, u_0) = & -\Delta \B{{Q}}_1 (u_1, u_0) + (u_1-u_0)\mathscr{D}^2 \B Q_0(u_0)\nonumber\\
    &+\int_{u_0} ^{u_1} \ud u_2\B{{\mathcal F}_1}  (u_2). 
\end{align}
\end{subequations}
The electric- and magnetic-parity part of the first moment contains the CM and spin memory, respectively. 
In Eq.~\eqref{eqn:first moment elec}, the first line on the right-hand side is the charge part of the CM memory.
The second is the integral of the displacement memory, and the third is the flux contribution to the CM memory.
In Eq.~\eqref{eqn:first moment mag}, both charge and flux terms on the right-hand side contribute to the spin memory.

Following a similar procedure, one can obtain the second moment of news using Eqs.~\eqref{eqn:dot_E0}, \eqref{eqn:dot_m}, and~\eqref{eqn:dot_N} (further details again are given in~\cite{Siddhant:2024nft}):
\begin{widetext}
\begin{subequations} \label{eq:secondMoments}
\begin{align}
    \frac{1}{24} (\mathscr{D}^2 +2)\mathscr{D}^2 \E {\ord{2}{\mathcal N}} (u_1, u_0) = {} & \Delta \E Q_2(u_1, u_0) - (u_1-u_0)\frac{1}{6}(\mathscr{D}^2+2) \E Q_1(u_0)- \frac{1}{12}(u_1-u_0)^2 \mathscr{D}^2(\mathscr{D}^2+2) \E Q_0(u_0)\nonumber \\
    &  -\frac{1}{6} (\mathscr{D}^2+2) \int_{u_{0}} ^{u_1} \ud u_2\int_{u_{0}} ^{u_2}   \ud u_3 \bigg[\E{\mathcal{F}_1}(u_3) + \mathscr{D}^2 \int_{u_{0}} ^{u_3} \ud u_4 \mathcal{F}_0(u_4)\bigg] \nonumber \\
    & - \int_{u_{0}}^{u_1} \ud u_2 \bigg[\E{\mathcal{F}_2}(u_2) + \E{\mathcal{G}_2} (u_2)\bigg] ,\label{eqn: second moment elec} \\
     \frac{1}{24} (\mathscr{D}^2 +2)\mathscr{D}^2 \B{\ord{2}{\mathcal N}}(u_1, u_0) = & -\Delta \B Q_2 (u_1, u_0)  + (u_1-u_0)\frac{1}{6}(\mathscr{D}^2+2) \B Q_1(u_0)+ \frac{1}{12}(u_1-u_0)^2 \mathscr{D}^2(\mathscr{D}^2+2) \B Q_0(u_0)\nonumber \\
     &+ \int_{u_{0}} ^{u_1} \ud u_2 [\B{\mathcal{F}_2}  (u_2) + \B{\mathcal{G}_2}(u_2)]+\frac{1}{6} (\mathscr{D}^2+2) \int_{u_{0}} ^{u_1} \ud u_2\int_{u_{0}} ^{u_2}   \ud u_3 \B{\mathcal{F}_1} (u_3).  \label{eqn:second moment mag}
\end{align}
\end{subequations}
\end{widetext}
For the second moment, the charge piece depends on the first higher Bondi aspect (as well as pieces linear and quadratic in retarded time that are proportional to the angular derivatives of the initial zeroth and first charges), and it contributes to the offset in the second moment.
There also are both fluxes and pseudofluxes that can contribute to the constant offset in the second moment. 

Below we focus on the effects of the different charge and flux terms on the GW strain, because CE is calibrated to measure the strain rather than the higher moments of the news.

\subsubsection{Electric-parity displacement and higher memory signals}

When written in terms of the change in the shear rather than the zeroth moment of the news, Eq.~\eqref{eqn: zeroth moment} has the form
\begin{equation} \label{eq:DeltaCab}
    \frac 14 \mathscr{D}_A \mathscr{D}_B \Delta C^{AB}(u,u_0) = \Delta \E Q_0(u,u_0) - \int_{u_0}^u du' \E{\mathcal F}_0 .
\end{equation}
While Eq.~\eqref{eq:DeltaCab} or~\eqref{eqn: zeroth moment} is just an integral form of Eq.~\eqref{eqn: evol Q_0}, we will use it to approximately compute a memory signal by specifying portions of the full waveform that enter into $\E{\mathcal F}_0$ and portions of a charge $\E Q_0$ that generate a nonoscillatory memory signal in $\Delta C_{AB}$.
More details about how we choose the relevant data to use in this procedure will be given in Secs.~\ref{sec:stationary} and~\ref{sec:computingMemory}.

One can obtain an alternate constraint on the shear by solving for $\E Q_0$ in Eq.~\eqref{eq:DeltaCab} and substituting the result into Eq.~\eqref{eqn: evol Q_1}, or appropriately differentiating Eq.~\eqref{eqn:first moment elec}.
Doing so gives
\begin{align} \label{eq:DeltaCabQ1}
    \frac 14 \mathscr D^2 \mathscr{D}_A \mathscr{D}_B \Delta C^{AB}(u,u_0) = {} & \E{\dot Q}_1(u) - \mathscr D^2 \E Q_0(u_0) \nonumber \\
    & - \mathscr D^2 \int_{u_0}^u du' \E{\mathcal F}_0 - \E{\mathcal F}_1(u).
\end{align}
The second term $\E{\mathcal F}_1(u)$ on the second line vanishes in nonradiative regions; therefore, it will not produce an offset in $\Delta C_{AB}$.
When it is integrated, as in Eq.~\eqref{eqn:first moment elec}, it contributes to the first moment of the news (as noted above).
There is a portion of the waveform associated with this offset in the first moment of the news, which is what we will define to be the corresponding GW signal associated with this moment of the news.
The charge term $\E{\dot Q}_1(u) - \mathscr D^2 \E Q_0(u_0)$ on the first line generically will have terms that lead to an offset in both the shear and the first moment of the news.
We will discuss how we compute the GW signal associated with this charge contribution in Sec.~\ref{sec:stationary}.
Finally, the flux term at the start of the second line of Eq.~\eqref{eq:DeltaCabQ1} contains an equivalent constraint for computing the memory as in Eq.~\eqref{eq:DeltaCab}, because the operator $\mathscr D^2$ removes the $l=0$ spherical harmonic in the flux; however, the $l=0$ harmonic does not contribute to the memory, which contains only $l\geq 2$ spherical harmonics.\footnote{\label{fn:momentCaveat}It is conceivable that there could be specific terms in the waveform such that the integrand $\E{\mathcal F}_0$ is a total derivative and the value at the endpoints vanishes, which would produce an offset in its integral (the first moment of the news).
Unlike the charge terms which do produce offsets in the zeroth and first moments of the news, we have not identified such behavior in the flux terms, which is why we discuss them as contributing to just a single moment of the news (specifically the highest moment at which they can contribute).
The moment of the news to which the pseudoflux terms contribute is more subtle, and it will be discussed in more detail in Sec.~\ref{sec:stationary}.}

Differentiating Eq.~\eqref{eqn: evol Q_2} gives an equation that involves the derivative of $\E Q_1$.
With Eq.~\eqref{eq:DeltaCabQ1}, the derivative of the evolution of $\E Q_2$ gives an alternate expression for the offset in the shear:
\begin{align}\label{eq:DeltaCabQ2}
    \frac 14 (\mathscr D^2 + 2)\mathscr D^2 \mathscr{D}_A \mathscr{D}_B \Delta C^{AB} &= 6\E{\ddot Q}_2 - (\mathscr D^2 + 2) \mathscr D^2 \E Q_0(u_0) \nonumber\\
    &- (\mathscr D^2 + 2) \E{\mathcal F}_1 - 6 \E{\dot{\mathcal G}}_2 - 6 \E{\dot{\mathcal F}}_2\nonumber\\
    &- (\mathscr D^2 + 2) \mathscr D^2 \int_{u_0}^u du' \E{\mathcal F}_0 
     .
\end{align}
For brevity, we have dropped the $u$ dependence of all terms except $\E Q_0(u_0)$, because it is the one term that depends only on $u_0$.
This result can also be obtained from appropriately differentiating Eq.~\eqref{eqn: second moment elec} twice.
The first two terms on the right-hand side contain the charge term $6\E{\ddot Q}_2 - (\mathscr D^2 + 2) \mathscr D^2 \E Q_0(u_0)$, which will generically have terms that produce an offset in the zeroth, first, and second moments of the news.
The third is the flux term that arises in Eqs.~\eqref{eq:DeltaCab} and~\eqref{eq:DeltaCabQ1}.
Again, it gives an equivalent constraint on the memory signal to those in Eqs.~\eqref{eq:DeltaCab} and~\eqref{eq:DeltaCabQ1}, because the operator $(\mathscr D^2+2)\mathscr D^2$ removes the $l=0$ and $l=1$ spherical harmonics in the flux, but these terms do not contribute to the memory which contains just $l\geq 2$ spherical harmonics.
Similarly, the fourth term $(\mathscr D^2+2) \E{\mathcal F}_1$ produces an equivalent memory signal associated with the first moment to that discussed in Eq.~\eqref{eq:DeltaCabQ1}, because the operator $(\mathscr D^2+2)$ removes the $l=1$ spherical harmonics from $\E{\mathcal F}_1$, which do not contribute to the memory signal (see also the discussion in Footnote~\ref{fn:momentCaveat}).

The ``new'' terms are the final two terms in Eq.~\eqref{eq:DeltaCabQ2}.
As noted above, the pseudoflux $\E{\mathcal G}_2$ does not vanish in a nonradiative region, but its first derivative does.
Thus, generically the term $-6\E{\dot{\mathcal G}}_2$ will produce an offset in the first moment of the news, and there will also be a part that contributes to the second moment.
We discuss how we separate these two parts in Sec.~\ref{sec:stationary}.
Finally, because $\E{\mathcal F}_2$ vanishes in a non-radiative region, the term $-6 \E{\dot{\mathcal F}}_2$ will produce an offset in the second moment of the news (second integral of the change in the shear).
A portion of the charge term will also contribute to the second moment of the news.

It is perfectly valid to keep iterating this procedure of differentiating the evolution equations for the higher-order Bondi aspects and writing the resulting equation as a change in time derivatives of higher-order charges and an increasing sum of fluxes and pseudofluxes.
However, as noted in~\cite{Grant:2021hga,Siddhant:2024nft}, the angular operators that act on the $n^\textrm{th}$-order Bondi aspect in the evolution equations set to zero the $l=n+2$ multipole moments.
Therefore, for the quadrupole moment of the strain, no other terms besides those in Eq.~\eqref{eq:DeltaCabQ2} contribute to the higher memory effects.
As a practical matter, too, the higher multipoles and higher fluxes for comparable-mass BBH sources typically are smaller than the lower multipoles and fluxes.
For this reason, we will stop at the second moment of the news and the evolution equation for the quadrupole aspect.
We will, however, use the next higher evolution equation for the octopole Bondi aspect in the approximation discussed in Sec.~\ref{sec:stationary}.

\subsubsection{Magnetic-parity higher memory signals}

A similar procedure can be applied to the magnetic part of the shear.
We can start from Eq.~\eqref{eqn: evol Q_1}, and we substitute in the expression for $\B Q_0$ given in Eq.~\eqref{eqn:Def Bcharges}, or, alternately, differentiate Eq.~\eqref{eqn:first moment mag}.
The resulting expression implies that the instantaneous value of the magnetic part of the shear (rather than its change) must satisfy
\begin{equation} \label{eq:starCABQ1}
    \frac{1}{4} \mathscr{D}^2 \mathscr{D}_A \mathscr{D}_B (^*C)^{AB} = -\B{\dot Q}_1 + \B{\mathcal F}_1 .
\end{equation}
Note that because every quantity in Eq.~\eqref{eq:starCABQ1} is a function of $u$, we drop the $u$ dependence in all equations in this part.
While stress-energy sources have been proposed that produce an offset in the magnetic part of the shear (see, e.g.,~\cite{Satishchandran:2019pyc}), they would not arise from the flux term $\B{\mathcal F}_1$, because it vanishes in nonradiative regions.
The charge contribution $\B{\dot Q}_1$, in principle, could be nonvanishing in nonradiative regions, which would produce a magnetic-parity contribution to the zeroth moment.
As we review in Sec.~\ref{subsec: st Bondi func}, in regions that are stationary, this term vanishes and does not contribute to the zeroth moment (namely, they have zero magnetic-parity memory).
We discuss in Sec.~\ref{subsec:accuracy} the extent to which the infinite past and future of BBH systems approximately satisfy this condition of stationarity.

Differentiating the evolution equation for $\B{\dot Q}_2$ with respect to $u$ and using Eq.~\eqref{eq:starCABQ1} gives 
\begin{align} \label{eq:starCABQ2}
    \frac{1}{4} (\mathscr{D}^2+2) \mathscr{D}^2 \mathscr{D}_A \mathscr{D}_B (^*C)^{AB} = {} & - 6\B{\ddot Q}_2 + (\mathscr{D}^2 + 2) \B{\mathcal F}_1 \nonumber \\
    & + 6 \B{\dot{\mathcal G}}_2 + 6 \B{\dot{\mathcal F}}_2.
\end{align}
This expression can also be obtained from differentiating Eq.~\eqref{eqn:second moment mag} twice.
As with the electric-parity case, the charge term $- 6\B{\ddot Q}_2$ can contribute to the zeroth, first, and second moments of the news (but does not contribute to the zeroth moment in the context discussed in the previous paragraph).
The first flux term $(\mathscr{D}^2 + 2) \B{\mathcal F}_1$ is equivalent to that in Eq.~\eqref{eq:starCABQ1}, because the $l=1$ harmonics do not contribute.
The pseudoflux $6 \B{\dot{\mathcal G}}_2$ is not required to vanish in nonradiative regions when there is magnetic-parity memory, but for the nonprecessing BBH systems that we consider in this paper, it will vanish. 
In this case, both it and $6 \B{\dot{\mathcal F}}_2$ contribute to the second moment of the news and not lower moments.

\section{Multipolar expansion of the higher memory signals} \label{sec:multipole}

Computing the fluxes (and pseudofluxes) is a long, but otherwise straightforward calculation, which has been performed in several sources elsewhere (e.g.,~\cite{Nichols:2017rqr,Nichols:2018qac,Grant:2022bla,Siddhant:2024nft}) in different ways.
In this paper, we will do this by means of a multipolar expansion, which we will review in the first section.
Once these methods are established, we turn to computing full expressions for the expansions of the fluxes and pseudofluxes in the following section.
Apart from the pseudoflux $\mathcal G_3^{AB}$ (which we will use in Sec.~\ref{sec:stationary} and subsequently), these were previously computed in~\cite{Siddhant:2024nft}; here we give expressions in a slightly more concise form.

\subsection{Multipolar expansions} \label{subsec:multipoles}

Solving for the shear from Eq.~\eqref{eq:DeltaCabQ2} or~\eqref{eq:starCABQ2} requires inverting an angular differential operator, such as $(\mathscr{D}^2 + 2) \mathscr{D}^2$.
As the spherical harmonics are eigenfunctions of the Laplacian $\mathscr D^2$ on the sphere, these angular operators can be straightforwardly inverted on this basis.
Thus,  it will be convenient to expand the shear in spherical harmonics and to write the corresponding charge and flux terms as an expansion in the spherical harmonic modes.
Two types of spherical harmonics are commonly used in the multipole expansion of the shear: STF tensor spherical harmonics and spin-weighted spherical harmonics of weight $s = -2$ (see Appendix~\ref{app:harmonics} for more detail about our conventions for the harmonics). 
We will use the notation $\left(T^{\mathrm I}_{l m}\right)_{AB}$ for the STF tensor spherical harmonics (similar to the ``pure spin'' harmonics in~\cite{Thorne:1980ru} for spin two).
The additional index $\mathrm{I}$ is a discrete label ($\mathrm E$ or $\mathrm B$), which labels the parity of the harmonics (electric or magnetic).
The multipole expansion of the shear is given by
\begin{equation} \label{eqn:shear_TH_exp}
C_{AB}(u,\theta,\phi) = \sum_{l=2}^{\infty} \sum_{m=-l}^{l} \sum_{\mathrm I \in \{\mathrm E,\mathrm B\}}
C^{\mathrm I}_{l m}(u)\, \left(T^{\mathrm I}_{l m}\right)_{AB}(\theta,\phi) ,
\end{equation}
where the mode coefficients $C^{\mathrm I}_{l m}(u)$ also have an $\mathrm I = \mathrm E$ or $\mathrm B$ parity label.

For some expressions, we will also find it useful to introduce the multipolar expansion of the dual of the shear
\begin{equation} \label{eqn:shear_TH_dual}
(^*C)_{AB} = \sum_{l=2}^{\infty} \sum_{m=-l}^{l} \sum_{\mathrm I \in \{\mathrm E,\mathrm B\}}
{}^*C^{\mathrm I}_{l m} \left(T^{\mathrm I}_{l m}\right)_{AB}.
\end{equation}
As in~\cite{Siddhant:2024nft}, we introduce a Levi-Civita symbol for the parity indices by
\begin{equation}
    \epsilon^\mathrm{IJ} =
    \begin{cases}
         1 & \mathrm{I=E, \, J=B} ,\\
         -1 & \mathrm{I=B, \, J=E} ,\\
         0 & \mathrm{I=J=E,B} .
    \end{cases}
\end{equation}
Because the tensor harmonics satisfy
\begin{equation}
     \left(^*T^{\mathrm I}_{l m}\right)_{AB} = \sum_{\mathrm J \in \{\mathrm E,\mathrm B\}} \epsilon^\mathrm{IJ} \left(T^{\mathrm J}_{l m}\right)_{AB} ,
\end{equation}
it follows that the tensor harmonic modes of the dual shear are related to those of the shear by
\begin{equation}\label{eqn: dual relation}
    {}^*C^{\mathrm I}_{l m} = \sum_{\mathrm J \in \{\mathrm E,\mathrm B\}} \epsilon^\mathrm{JI} C^{\mathrm J}_{l m} .
\end{equation}

For the computations of the fluxes and pseudofluxes, once we are working on the basis of spherical harmonics, we will encounter products of spherical harmonics, which we will need to re-expand as sums of harmonics.
To perform this re-expansion, one can first write the tensor harmonics in terms of spin-weighted spherical harmonics as in Eq.~\eqref{eq:Tlm2sYlm}.
As the spin-weighted spherical harmonics are orthogonal, expanding a product of two spin-weighted spherical harmonics involves computing integrals of three spin-weighted spherical harmonics:
\begin{equation} \label{eq:Cdef3Ylm}
    \mathscr C^{lm,s's''}_{l'm'l''m''} \equiv  \int d^2\Omega (_{s}\bar Y_{lm})(_{s'}Y_{l'm'})(_{s''}Y_{l''m''}) .
\end{equation}
This integral can be expressed in terms of Wigner 3-j symbols by
\begin{widetext}
    \begin{equation} \label{eq:int3Ylm}
        \int d^2\Omega (_{s} Y_{lm})(_{s'}Y_{l'm'})(_{s''}Y_{l''m''}) = \sqrt{\frac{(2l+1)(2l'+1)(2l''+1)}{4\pi}} \begin{pmatrix} l & l' & l''\\ m & m' & m''\end{pmatrix} \begin{pmatrix} l & l' & l''\\ -s & -s' & -s''\end{pmatrix} .
    \end{equation}
\end{widetext}
The left-hand side is the spin-weighted generalization of Gaunt's integral, and the right-hand side are the corresponding generalizations of the Gaunt coefficients (see,~\cite{Gaunt1929} and~\cite[\S{}34.3]{NIST:DLMF}); for simplicity, we will refer to the $\mathscr C^{lm,s's''}_{l'm'l''m''}$ as just ``Gaunt coefficients.''

The ``selection rules'' imposed by the 3-j symbols imply that the Gaunt coefficients in Eq.~\eqref{eq:Cdef3Ylm} are nonzero when $s = s'+s''$ and $m = m'+m''$.
We do not include the $s$ index on these coefficients because it can be inferred from the $s'$ and $s''$ values.
We do, however, keep the $m$ index (even though it can also be inferred from $m'$ and $m''$) because there are fluxes with cubic terms, the multipole coefficients of which can be computed from the integral of a product of four spin-weighted spherical harmonics.
These can be written as a sum over products of the Gaunt coefficients, and this sum would be more cumbersome to write without this additional index.
The Gaunt coefficients are nonzero when the $l$ index satisfies $\max(|l'-l''|,|m|, |s|) \leq l \leq |l'+l''|$.

In several of our calculations, we will consider the case when $s = 0$ (namely, we are expanding the product in terms of scalar spherical harmonics) and thus $s'' = -s'$.
In this case, we will denote this special case of the Gaunt coefficient by
\begin{equation}
    \mathscr C^{lm,s'}_{l'm'l''m''} \equiv \mathscr C^{lm,(-s')s'}_{l'm'l''m''} 
\end{equation}
so as to make the notation slightly more compact.

We found it convenient to use these results for products of spin-weighted spherical harmonics to derive corresponding results for products of tensor harmonics.
We explicitly compute the necessary identities to do this in Appendix~\ref{app:harmonics}.
Specifically, these identities show how outer products and contractions of tensor harmonics (with scalar harmonics being a special case) can be re-expanded in terms of tensor harmonics.
The key difference between the cases for spin-weighted and tensor spherical harmonics is that there is now a parity-dependent factor $\eta^{\mathrm I \mathrm I' \mathrm I''}_{l l' l''}$ for the tensor case, which is given by
\begin{equation} \label{eq:etaCoeff}
    \eta^{\mathrm I \mathrm I' \mathrm I''}_{l l' l''} = 
    \begin{cases}
        1+(-1)^{l+l'+l''} & \mathrm{I=E, \ I'=I''=E,B} ,\\
        i[1-(-1)^{l+l'+l''}] & \mathrm{I=E, \ I'=E, \ I''=B} ,\\
        -i[1-(-1)^{l+l'+l''}] & \mathrm{I=E, \ I'=B, \ I''= E} ,\\
        i[1-(-1)^{l+l'+l''}] & \mathrm{I=B, \ I'=I''=E,B} ,\\
        -[1+(-1)^{l+l'+l''}] & \mathrm{I=B, \ I'=E, \ I''=B} ,\\
        1+(-1)^{l+l'+l''} & \mathrm{I=B, \ I'=B, \ I''=E}.
    \end{cases}
\end{equation}
Explicit expressions for these identities are given in Eq.~\eqref{eq:TlmProducts}.

In this approach, it is also helpful to have ``raising'' and ``lowering'' relationships for the gradients and divergences of the tensor harmonics.
These are naturally written in terms of coefficients
\begin{equation}
    \mathcal A_{ls} = \sqrt{\frac{(l-s)!}{(l+s)!}}, \qquad 
    \mathcal B_{ls} = \frac{\mathcal A_{ls}}{\mathcal A_{l(s+1)}};
\end{equation}
For more details, see Eqs.~\eqref{eqn:tensor_raise}--\eqref{eqn:vector_lower_zero}.
We will use $\mathcal A_{ls}$ and $\mathcal B_{ls}$ when relating the scalar-harmonic moments of the electric and magnetic scalars to corresponding moments of a tensor-harmonic expansion.
For scalars $\E S$ and $\B S$ defined in terms of a tensor $S^{A_1 \cdots A_s}$ from
\begin{subequations}
    \begin{align}
        \E S &= \mathscr D_{A_1} \cdots \mathscr D_{A_s} S^{A_1 \cdots A_s}, \\
        \B S &= \mathscr D_{A_1} \cdots \mathscr D_{A_s} (^* S)^{A_1 \cdots A_s},
    \end{align}
\end{subequations}
the definitions of the tensor harmonics in Eq.~\eqref{eq:TlmAsDef} imply that
\begin{subequations}\label{eq:tensorscalarlm}
    \begin{align} 
    S^{\mathrm E}_{lm} &= -(-\sqrt 2)^{s-1} \mathcal A_{ls} (\E S)_{lm}, \\
    S^{\mathrm B}_{lm} &= (-\sqrt 2)^{s-1} \mathcal A_{ls} (\B S)_{lm}.
    \end{align}
\end{subequations}
As we perform calculations using both of these two different notions of multipolar harmonics, we will frequently use Eq.~\eqref{eq:tensorscalarlm} to relate the two.

\subsection{Flux and pseudoflux contributions} \label{subsec:fluxes}

Our expressions for the multipole moments of the fluxes will use a notation similar to that of~\cite{Siddhant:2024nft}, but our expressions are written in a somewhat more compact form than those appearing there.
The flux $\E{\mathcal F}_0$, used for computing the displacement memory, is given by
\begin{equation} \label{eq:F0lm}
    (\E{\mathcal{F}}_0)_{lm} = - \frac 1{16} \sum_{\mathrm I', \mathrm I''} \sum_{l', m'} \sum_{l'', m''} \eta^{\mathrm E \mathrm I' \mathrm I''}_{l l' l''} \mathscr C^{lm,2}_{l'm'l''m''} \dot{C}^{\mathrm I'}_{l'm'} \dot{C}^{\mathrm I''}_{l''m''} .
\end{equation}
The sums over $\mathrm I'$ and $\mathrm I''$ run over just $\mathrm E$ and $\mathrm B$, whereas the sums over $l'$ and $l''$ run over integers $l'\geq 2$ and $l''\geq 2$.
The $m'$ and $m''$ sums run over integers satisfying $-l'\leq m' \leq l'$ and $-l''\leq m'' \leq l''$.

The multipole moments of both the fluxes $\mathcal F_1^A$ (in a basis of vector spherical harmonics) or the scalar versions $\E{\mathcal F}_1$ and $\B{\mathcal F}_1$ (in a basis of scalar harmonics) have been computed in~\cite{Nichols:2017rqr,Nichols:2018qac} and~\cite{Siddhant:2024nft}, respectively.
We denote the multipoles of $\mathcal F_1^A$ by $(\mathcal F_1)_{lm}^\mathrm{I}$ and those of $\E{\mathcal F}_1$ and $\B{\mathcal F}_1$ by $(\E{\mathcal F}_1)_{lm}$ and $(\B{\mathcal F}_1)_{lm}$.
The two types of multipole coefficients can be related by Eq.~\eqref{eq:tensorscalarlm}.
The moments $(\mathcal F_1)_{lm}^\mathrm{I}$ are given by
\begin{align} \label{eqn:First moment flux exp elec}
    (\mathcal F_1)_{lm}^\mathrm{I} = & - \frac{1}{32} \sum_{\mathrm I', \mathrm I''} \sum_{l', m'} \sum_{l'', m''} \eta^\mathrm{I I' I''}_{l l' l''}\mathscr B^{lm}_{l'm'l''m''} \nonumber \\
    & \times [\dot{C}^{\mathrm I'} _{l'm'} {C}^{\mathrm I''} _{l''m''} - {C}^{\mathrm I'} _{l'm'} \dot{C}^{\mathrm I''} _{l''m''}] , 
\end{align}
where the coefficient $\mathscr B^{lm}_{l'm'l''m''}$ is given in terms of the Gaunt and $\mathcal B_{ls}$ coefficients by
\begin{equation}
    \mathscr B^{lm}_{l'm'l''m''} = \mathcal B_{l'2} \mathscr C^{lm,(-3)2}_{l'm'l''m''} + 3 \mathcal B_{l''1} \mathscr C^{lm,(-2)1}_{l'm'l''m''} .
\end{equation}

The multipolar expansion of the second flux was also given in~\cite{Siddhant:2024nft} for the scalar fluxes.
It is somewhat more involved because it is cubic in the shear and news.
Here we give the moments of the tensor flux instead:
\begin{align}
    (\mathcal{F}_2)_{lm}^\mathrm I = {} & \frac{1}{16} \sum_{\mathrm I', \mathrm I'', \mathrm I'''} \sum_{l',m'} \sum_{l'',m''} \sum_{\bar l,\bar m} \sum_{l''',m'''} \eta^{\mathrm E \mathrm I' \mathrm I''}_{\bar l l' l''} \eta^\mathrm{E I I'''}_{l \bar l l'''} \nonumber \\
    & \times  \mathscr C^{\bar l \bar m, 2}_{l' m' l'' m''} \mathscr C^{lm,02} _{\bar l \bar m l'''m'''} \dot{C}^{\mathrm I'} _{l'm'} {C}^{\mathrm I''} _{l''m''} {C}^{\mathrm I'''} _{l'''m'''} . \label{eqn:Second moment flux exp elec2}
\end{align}
The relation to the scalar moments is given by Eq.~\eqref{eq:tensorscalarlm}.

We now consider the pseudofluxes.
The scalars $\E{\mathcal G}_2$ and $\B{\mathcal G}_2$ were previously given in~\cite{Siddhant:2024nft}.
We present a combined expression for the tensor flux, in terms of the multipole moments of the mass, shear, and $\mathcal W^A$:
\begin{widetext}
\begin{equation} \label{eq:G2Ilm}
    (\mathcal G_2)^{\mathrm I}_{lm} = \frac{1}{4} \sum_{\mathrm I''} \sum_{l', m'} \sum_{l'', m''} \eta^{\mathrm{EII''}}_{ll'l''} \mathscr C^{lm,02}_{l'm'l''m''} \left[\mass_{l'm'} C^{\mathrm I''}_{l''m''} + \frac{\mathcal A_{l'(-2)}}{4 \sqrt 2} (^* C)^{\mathrm E}_{l'm'} (^* C)^{\mathrm I''}_{l''m''} \right] + \frac{1}{3 \sqrt 2} \mathcal B_{l1} \mathcal W^{\mathrm I}_{lm} .
\end{equation}
Here, $\mass_{lm}$ are terms in the multipolar expansion of the mass aspect, and
\begin{equation} \label{eq:WIlm}
    \mathcal W^{\mathrm I}_{lm} = -\frac{1}{32} \sum_{\mathrm{I', I''}} \sum_{l', m'} \sum_{l'', m''} \eta^{\mathrm{II'I''}}_{ll'l''} C^{\mathrm I'}_{l'm'} C^{\mathrm I''}_{l''m''} \mathscr B^{lm}_{l'm'l''m''}.
\end{equation}
Given some initial value $\mass(u_0)$, the mass aspect can be computed in a multipolar expansion using Eq.~\eqref{eq:mdotF}:
\begin{equation} \label{eqn:m_lm}
    \Delta \mass_{lm} (u, u_0) = \frac{1}{4 \sqrt 2} \mathcal A_{l(-2)} \Delta C_{lm}^\mathrm{E} (u, u_0) + \int_{u_0}^u du' (\E{\mathcal F}_0)_{lm}.
\end{equation}

For completeness (although we will not need the full expression in our forecasts in Sec.~\ref{sec:results}), we also give the multipolar expression for $\mathcal G_3^{AB}$ in terms of the multipole moments of the shear and the angular-momentum aspect,
\begin{align} \label{eqn:G3_lm}
    (\mathcal G_3)^{\mathrm I}_{lm} = -\frac{1}{4} \mathcal B_{l 2} \hspace{-0.5em} \sum_{l'',m'',\mathrm I''} \Bigg[ & \eta^{\mathrm{I'II''}}_{ll'l''} \mathscr C^{lm, 12}_{l'm'l''m''} L^{\mathrm I'}_{l'm'} \nonumber \\
    & + \frac{5}{64} \sum_{l',m', \mathrm I'} \sum_{l''',m''',\mathrm I'''} \mathcal B_{l'''2} \eta^{\mathrm{EI'I''}}_{\bar l l'l''} \eta^{\mathrm{EII'''}}_{l\bar l l'''} \mathscr C^{\bar l \bar m, 2}_{l'm'l''m''} \mathscr C^{lm, 03}_{\bar l \bar m l'''m'''} C^{\mathrm I'}_{l'm'} C^{\mathrm I'''}_{l'''m'''} \Bigg]C^{\mathrm I''}_{l''m''},
\end{align}
where
\begin{equation}
    L^{\mathrm I}_{lm} = -\frac{1}{3} N^{\mathrm I}_{lm} + \frac{1}{16} \sum_{l',m',\mathrm I'} \sum_{l'',m'',\mathrm I''} \left[\mathcal B_{l'' 1} \eta^{\mathrm{II'I''}}_{ll'l''} \mathscr C^{lm,(-2)1}_{l'm'l''m''} - \frac{3}{4} \mathcal B_{l0}\delta^{\mathrm{IE}} \eta^{\mathrm{EI'I''}}_{ll'l''} \mathscr C^{lm,2}_{l'm'l''m''}\right] C^{\mathrm I'}_{l'm'} C^{\mathrm I''}_{l''m''}.
\end{equation}
As with the mass aspect above, we can compute the change in $N^{\mathrm I}_{lm}$ using its evolution equation.
First, we write
%
    $N^{\mathrm I}_{lm} = \hat N^{\mathrm I}_{lm} + \mathcal W^{\mathrm I}_{lm}$.
%
Next, we can explicitly compute the expression for $\hat N^{\mathrm I}_{lm}$ from the evolution equations for its electric and magnetic pieces, $\E Q_1$ and $\B Q_1$, and Eq.~\eqref{eq:tensorscalarlm}.
From Eqs.~\eqref{eqn: evol Q_1}, we obtain
\begin{subequations}
\begin{align}
    \Delta (\E Q_1)_{lm} (u, u_0) & = \int_{u_0}^u du' \left[-\mathcal A_{l(-1)}^2 \mass_{lm} (u') + (\E{\mathcal F}_1)_{lm} (u')\right], \\
    \Delta (\B Q_1)_{lm} (u, u_0) & = \int_{u_0}^u du' \left[{-\frac{1}{4 \sqrt{2} } \mathcal A_{l(-1)}^2 \mathcal A_{l (-2)} C^{\mathrm B}_{lm} (u')} + (\B{\mathcal F}_1)_{lm} (u')\right] .
\end{align}
\end{subequations}

\end{widetext}

\section{Stationary-to-stationary transitions} \label{sec:stationary}

The results of the previous section are expressions for the fluxes and pseudofluxes, which are computed from the shear and initial values for the other Bondi metric functions (i.e., the mass, angular-momentum, and higher-Bondi aspects).
However, to completely determine the higher memories, we need expressions for the charges as well, and we \emph{cannot} directly derive their values from initial data and the evolution equations, as the evolution equations depend on the higher memories that we are trying to compute.
Even with the full charge data as a function of time (which can be obtained from numerical relativity simulations with CCE data, as reviewed in~\cite{Grant:2023ged}), there still remains the problem of determining which part of the charge contributes to a given moment of the news and the corresponding displacement or higher memory signal.
The goal of this section is to present a context in which computing this part of the charge is possible.

As a starting point for this discussion, note that in a generic nonradiative region of spacetime, the $u$ derivatives of the charges will not vanish (see, e.g.,~\cite{Flanagan:2015pxa}).
A more restricted class of spacetimes is one in which the $u$ derivatives of the charges do vanish in some region of the spacetime.
Spacetimes that undergo a transition between two such regions with intervening radiation are referred to as ``stationary-to-stationary transitions.''\footnote{These regions are called ``stationary'' because they possess a timelike Killing vector field, $\partial_u$, in both regions.
In a stationary-to-stationary transition, the Killing vectors in the two stationary regions are also the same, whereas generically they can differ by a Lorentz boost (a rotation or supertranslation of the two frames will not affect the form of this Killing vector).
In the context of BBH mergers, stationary-to-stationary transitions will arise when the merger produces no GW recoil of the remnant black hole.
Because most BBH mergers do have such a recoil, we will discuss the scale of the errors in this approximation in Sec.~\ref{subsec:accuracy}.}
These are a subclass of ``nonradiative-to-nonradiative transitions'' in which only the news vanishes.

Such transitions were considered in~\cite{Nichols:2018qac} for arguing that there should be a charge contribution to the CM memory for BBH systems.
Approximating such systems as stationary-to-stationary transitions gave a partial estimate of the total charge contribution to the CM memory, because it neglected possible contributions from memory effects that arise in more general nonradiative-to-nonradiative transitions.

We will similarly adopt this approximation to compute charge and pseudoflux contributions to the higher memory effects, and we will generalize the results of~\cite{Nichols:2018qac} in the process.
The crux of this approximation is that between two stationary regions, there is a natural way of splitting the shear such that each piece contributes to a given moment of the news.
To describe the split, we first (in Sec.~\ref{subsec: st Bondi func}) discuss how Bondi metric functions ($\mass$, $C_{AB}$, $N^A$, and $\ord{0}{\mathcal E}_{AB}$) behave in the stationary regions.
We show that most $l$ modes are completely constrained, with the remaining pieces constrained (except in the case of $\ord{0}{\mathcal E}_{AB}$) by their corresponding evolution equations.
To provide a prescription for defining a higher memory signal associated with the part of the charge contributing to the stationary-to-stationary value of the moments of the news, we extend the stationary constraints into a radiative region so as to define a ``stationary-to-stationary part'' of each of these Bondi metric functions.
With these expressions, we then describe the procedure for defining higher memory signals within the stationary-to-stationary framework in Sec.~\ref{subsec: st-st charge}.

\subsection{Bondi metric functions} \label{subsec: st Bondi func}

We will first define the \emph{stationary-to-stationary parts} of the Bondi metric functions and then explicitly compute these parts for $\mass$, $C_{AB}$, $N^A$, and $\ord{0}{\mathcal E}_{AB}$.
This approach will be iterative: at each step, we use the assumption of stationarity to compute conditions that some quantity must satisfy in a stationary region.
We denote these constraints using $\seq$ to indicate that such equations hold only in a stationary region of spacetime.

The stationary constraints take the form of elliptic differential equations involving the angular derivative $\mathscr D_A$, which we will solve for the stationary values of the metric functions.
The angular operators in the elliptic equation have a nontrivial kernel (in terms of spherical harmonics, the kernel will contain harmonics with an $l$ index not equal to a given integer).
We compute the stationary-to-stationary part by taking these constraints and enforcing that this part of the quantity obeys these constraints at all times (including in radiative regions).
For the remaining parts which are unconstrained, we use appropriate evolution equations to determine the values of the remainders.

To make this abstract description more concrete, we start by computing the stationary-to-stationary part of the mass aspect, $\mass^{\mathrm S}$.
We begin with the constraints on the evolution equations in Eq.~\eqref{eqn:Q_n_PDEs} in regions where $N_{AB} = 0$ and the time derivatives of the charges vanish.
Requiring that $\E Q_1$ is time-independent produces the following condition on the mass aspect:
\begin{equation} \label{eqn:mass_stat}
    \mathscr D^2 \mass \seq 0.
\end{equation}
This constraint implies that $\mass$ is a constant function on the sphere in the stationary regions; as such, it is characterized entirely by its $\ell = 0$ harmonic.
We now define $\mass^{\mathrm S}$ such that the constraint in this equation holds for all time [i.e., $\mass^\mathrm{S}(u) = \mass^S_{00}(u) Y_{00}$].
Equation~\eqref{eqn:mass_stat} does not constrain the time dependence of the $l = 0$ part of $\mass^{\mathrm S}$.
To obtain its time dependence, we use the $l = 0$ part of the evolution equation for $\mass$, and we define $\mass^{\mathrm S}$ by the change in its value through
\begin{equation}
    \Delta \mass^{\mathrm S} (u, u_0) \equiv \mathcal P_{l = 0} \int_{u_0}^u du' \E{\mathcal F}_0.
\end{equation}
We use the notation $\mathcal P_{(\cdots)}$ to denote a projector that picks out the terms in the spherical harmonic expansion of an expression where the condition $(\cdots)$ on $l$ holds.
In terms of a multipolar expansion of the shear, we have that
\begin{equation} \label{eq:mSofu_lm}
    \Delta \mass^S (u, u_0) = -\frac{1}{32\pi} \sum_{l',m',\mathrm I'} \int_{u_0}^u du' |\dot C^\mathrm{I'}_{l'm'}|^2.
\end{equation}
This expression is simpler than that in Eq.~\eqref{eqn:m_lm}, because we are only interested in $l = 0$; thus, we can use the orthogonality of the spin-weighted (or tensor) spherical harmonics instead of the more general expression for their product.
Moreover, we have used the fact that $C^{\mathrm I}_{l(-m)} = (-1)^m \overline{C^{\mathrm I}_{lm}}$, as well as that $\eta^{\mathrm{EII}}_{0ll}=2$.

An analogous constraint to Eq.~\eqref{eqn:mass_stat} holds for $\B Q_0$, which comes from requiring that $\B Q_1$ be time-independent:
\begin{equation} \label{eqn:Q0S}
    \mathscr D^2 \B Q_0 \seq 0,
\end{equation}
or equivalently
\begin{equation} \label{eqn:CS_mag}
    \mathscr D_A \mathscr D_B (^* C)^{AB} \seq 0.
\end{equation}
This, along with the fact that it contains $l \geq 2$ harmonics, is the only constraint on the value of the shear in the stationary regions.
It is a condition on the stationary-to-stationary part of the magnetic-parity shear, which we define to be vanishing:
\begin{equation}
    \mathscr D_A \mathscr D_B (^* C)_{\mathrm S}^{AB} = -\frac{1}{\sqrt{2}}\sum_{l,m}\mathcal A_{l(-2)}C^{\mathrm{S, B}}_{lm} Y_{lm} \equiv 0.
\end{equation}
As with the mass aspect, we fix the remainder of $C_{\mathrm S}^{AB}$ [the electric-parity part, which is in the kernel of the angular operator in Eq.~\eqref{eqn:CS_mag}, when the dual operation is considered to be part of the angular differential operator] by using the integral of Eq.~\eqref{eqn: evol Q_0}:
\begin{equation} \label{eqn:CS_elec}
    \mathscr D_A \mathscr D_B \Delta C_{\mathrm S}^{AB} (u, u_0) \equiv -4 \mathcal P_{l \geq 2} \int_{u_0}^u \ud u' \E{\mathcal F}_0.
\end{equation}
Equations~\eqref{eqn:CS_mag} and~\eqref{eqn:CS_elec} completely define what we refer to as the stationary-to-stationary part of the shear.
In terms of a multipolar expansion, we find that the multipole moments of the (nonvanishing) electric part are given by
\begin{equation}\label{eqn:C_S multipole}
    \Delta C^{\mathrm{S, E}}_{lm} (u, u_0) = -4 \sqrt 2 \mathcal A_{l2} \int_{u_0}^u du' (\E{\mathcal F}_0)_{lm}.
\end{equation}

We next turn to the stationary-to-stationary part of $N^A$, which can be computed from the stationary-to-stationary part of $\hat N^A$ (or equivalently $\E Q_1$ and $\B Q_1$).
Similar to the calculation of $\E Q_{0, \mathrm S}$ above, the constraint on $\p Q_{1, \mathrm S}$ (for $p=e$ or $b$) is obtained from the evolution equation at one higher order.
Specifically, by setting $\p{\dot Q}_2 \seq 0$ in Eq.~\eqref{eqn: evol Q_2}, we find
\begin{equation} \label{eq:Q1stationary}
    (\mathscr D^2 + 2) \p Q_1 \seq -6 \p{\mathcal G}_2.
\end{equation}
This expression constrains the $l \geq 2$ spherical harmonics of $\p Q_1$.
By using the fact that, in stationary regions, $\mass$ is constant and $\mathscr D^A \mathscr D^B (^* C)_{AB}$ vanishes, this implies that we can define the $l \geq 2$ parts of these charges by
\begin{subequations} \label{eq:Q1shear}
    \begin{align}
        (\mathscr{D}^2 + 2) \E Q_{1, \mathrm{S}} &\equiv -6 \E{\mathcal G}_{2, \mathrm S} \nonumber \\
        &= -3 \mass^\mathrm{S} \mathscr{D}_A \mathscr{D}_B C^{AB}_\mathrm{S} - \mathscr{D}_A \mathcal W^A_\mathrm{S}, \\
        (\mathscr{D}^2 + 2) \B Q_{1, \mathrm S} &\equiv -6 \B{\mathcal G}_{2, \mathrm S} = -\mathscr D_A (^*\mathcal W)^A_{\mathrm S} .
    \end{align}
\end{subequations}
Here $\mathcal W^A_{\mathrm S}$ is the stationary-to-stationary value of $\mathcal W^A$, which is defined by replacing $C_{AB}$ with $C^{\mathrm S}_{AB}$ for all terms in the expression for $\mathcal W^A$.
As compared with Eq.~\eqref{eq:G2AB}, Eq.~\eqref{eq:Q1shear} contains simplified expressions for $\E{\mathcal G}_{2, \mathrm S}$ and $\B{\mathcal G}_{2, \mathrm S}$ that make use of the properties of the Bondi metric functions in stationary regions.

Since these equations do not constrain the $l = 1$ part of these charges, as we did with the mass aspect and the shear above, we use evolution equations~\eqref{eqn: evol Q_1} to fix the remaining $l=1$ parts.
The expressions are given by
\begin{widetext}
\begin{subequations} \label{eqn:N_kernel}
\begin{align} 
    \mathcal P_{l=1} \Delta \E Q_{1, \mathrm S} (u, u_0) &= \mathcal P_{l=1} \int_{u_0}^u  du' \left\{\E{\mathcal F}_1 (u') - 2 \left[\mass(u_0) + \int_{u_0}^{u'} du'' \E{\mathcal F}_0(u'')\right]\right\}, \\
    \mathcal P_{l=1} \Delta \B Q_{1, \mathrm S} (u, u_0) &= \mathcal P_{l=1} \int_{u_0}^u  du' \B{\mathcal F}_1,
\end{align}
\end{subequations}
\end{widetext}
where we have explicitly replaced a factor of $\mathscr D^2$ with $-2$, given that that is the action of the spherical Laplacian on an $l = 1$ spherical harmonic.

We also compute the stationary-to-stationary part of the angular-momentum aspect itself, which will be needed subsequently.
Because $N^A$ and $\hat N^A$ (which is used to construct the electric and magnetic scalars $\E Q_1$ and $\B Q_1$) are related by $\mathcal W^A$, it follows from Eq.~\eqref{eq:Q1shear} that, for $l \geq 2$, the stationary-to-stationary part of the angular-momentum aspect is given by
\begin{subequations} \label{eq:NAstationary}
    \begin{align}
        (\mathscr D^2 + 2) \mathscr D_A N_{\mathrm S}^A &\equiv -3 \mass^{\mathrm S} \mathscr D^A \mathscr D^B C^{\mathrm S}_{AB} , \\
       (\mathscr D^2 + 2) \mathscr D_A (^*N)_{\mathrm S}^A &\equiv 0 .
    \end{align}
\end{subequations}
The remainder (i.e., the $l = 1$ part) of $N_{\mathrm S}^A$ can be obtained from Eq.~\eqref{eqn:N_kernel}.
Because the Wald-Zoupas term $\mathcal W^A$ has the property that $\mathcal P_{l=1} \mathcal W^A$ vanishes when $C_{AB}$ is purely electric~\cite{Compere:2016jwb, Elhashash:2021iev}, it follows that $\mathcal P_{l = 1} \mathcal W_{\mathrm S}^A = 0$.
Thus, $\mathcal P_{l=1} N^A_\mathrm{S} = \mathcal P_{l=1} \hat N^A_\mathrm{S}$, and Eq.~\eqref{eqn:N_kernel} directly gives the change in the $l = 1$ part of $N^A_{\mathrm S}$ as well.

We next consider multipolar expansions of $\E Q_{1, \mathrm S}$ and $\B Q_{1, \mathrm S}$.
The $l \geq 2$ harmonics are given by
\begin{subequations}\label{eqm:drift charge STS}
    \begin{align}
        (\E Q_{1,\mathrm{S}})_{lm} &= \frac{\mathcal B_{l0}}{\mathcal B_{l1}} \Bigg[ \frac{3}{\sqrt 2} \mass^\mathrm{S} C^\mathrm{S, E}_{lm} - \frac{1}{\mathcal B_{l1}} \mathcal W^{\mathrm{S, E}}_{lm}\Bigg], \label{eq:Q1elm} \\
        (\B Q_{1,\mathrm{S}})_{lm} &= \frac{\mathcal B_{l0}}{(\mathcal B_{l1})^2}\mathcal W^{\mathrm{S, B}}_{lm}, \label{eq:Q1blm}
    \end{align}
\end{subequations}
where $\mathcal W^{\mathrm{S, I}}_{lm}$ is computed from taking Eq.~\eqref{eq:WIlm} and replacing the shear with its stationary part.
Another simplification in Eq.~\eqref{eq:Q1elm} comes from the fact that the first term does not involve products of spherical harmonics, because it involves the angle-independent $\mass^{\mathrm S}$ rather than the Bondi mass aspect.
The $l = 1$ harmonics follow from a straightforward expansion of Eq.~\eqref{eqn:N_kernel}, so we do not reproduce the multipolar expression here.
For the $l \geq 2$ parts of $N_{\mathrm S}^A$, we find that these expressions simplify even further, as they do not have a contribution from the Wald-Zoupas correction term $\mathcal W^A$:
\begin{equation} \label{eqn:N_stationary}
    N^{\mathrm{S, E}}_{lm} = -\frac{3}{\sqrt 2 \mathcal B_{l1}} \mass^{\mathrm S} C^{\mathrm{S, E}}_{lm},
\end{equation}
Note that $N^{\mathrm{S, B}}_{lm} = 0$ for $l \geq 2$.
Again, the $l = 1$ harmonics can be straightforwardly determined from Eq.~\eqref{eqn:N_kernel}.

Finally, we compute the stationary-to-stationary parts of $\E Q_2$ and $\B Q_2$, which we describe equivalently in terms of $\ord{0}{\mathcal E}_{AB}$.
Analogously to the calculations above, we obtain a constraint by setting $\partial_u \ord{1}{\mathcal E}_{AB} = 0$ in Eq.~\eqref{eq:EAB1dotG} and solving for $\ord{0}{\mathcal E}_{AB}$:
\begin{equation}\label{eqn:Q2shear}
    (\mathscr D^2 + 2) \ord{0}{\mathcal E}_{AB} \seq 4 \mathcal G^3_{AB}.
\end{equation}
We use this to define the stationary-to-stationary part of $\ord{0}{\mathcal E}_{AB}$:
\begin{equation} \label{eqn:Q2stationary}
    (\mathscr D^2 + 2) \ord{0}{\mathcal E}^{\mathrm S}_{AB} \equiv 4 \mathcal G^{3, \mathrm S}_{AB}.
\end{equation}
We define $\mathcal G^{3, \mathrm S}_{AB}$, the stationary-to-stationary part of $\mathcal G^3_{AB}$, by replacing $N^A$ and $C_{AB}$ in the original expression in Eq.~\eqref{eq:G3AB} with $N^A_\mathrm{S}$ and $C_{AB}^\mathrm{S}$.
Equation~\eqref{eqn:tensor_D2} implies that the angular operator on the left-hand side annihilates $l = 2$ harmonics; therefore, Eq.~\eqref{eqn:Q2stationary} is a constraint on the $l \geq 3$ part of the stationary-to-stationary part of $\ord{0}{\mathcal E}^{\mathrm S}_{AB}$.
However, \emph{unlike} before, there is no evolution equation which we can use to fix the $l = 2$ part, because the $l = 2$ part of the evolution equation for $\ord{0}{\mathcal E}_{AB}$ depends on precisely the part of the shear which we are trying to compute.
Namely, even under the assumption of a stationary-to-stationary transition, there is not a way to compute the $l = 2$ part of the charge contribution to the second moment of the news (without some prior or independent knowledge of $\ord{0}{\mathcal E}_{AB}$).
Restricting to $l \geq 3$, Eq.~\eqref{eqn:Q2stationary} defines a stationary-to-stationary part whose multipolar expansion is given by
\begin{equation}\label{eqn:E0stationary}
    \ord{0}{\mathcal E}^{\mathrm{S, I}}_{lm} = -\frac{4}{\mathcal B_{l2}^2} (\mathcal G_3^{\mathrm S})^{\mathrm I}_{lm}.
\end{equation}
The multipoles of the scalar charges $(Q_{2}^{(p)})_{lm}$ can be obtained using Eqs.~\eqref{eq:tensorscalarlm}. 
As there are no significant simplifications in the expression for {$\ord{0}{\mathcal E}^{\mathrm{S, I}}_{lm}$} in a stationary region, we do not repeat Eq.~\eqref{eqn:G3_lm} here [though we do note that the $N^{\mathrm{S, I}}_{lm}$ in $(\mathcal G_3^{\mathrm S})^{\mathrm I}_{lm}$ can be computed straightforwardly from Eq.~\eqref{eqn:N_kernel} and~\eqref{eqn:N_stationary}].
This is the last of the multipolar coefficients that we will need in a stationary region in this paper.

\subsection{Higher memory signals} \label{subsec: st-st charge}

We now show how the displacement and higher memory signals, including charge and flux contributions, can be obtained in the context of a stationary-to-stationary transition.
We write the shear as a sum of contributions, which we call $C^0_{AB}$, $C^1_{AB}$, $C^2_{AB}$, and so on, where each $C^n_{AB}$ is associated with an offset in the $n^\mathrm{th}$ moment of the news. 
This will be an iterative approach: at the lowest order, we can write the total change in shear as
\begin{equation} \label{eqn:shear_iter_0}
    \Delta C_{AB} (u, u_0) = C^0_{AB} (u, u_0) + C^{0, \mathrm R}_{AB} (u, u_0).
\end{equation}
Here $C^0_{AB}$ captures the total offset in the shear in the stationary-to-stationary approximation, so that only $C^0_{AB}$ contributes to the zeroth moment of the news.
The residual piece $C^{0, \mathrm R}_{AB}$ vanishes in the past and future stationary regions.
At each subsequent stage (for $n \geq 1$), we split the residual piece into a piece which (in this approximation) contributes to the $n^\mathrm{th}$ moment, and another residual piece that contributes to higher moments:
\begin{equation} \label{eqn:shear_iter_n}
    C^{n - 1, \mathrm R}_{AB} (u, u_0) = C^n_{AB} (u) + C^{n, \mathrm R}_{AB} (u, u_0).
\end{equation}
We will associate each $C^n_{AB}$ with a (higher) memory signal in the remainder of this paper.
Unlike $C^0_{AB}$ and all of the $C^{n, \mathrm R}_{AB}$ terms,  $C^n_{AB}$ is a function of $u$ only, not $u_0$, for $n \geq 1$.

The lowest order of this iterative procedure is a special case and is the simplest.
In Eq.~\eqref{eqn:shear_iter_0}, we want $C^0_{AB}$ to be the part of $\Delta C_{AB}$ that captures the entire change in the shear when $u$ and $u_0$ are in stationary regions.
To satisfy this requirement, $C^0_{AB}$ should be the same as the change in the stationary part of the shear:
\begin{equation}\label{eqn:C0=CS}
    C^0_{AB} (u, u_0) \equiv \Delta C^{\mathrm S}_{AB} (u, u_0).
\end{equation}
From Eq.~\eqref{eqn:CS_elec}, the electric part of $C^0_{AB}$ is given by

\begin{equation}\label{eq:CAB0stationary}
    \mathscr D^A \mathscr D^B C^0_{AB} = -4 \mathcal P_{l \geq 2} \int_{u_0}^u du' \E{\mathcal F}_0(u'),\\
\end{equation}
and from Eq.~\eqref{eqn:CS_mag}, the magnetic part vanishes:
\begin{equation} \label{eq:CAB0stationary_mag}
    \mathscr D^A \mathscr D^B (^* C)^0_{AB} = 0.
\end{equation}

While the zeroth moment of the news (displacement memory) contains only a flux contribution for a stationary-to-stationary transition, this will not be true for the higher memory signals.
For the first moment, there is a charge contribution that we also must compute.
To obtain the expression that we use to calculate the charge and flux contributions, we insert Eqs.~\eqref{eqn:shear_iter_0} and~\eqref{eqn:shear_iter_n} (with $n = 1$) into Eq.~\eqref{eq:DeltaCabQ1}, and use Eq.~\eqref{eq:CAB0stationary} to eliminate parts that are already contained in the displacement memory ($C^0_{AB}$):
\begin{equation} \label{eqn:C0R}
    \begin{split}
        \frac{1}{4} \mathscr D^2 \mathscr D^A \mathscr D^B &[C^1_{AB} (u) + C^{1, \mathrm R}_{AB} (u, u_0)] \\
        &= \E{\dot Q}_1 (u) - \mathscr D^2 \E Q_0 (u_0) - \E{\mathcal F}_1 (u).
    \end{split}
\end{equation}
As we did with $C^0_{AB}$, we define $C^1_{AB}$ to be the part of this expression that contributes to the first moment of the news, and $C^{1, \mathrm R}_{AB}$ to be the part which does not contribute, when the spacetime at $u_0$ and $u$ is stationary.
To make this separation, we therefore need to consider the integral of this equation.
Upon integrating, the second term on the left-hand side of Eq.~\eqref{eqn:C0R} vanishes by construction, because it is defined such that it does not contribute to the first moment.
The first term on the right-hand side becomes the difference in $\E Q_1$ evaluated at the endpoints.
Because we required that those two endpoints are in stationary regions, we can replace $\E Q_1$ with $\E Q_{1, \mathrm S}$.
Moreover, we have that $\mathscr D^2 \E Q_0 =0$, because in a stationary region $\E Q_0 = \mass^\mathrm{S}$ is independent of angular coordinates.
Finally, the third term turns into an integral of the flux, which does not simplify further.
In summary, these conditions imply that we can define $C^1_{AB}$ by
\begin{equation}\label{eqn:C1E}
        \frac{1}{4} \mathscr D^2 \mathscr D^A \mathscr D^B C^1_{AB} \equiv \E{\dot Q}_{1, \mathrm S} - \E{\mathcal F}_1,
\end{equation}
as this expression contains all the terms that contribute to the first moment.
Similar arguments imply that an analogous expression holds for the magnetic-parity case:
\begin{equation}\label{eqn:C1B}
    \frac{1}{4} \mathscr D^2 \mathscr D^A \mathscr D^B (^* C)^1_{AB} \equiv -\B{\dot Q}_{1, \mathrm S} + \B{\mathcal F}_1.
\end{equation}

To determine the part of the shear $C^2_{AB}$, which contributes to the second moment of the news for stationary-to-stationary transitions, we now plug Eqs.~\eqref{eqn:shear_iter_0} and~\eqref{eqn:shear_iter_n} (with $n = 1$ and $n = 2$) into Eq.~\eqref{eq:DeltaCabQ2}.
Upon using the expressions for $C^0_{AB}$ and $C^1_{AB}$, the result simplifies to
\begin{equation}
    \begin{split}
        \frac{1}{24} (\mathscr D^2 + 2) &\mathscr D^2 \mathscr{D}^A \mathscr{D}^B [C^2_{AB} (u) + C^{2, \mathrm R}_{AB} (u)] \\
        &= \E{\ddot Q}_\mathrm{2,S}(u) - \frac{1}{6} (\mathscr D^2 + 2) \mathscr D^2 \E Q_0 (u_0) \\
        &\hspace{1.1em}- \E{\dot{\mathcal F}}_2 - \E{\dot{\mathcal G}}_\mathrm{2,R},
    \end{split}
\end{equation}
where
\begin{equation}\label{eqn:G2R}
    \begin{split}
        \mathcal G_{2, \mathrm R}^{AB} &\equiv \mathcal G_2^{AB} - \mathcal G_{2, \mathrm S}^{AB} \\
        &= \frac{1}{2} \left(\mass C^{AB} - \mass^{\mathrm S} C_{\mathrm S}^{AB}\right) + \frac{1}{3} \STF \mathscr D^A (\mathcal W^B - \mathcal W_{\mathrm S}^B) \\
        &\hspace{1.1em}+ \frac{1}{8} [\mathscr D_C \mathscr D_D (^*C)^{CD}] (^*C)^{AB} .
    \end{split}
\end{equation}
Much like with the discussion below Eq.~\eqref{eqn:C0R}, we can integrate this equation in time (now doing so \emph{twice}) and make the assumption of stationarity at $u$ and $u_0$.
This allows us to define $C^2_{AB}$ such that it gives rise to the second moment of the news for stationary-to-stationary transitions.
The electric and magnetic-parity parts are obtained from
\begin{subequations}\label{eqn:C2E/B}
    \begin{align}
        \frac{1}{24} (\mathscr D^2 + 2)\mathscr D^2 \mathscr{D}^A \mathscr{D}^B C^2_{AB}\equiv {} & \E{\ddot Q}_{2, \mathrm S} - \E{\dot{\mathcal F}}_2 -\E{\dot{\mathcal G}}_{2, \mathrm R}, \\
        \frac{1}{24} (\mathscr D^2 + 2)\mathscr D^2 \mathscr{D}^A \mathscr{D}^B (^*C)^2_{AB}\equiv & -\B{\ddot Q}_{2, \mathrm S} + \B{\dot{\mathcal F}}_2 +\B{\dot{\mathcal G}}_{2, \mathrm R} .
    \end{align}
\end{subequations}

It is important to note here that it is only this ``residual'' part of the pseudoflux, $\mathcal G_{2, \mathrm R}^{AB}$, which contributes to the second moment; the contribution of the rest of the pseudoflux, $\mathcal G_{2, \mathrm S}^{AB}$, has already been accounted for in the first moment.
An alternate way of understanding this comes from the fact that pseudofluxes generically have a nonzero value in stationary regions.
From Eq.~\eqref{eq:secondMoments}, it is the integral of the pseudoflux which contributes to the second moment, and in a stationary region this contribution would therefore grow linearly in time; thus,  $\mathcal G_{2, \mathrm S}^{AB}$ is contributing to the first, rather than the second moment of the news.
By replacing the pseudoflux with its residual part, which \emph{does} vanish in stationary regions (by definition), $\mathcal G_{2, \mathrm R}^{AB}$ contributes to just the second moment.

We can obtain the multipolar expression for the parts of shear associated with the displacement and higher memory effects, which can be expressed in terms of the multipoles of the stationary parts of the Bondi aspects (or charges) in Sec.~\ref{subsec: st Bondi func} and the multipoles of the fluxes in Sec.~\ref{subsec:fluxes}:
\begin{subequations}\label{eqn:memorysignals}
    \begin{align}
        \label{eq:Clm0S2S}
        C_{lm}^{0,\mathrm I}&\equiv -4\sqrt{2}\mathcal{A}_{l2} \delta^{\mathrm{EI}}\mathcal P_{l \geq 2} \int_{u_0}^u du' (\E{\mathcal F}_{0})_{lm}(u'), \\
        \label{eq:Clm1S2S}
        C^{1,\mathrm I}_{lm} &\equiv 4\sqrt{2}\;\mathcal A_{l1} \mathcal{A}_{l2} 
        (\delta^\mathrm{EI} - \delta^\mathrm{BI})
        [\dot {\hat{N}}^{\mathrm S, \mathrm I}_{lm} - {(\mathcal F_1)}^{ \mathrm I}_{lm}],\\
        \label{eq:Clm2S2S}
        C^{2,\mathrm I}_{lm}&\equiv 24 \mathcal A_{l2}^2
        (\delta^\mathrm{EI} - \delta^\mathrm{BI})
        [ \ord{0}{\ddot{\mathcal E}}^{\mathrm S,\mathrm I}_{ lm} - {(\dot{\mathcal F}_2)}^{\mathrm I}_{lm} - {(\dot{\mathcal G}_2)}^{\mathrm R,\mathrm I}_{ lm}].
    \end{align}
\end{subequations}
The factor of $(\delta^\mathrm{EI} - \delta^\mathrm{BI})$ accounts for the relative minus sign in the expressions for the electric and magnetic parts.

\section{Computing memory signals from black-hole mergers} \label{sec:computingMemory}

This section summarizes the details of and the justifications for the approximations that we make in computing and defining the GW memory signals that we associate with a given moment of the news.
We also discuss the waveform models and NR data we use to assess the quality of these approximation.
Finally, we illustrate the signals for a few BBH systems.

\subsection{Oscillatory versus nonoscillatory memory signals} \label{subsec:nonosc}

While Eq.~\eqref{eqn:memorysignals} defined our prescription for computing displacement and higher memory signals, there remains the question of which $(l,m)$ modes should be included in our forecasts of the memory signals.
A closely related issue, of whether to average the nonlinear flux term $\E{\mathcal F}_0$ over the instantaneous period of the GWs, has been discussed recently in the literature~\cite{Inchauspe:2024ibs,Zosso:2026czc,Cogez:2026frh}.
When the averaging is performed, the resulting memory signal does not oscillate (and hence is described as ``nonoscillatory'') whereas when averaging is not performed, there will be oscillatory and nonoscillatory portions of the memory signal.

In this paper, we will focus on nonprecessing BBH systems, for which a frame is chosen so that the nonoscillatory modes predominantly arise in the $m=0$ spherical harmonics, and the oscillatory modes in the $m\neq 0$ harmonics.
This relationship of the $m=0$ modes with the nonoscillatory part and the $m\neq 0$ modes with the oscillatory part holds most precisely during the inspiral of the BBH system.
During the ringdown, however, there are damped oscillations in the waveform that make even the $m=0$ modes have an oscillatory component.
While we often use the $m=0$ modes as a rough proxy for the nonoscillatory part of the signal, in this context, we will have to make a further distinction between the oscillatory and the nonoscillatory part of the $m=0$ modes.\footnote{There also are small nonoscillatory components of the $m\neq 0$ waveform modes that arise during the merger phase of the BBH waveform.
We will neglect these portions of the memory signal in this paper.}

From a practical perspective, the distinction between the ``oscillatory versus nonoscillatory'' approaches is small for the displacement memory signal from BBH mergers, because the nonoscillatory part is larger than the oscillatory part of the full flux contribution to the memory signal.
In the post-Newtonian (PN) approximation, for example, after the flux integral is evaluated, the $m=0$ modes arise at leading Newtonian order in the waveform, whereas the $m\neq 0$ modes of the memory signal enter at 2.5 PN orders higher (namely, with a multiplicative factor of $(v/c)^5$~\cite{Favata:2008yd}).
From a theoretical perspective, there are arguments that can be made both in favor of and against using the oscillatory portions of the memory signal (see, e.g.,~\cite{Siddhant:2024nft,Inchauspe:2024ibs,Zosso:2026czc}).
Next, we will discuss some considerations pertaining to why we compute the memory signal from just the nonoscillatory parts of the waveform for all the higher memory effects in this paper (although we had previously computed oscillatory memory signals in the PN approximation in~\cite{Siddhant:2024nft}).

For the higher memory signals, including the nonoscillatory versus both oscillatory and nonoscillatory signals makes more of a difference.
As an example, for the spin memory, the relative size of the oscillatory and nonoscillatory contributions from the flux terms in the memory signal are comparable (though in the first moment of the news, the nonoscillatory contribution is still larger).
In the PN approximation again, the $m\neq 0$ and $m=0$ modes both enter at the same order in the waveform (2.5 PN orders above the leading Newtonian waveform)~\cite{Siddhant:2024nft}.
For the CM memory signal, the $m\neq 0$ contributions were larger than the $m=0$ ones (and even in the first moment of the news, the $m=0$ contribution was just 0.5PN orders larger than the $m\neq 0$ contribution)~\cite{Siddhant:2024nft}.

Another consideration related to this discussion about oscillatory and nonoscillatory contributions to the memory signals pertains to the relative size of the charge and flux contributions to a given spherical-harmonic mode of the signal.
Specifically, it seems more reasonable to focus on the flux contribution to a memory signal and neglect the charge contribution when the charge contribution can be shown to be smaller than the flux.
For the displacement and spin memory signals, the $m=0$ multipole moments of the memory signal have a flux contribution that is larger than the charge contribution (see, e.g.,~\cite{Mitman:2020pbt}).
However, from the relationships between the harmonic-coordinate-PN and Bondi-coordinate metric functions derived in~\cite{Blanchet:2020ngx,Blanchet:2023pce}, one can see that the flux terms and the nonlinear contributions to the charge enter at the same PN order for the $m\neq 0$ multipole moments.
Thus, it is not clear, \emph{a priori}, whether the charge and flux contributions would add or subtract.
For this reason, when computing the multipole moments of the flux contributions to the displacement and spin memory signals, we will use just the $m=0$ modes and flux contributions in our computations.

For the CM and ballistic memory signals, the charge contributions can be comparable or larger than the flux or pseudoflux contributions in some waveform modes~\cite{Grant:2023jhd}; this is why we will use the stationary-to-stationary approximation in Sec.~\ref{subsec: st-st charge}.\footnote{In fact, it was noted in~\cite{Siddhant:2024nft}, for example, that there was a portion of the pseudoflux contribution to the ballistic memory signal that was precisely canceled by a charge contribution at the same order in the PN approximation.
This highlights the importance of modeling both charge and flux when they are of comparable sizes.}
We will still focus on the $m=0$ spherical-harmonic waveform modes of the charge contributions to these signals, so as to avoid oscillatory signals during the inspiral.
However, as noted above, the $m=0$ modes have both oscillatory and nonoscillatory components (because of the damped oscillations during the ringdown), and in the charge term the oscillatory part of the $m=0$ mode can exceed the nonoscillatory part.
The stationary-to-stationary approximation that we introduced in Sec.~\ref{subsec: st-st charge} captures a part of the signal that leads to the offset in the moment of the news and does not have as significant oscillatory components.
We provide simplified expressions for some of the higher memory signals restricted to the $m=0$ and nonoscillatory parts in the next subsection.

\subsection{Nonoscillatory higher memory signals for nonprecessing binaries} \label{subsec:signals_nonprecessing}

When specializing to the $m=0$ modes of the higher memory signals of nonprecessing BBHs, the expressions for these signals in Eq.~\eqref{eqn:memorysignals} simplify to varying degrees, and certain choices of initial or final data need to be made for the Bondi metric functions.
We discuss these simplifications, choices, and a few additional approximations that we make in this subsection.

First, we introduce the following notation to distinguish between the $m\neq 0$ and $m=0$ waveform modes: 
\begin{equation}
    C^{\mathrm I}_{lm}=
    \begin{cases}
        C^{\mathrm{osc},\mathrm I}_{lm} \;\; & \mathrm{if} \;\; m\neq0 , \\
        C^{\mathrm{nonosc},\mathrm I}_{lm} \;\; & \mathrm{if} \;\; m=0 .
    \end{cases}
\end{equation}
As noted in Sec.~\ref{subsec:nonosc}, while the ``oscillatory'' and ``nonoscillatory'' terminology is used, it is a bit of a misnomer, as the $m=0$ modes have decaying quasinormal-mode oscillations, and the $m\neq 0$ modes can develop small memory offsets that are not manifestly oscillatory.
For nonprecessing binaries, the oscillatory modes $(m\neq0)$ are much larger than the nonoscillatory ones ($m=0$).
For this reason, we use just the oscillatory modes in the computation of the fluxes and pseudofluxes.
It will also be convenient to make this approximation, as we employ the \textsc{NRSur7dq4} waveform model to generate memory signals for the detection forecasts discussed in Sec.~\ref{sec:forecastMethods}.
This surrogate is fast and spans a large parameter space; however, it does not contain memory signals in the $(m=0)$ modes of the waveform in the co-precessing frame.

Thus, the main simplification we use in computing the displacement memory in Eq.~\eqref{eq:Clm0S2S} is that we replace the $\dot C^{\mathrm I'}_{l'm'}$ in $(\E{\mathcal F_0})_{lm}$ with $\dot C^\mathrm{osc, I'}_{l'm'}$ (and we restrict to oscillatory modes with $l \leq 4$ for computational efficiency).
This approximation, therefore, neglects the contributions of the $m=0$ modes to the memory signal.
Because we restrict to the $m=0$ modes of the memory signal, the corresponding multipoles of $C^\mathrm{S}_{AB}$ will have even $l$ and $m=0$ for nonprecessing BBH systems.
This is a special case of the fact that for nonprecessing BBH systems, the electric-parity waveform modes have $l+m$ even and the magnetic-parity modes have $l+m$ odd.

A similar approximation of replacing $\dot C^{\mathrm I'}_{l'm'}$ in the fluxes $(\p{\mathcal F_1})_{lm}$ with $ \dot C^\mathrm{osc, I'}_{l'm'}$ (and similarly without the $u$ derivative) holds for the drift memory.
To compute the charge contribution to the nonoscillatory multipoles of the drift-memory signal from $\p Q_\mathrm{1,S}$ (equivalently $\hat N_A^\mathrm{S}$) in Eq.~\eqref{eqm:drift charge STS}, we must specify initial values of $\mass^\mathrm{S}$ and $C^\mathrm{S}_{AB}$ that enter into $\dot {\hat{N}}^{\mathrm S, \mathrm I}_{lm}$ in Eq.~\eqref{eq:Clm1S2S}.
For the initial value of $\mass^\mathrm{S}(u_0)$, we use the total Arnowitt-Deser-Misner mass~\cite{Arnowitt:1960es} of the BBH system at the initial $u_0$ in our calculations.\footnote{In the limit $u_0 \rightarrow -\infty$ and in the rest frame of the system, the stationary value of the mass aspect $\mass^\mathrm{S}(u_0)$ will coincide with the full value of the mass aspect, which will coincide with the total Arnowitt-Deser-Misner mass.}
For the initial value of $C^\mathrm{S}_{AB}$, we choose it to be zero as $u_0 \rightarrow-\infty$.

Similarly, the fluxes and pseudofluxes of the ballistic memory are computed using the same approximations as described for the flux of the drift memory.
The nonoscillatory multipoles of the charge part of the ballistic memory are computed from Eqs.~\eqref{eqn:E0stationary} and~\eqref{eqn:G3_lm} in the stationary limit. 
The charge computation requires the multipoles $(N_A^\mathrm{S})_{lm}$ both for $l>1$ in Eq.~\eqref{eqn:N_stationary} and for $l=1$ in Eq.~\eqref{eqn:N_kernel}. 
Again, because we restrict to $m=0$ modes of the ballistic memory, and because we restrict to $m=0$ modes for the stationary shear $C^{\mathrm{S,E}}_{lm}$, only the $m=0$ modes of $(N_A^\mathrm{S})_{lm}$ will be used in the computation of the charge in Eq.~\eqref{eqn:G3_lm}. 
The multipoles of the electric-parity part of ballistic memory are nonvanishing for $l+m$ being even, and the magnetic-parity parts are nonzero for $l+m$ odd (as with any other waveform mode of a nonprecessing BBH system).
Given the properties of the Gaunt coefficients appearing in Eq.~\eqref{eqn:G3_lm}, when restricting to $m=0$ multipoles of the charge (and given that only $m=0$ modes are used in the stationary shear and angular-momentum aspect), the electric-parity part of Eq.~\eqref{eqn:N_kernel} will not be required for the computation of the ballistic charge.
However, the $l=1$ magnetic-parity part of the angular-momentum aspect $N_A^\mathrm{S}$ will enter into the computation of the charge contribution to the ballistic memory.

Near the peak of the waveform for BBH mergers, the $l = 1$ pieces of $N_A^\mathrm{S}/\mass^2_\mathrm{S}$ are order unity, whereas the $l \geq 2$ pieces of $C^{\mathrm S}_{AB}/\mass^\mathrm{S}$ are order $10^{-1}$.
Thus, the largest multipole of the angular-momentum aspect $N_A^\mathrm{S}$ will come from the $l = 1$ part, but it contributes just to the magnetic-parity part of the ballistic memory signal for the $m=0$ modes of nonprecessing BBHs.
We discuss how we compute the initial data for the $(l,m)=(1,0)$ part of $N_A^\mathrm{S}$ in Sec.~\ref{subsec:accuracy}.
Although smaller, we will use the $l \geq 2$ parts of $N_A^\mathrm{S}$, because they will be relevant for computing the electric-parity part of the ballistic memory signal.

The relative size of the contributions of the different parts of the charge and flux terms merits a brief discussion.
As noted above, the part of the charge contribution involving the product of the $l=1$ part of $N_A^\mathrm{S}$ multiplied by the nonoscillatory part of the shear will be the largest part of the charge (the $l > 1$ parts of $N_A^\mathrm{S}$ times the nonoscillatory shear will be an order of magnitude or more smaller).
The flux term is cubic in the shear and its time derivatives, whereas the pseudoflux $\mathcal G_\mathrm{2,R}^{AB}$ is quadratic in the shear.
Given that we restrict to $m=0$ modes of the ballistic memory, the largest terms in the flux will be of the form of two oscillatory modes of the shear times one nonoscillatory mode, whereas the largest terms in the pseudoflux will involve products of two oscillatory modes.
Because the stationary shear normalized by the mass is a quantity less than one, the cubic terms in the flux are small, and we will neglect them in our forecasts.
Thus, for the ballistic memory, we use the terms that are quadratic in the oscillatory shear to compute the residual pseudoflux term (and one term involving the $l=1$ moments of the mass aspect).
We give the multipolar expansion $(\mathcal{G}_{2,\mathrm R})_{lm}^\mathrm{I}$ as it undergoes some simplifications after applying the approximations noted above: 
\begin{widetext}
\begin{align} \label{eq:G2Rl0}
     (\mathcal{G}_{2})_{l0}^{\mathrm R,\mathrm{I}} = \frac{1}{16\sqrt{2}}\sum_{\mathrm I'', l''} \sum_{l',m'} &  \Bigg[ \Big\{{\mathcal A_{l(-2)}} ({^* C}^{\mathrm{osc},\mathrm E} _{l'm'} \;{^* C}^{\mathrm{osc},\mathrm I''} _{l''(-m')}+ {C}^{\mathrm{osc},\mathrm E} _{l'm'}{C}^{\mathrm{osc},\mathrm I''}_{l''(-m')})+4\sqrt{2}\delta_{l'1} \mass_{1m'} C^{\mathrm{osc},\mathrm I''}_{l''(-m')}\Big\} \eta^\mathrm{E I I''}_{ll'l''} \mathscr{C}^{lm,02}_{l'm'l''(-m'),I''} \nonumber \\
     & - \frac{\mathcal B_{l1}}{6} \sum_{\mathrm I'} {C}^{\mathrm{osc},\mathrm I'} _{l'm'}{C}^{\mathrm{osc},\mathrm I''}_{l''(-m')} \eta^\mathrm{I I' I''}_{l l' l''}  \mathscr B^{lm}_{l'm'l''(-m')} \Bigg]+ O(C_{\mathrm{osc}}^3) .
\end{align}
\end{widetext}
The relative size of the charge and flux terms is less straightforward to determine from first principles, and we instead compute the numerical values of the related moments of the news and memory signals in the next two subsections.

\subsection{Accuracy of the stationary-to-stationary approximation and relative scale of charge and flux terms for nonprecessing binaries} \label{subsec:accuracy}

To determine how well the stationary-to-stationary approximation can capture the offset in the first and second moments of the news, we require NR simulations that include asymptotic Bondi data not just related to the gravitational waveform (which many NR codes extract), but also the Bondi mass, angular-momentum, and higher-multipole aspects.
This data can be obtained from the Simulating eXtreme Spacetimes (SXS) CCE catalog, which contains simulations that include the leading-order parts of the Newman-Penrose Weyl-curvature scalars~\cite{Boyle:2019kee,Scheel:2025jct}.\footnote{The current version of the main SXS catalog~\cite{SXScatalog} contains waveform and horizon data, but not all the Weyl scalars. 
The CCE catalog contains the relevant Weyl scalars, but warns that it is an older version of the catalog that is deprecated.
We use this data nevertheless, given the lack of the relevant data in the main SXS catalog.}

At sufficiently large separations, the inspiral of a BBH system is well approximated by a stationary-to-stationary transition.
Specifically, the velocity of the black holes goes to zero inversely with the square root of the separation at early times (thereby making the initial state approximately stationary). 
At late times, the final black hole is precisely stationary only for nonspinning, equal-mass binaries, because there is no gravitational-wave recoil for such a merger.
For unequal mass ratios, the remnant black hole does have a kick velocity, which is directed in the initial orbital plane of the binary, and has a magnitude of order $10^{-3} c$.
Because the kick in this case is in the orbital plane, it will have a small impact on the $m=0$ spherical harmonic modes of the memory signals.
The addition of nonprecessing spins is unlikely to change this significantly, whereas systems with spins in the orbital plane can have larger kicks perpendicular to the orbital plane, which will have a larger impact on the $m=0$ waveform modes.
For this reason, we specialize to nonprecessing binaries in our calculations in the rest of the paper.

In the late inspiral, merger, and early ringdown stages of the waveform, we do \emph{not} expect the stationary-to-stationary approximation to accurately represent the \emph{full} charge contribution to the first or second moment of the news.
However, we do expect that it will serve as a good approximation to the slowly varying, nonoscillatory part of the signal that gives rise to the offset in the first or second moment of the news.
This is analogous to how the flux contribution to the displacement memory signal captures the nonoscillatory accumulation of a nonzero offset in the $(l,m)=(2,0)$ gravitational waveform, but there is additional quasinormal-mode ringing in the waveform that does not contribute to the offset (and similar statements hold about the spin memory signal in the $(l,m)=(3,0)$ waveform mode).
Unlike the displacement or spin-memory cases, where the memory portion of the signal is larger than the quasinormal-mode portion, for the higher moments of the news, the charge contribution to the memory signal can be smaller than the other oscillatory, zero-mean portions of these signals.
For this reason (and because the displacement and spin memory effects have been studied in more detail elsewhere~\cite{Mitman:2020pbt,Grant:2023jhd}), we focus on the CM memory and the electric and magnetic-parity parts of the ballistic memory.

\begin{figure*}
  \centering
  \includegraphics[width=\columnwidth]{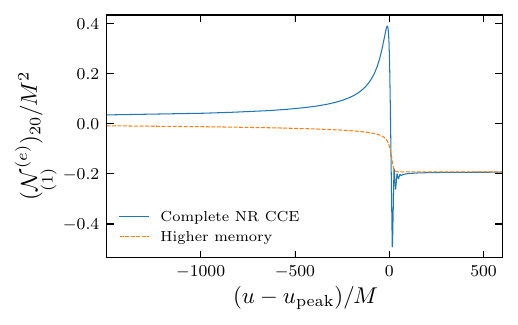}
  \includegraphics[width=\columnwidth]{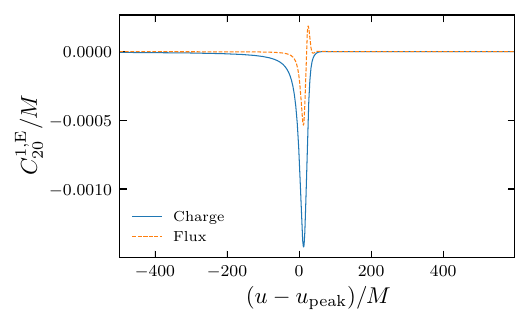}
  \includegraphics[width=\columnwidth]{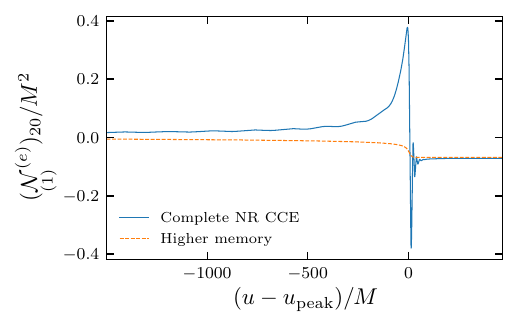}
  \includegraphics[width=\columnwidth]{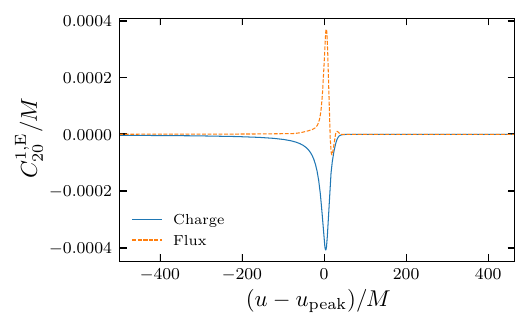}
  \caption[Higher memory signals for two black-hole mergers]{\textbf{Higher memory signals for two black-hole mergers.}
  \emph{Left column}: The time dependence of the $(l,m)=(2,0)$ charge contribution to the first moment of the news, $(\E{\ord{1}{\mathcal N}})_{20}$.
  The solid blue curve is computed directly from the full NR CCE data from~\cite{SXScatalogCCE}, which contains both oscillatory and nonoscillatory parts.
  The dashed orange curves are computed in the stationary-to-stationary approximation described in Sec.~\ref{sec:stationary}.
  Both cases use the additional approximations described in Sec.~\ref{subsec:signals_nonprecessing}.
  The stationary-to-stationary approximation captures the accumulation of the offset in the first moment of the news without capturing the oscillatory features that do not produce such an offset. 
  \emph{Right column}: The time dependence of the flux $(\mathcal F_1)^{\mathrm E}_{20}$ and charge ${({\dot Q}_1)^{\mathrm S, \mathrm E}_{20}}$ contributions to the higher memory signal $C^\mathrm{1,E}_{20}$.
  The blue solid curve is the charge contribution and the dashed orange curve is the flux.
  \emph{Top row}: Results for a $q=1$, nonspinning BBH merger.
  \emph{Bottom row}: Results for a $q=4$, nonspinning BBH merger.
  The smaller change in the moment on the news for the mass ratio $q=1$ versus $q=4$ in the left column arises because the displacement memory signal $C^{0, \mathrm E}_{20}$ decreases with increasing $q=m_1/m_2$.
  The relative contributions of the charge and flux terms depend on the mass ratio, as illustrated in the right column.
  }
  \label{fig:Q1_e_nospin}
\end{figure*}

To illustrate this, we calculate, for binaries with two different mass ratios, the $(l, m) = (2, 0)$ spherical harmonic moment of the charge contribution to the first moment of the news, $(\E{\ord{1}{\mathcal N}})_{20}$.
We also compute both charge and flux contributions to the corresponding memory signal in the shear, $C^\mathrm{1,E}_{20}$, in the stationary-to-stationary approximation.
We use the convention that $m_1$ is the primary mass, $m_2$ is the secondary, and $q = m_1/m_2 \geq 1$ is the mass ratio (and we denote $M=m_1+m_2$ as the total mass).
The results appear in Fig.~\ref{fig:Q1_e_nospin} for the mass ratios $q=1$ (top row) and $q=4$ (bottom row); the left column is the moment of the news $(\E{\ord{1}{\mathcal N}})_{20}$ and the right is the shear $C^\mathrm{1,E}_{20}$.

In the left panels, the solid blue curves are the full charge computed from the shear and Weyl scalars using the relationships between the Bondi metric functions and the Newman-Penrose scalars and tetrad given in~\cite{Grant:2023jhd}.
More details about this are given in Appendix~\ref{app:Bondi2NP}.
The NR data (available at~\cite{SXScatalogCCE}) come from the simulation SXS:BBH\_ExtCCE:0001 for the $q=1$ case (top) and SXS:BBH\_ExtCCE:0010 for the $q=4$ case (bottom).
BMS frame fixing~\cite{Mitman:2024uss}, implemented in the \textsc{scri} package~\cite{Boyle:2013nka,Boyle:2014ioa,Boyle:2015nqa}, was used to transform the data to the super rest frame of the system, which eliminated some additional oscillations that appear in the data without frame fixing.

The dashed orange curves are computed using the stationary-to-stationary approximation using the prescriptions described in Secs.~\ref{sec:stationary} and~\ref{subsec:signals_nonprecessing}.
Aside from the initial data for $m^\mathrm{S}$ and $N^\mathrm{S}_A$, the stationary-to-stationary part of the charges can be computed from the Bondi shear.
We use the oscillatory waveform modes of the corresponding CCE simulation to compute the stationary-to-stationary part of the charge (including the mass aspect $\mass^\mathrm{S}$ that enters into it) and the memory waveform modes $C_{lm}^\mathrm{0,E}$ in Eqs.~\eqref{eq:F0lm} and~\eqref{eq:CAB0stationary}.
We also use the CCE data to obtain the initial value of $m^\mathrm{S}$.
Just the $(l, m) = (2, 0)$ modes for the shear $C_{lm}^\mathrm{S,E}$ were used in the computation of $\Delta (\E{\mathcal Q}_\mathrm{1,S})_{20}(u)$, though $C_{lm}^\mathrm{S,E}$ and $\mass^\mathrm{S}$ are both computed from multiple oscillatory CCE modes.

There is good agreement between the final offset in the stationary-to-stationary calculation and the full NR charge at late times, both when there is no kick ($q=1$; truly stationary) and when there is a kick ($q=4$; approximately stationary).
There is a somewhat larger mismatch at early times, because the individual black holes in the binary have more relativistic velocities when the binary is of order $10^3M$ from merger.
Nevertheless, our prescription for computing the charge contribution described in Sec.~\ref{sec:stationary} does capture the nonoscillatory accumulation of the first moment of the news from the charge.

In the right column of Fig.~\ref{fig:Q1_e_nospin}, the flux contributions, $(\mathcal F^{\mathrm E}_1)_{20}$, to the higher memory signal $C^\mathrm{1,E}_{20}$ are the dashed orange curves.
The charge contribution in the stationary-to-stationary approximation, $(\E{\dot Q}_{1,\mathrm S})_{20}$, are the solid blue curves (and they are equivalent to the $u$ derivative of the orange dashed curves on the left).
Although the charge contribution to the signal is somewhat larger for both mass ratios ($q=1$ top and $q=4$ bottom), the flux contribution is comparable.
Notably, the relative sign of the flux and charge terms changes between the $q=1$ and $q=4$ cases (similar behavior of different angular momentum charges was observed in~\cite{Elhashash:2021iev}).
It, therefore, will be important to include both charge and flux terms when computing the memory signal associated with the first moment of the news.

\begin{figure*}
  \centering
  \includegraphics[width=\columnwidth]{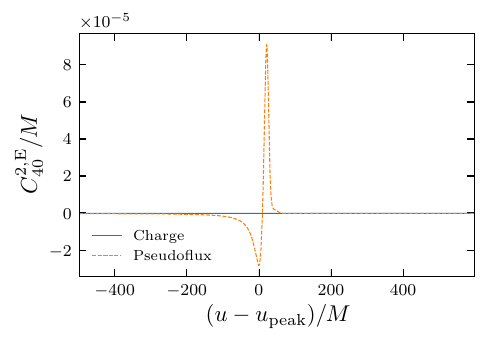}
  \includegraphics[width=\columnwidth]{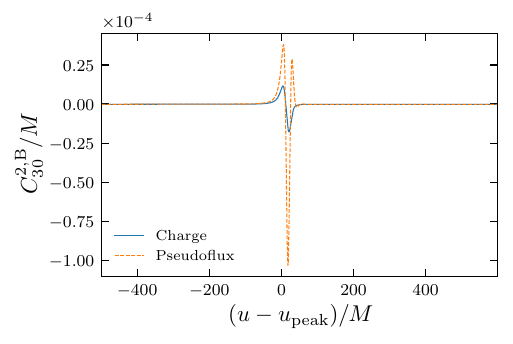}
  \includegraphics[width=\columnwidth]{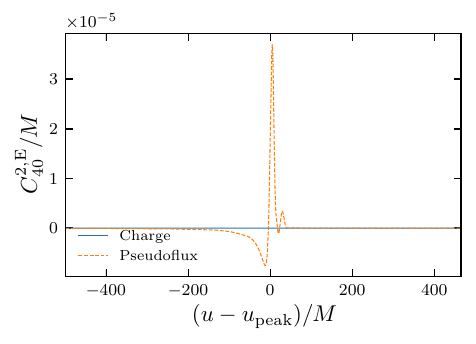}
  \includegraphics[width=\columnwidth]{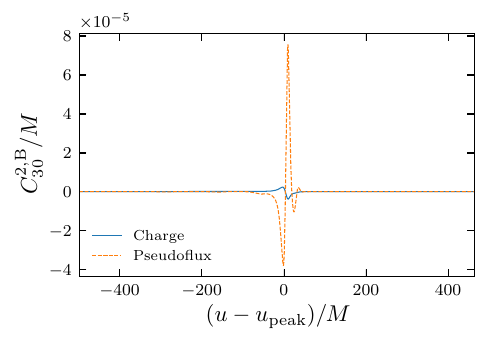}
  \caption[Charge and flux contributions to the ballistic memory signal for two BBH mergers]{\textbf{Charge and flux contributions to the ballistic memory signal for two BBH mergers.}
  In all panels, the dashed orange curves are pseudoflux contributions to the higher memory signal associated with the second moment of the news, and the solid blue curves are the charge contributions computed in the stationary-to-stationary approximation.
  The charge and pseudoflux contributions to the shear are computed using the prescriptions given in Secs.~\ref{sec:stationary} and~\ref{subsec:signals_nonprecessing}; we neglect the flux contribution as it is cubic in the shear.
  \emph{Left column}: The $(l,m)=(4,0)$ spherical harmonic mode of the higher memory signal associated with the second moment of the news.
  The pseudoflux term is substantially larger than the charge term for this mode.
  \emph{Right column}: The $(l,m)=(3,0)$ spherical harmonic mode of the higher memory signal associated with the second moment of the news.
  Here the charge and pseudoflux contributions are more comparable than they are in the left column.
  However, the pseudoflux terms are somewhat more oscillatory than they are in the left column.
  \emph{Top}: Results for an equal-mass ($q=1$), nonspinning BBH merger.
  \emph{Bottom}: Results for an unequal-mass ($q=4$), nonspinning BBH.
  }
  \label{fig:Q2_e_nospin_shear_contri}
\end{figure*}

We have also compared the charge contribution to the second moment of the news for the complete NR with CCE case and for the stationary-to-stationary approximation.
These results are given in Appendix~\ref{app:moreCharges}.
The stationary-to-stationary approximation still works well for capturing the final offset in the second moment; however, for the magnetic-parity part, the stationary-to-stationary approximation agrees less well with the NR CCE results.
As is shown in more detail in Appendix~\ref{app:moreCharges}, there are some irregularities in the NR data, which make it difficult to determine whether the differences arise from limitations in the approximation or in the accuracy of the NR simulations.

For the ballistic memory, given the approximation of neglecting terms that are cubic in the shear (see Sec.~\ref{subsec:signals_nonprecessing}), we will not compute the moments of the fluxes $(\mathcal F_2)^\mathrm{I}_{l0}$, which are strictly cubic, and instead compute the quadratic part of the pseudoflux $ (\mathcal{G}_{2})_{l0}^{\mathrm R,\mathrm{I}}$ in Eq.~\eqref{eq:G2Rl0}.
As we show in Fig.~\ref{fig:Q2_e_nospin_shear_contri}, however, the contributions of the charge terms to the ballistic memory signals are smaller than those of the pseudoflux term for the $(l,m)=(4,0)$ modes (left column) and $(l,m)=(3,0)$ modes (right column).
Aside from the different modes plotted, the panels in this figure are depicted in the same way as those in the right column of Fig.~\ref{fig:Q1_e_nospin} (though we also require the initial value of the angular-momentum aspect, which we obtain from the CCE data, as well).
These results show that the issue of the quality of the stationary-to-stationary approximation will have a much smaller effect on the $(l,m)=(4,0)$ mode than it will on the $(l,m)=(3,0)$ mode.
For this reason, we will compute only the electric-parity $(l,m)=(4,0)$ mode in our forecasts.

There are a few features of the pseudoflux terms that we discuss.
In the left column, the pseudoflux is more oscillatory for the $q=4$ case (bottom) than the $q=1$ case (top) after the peak time $u_\mathrm{peak}$.
We suspect this is related to the fact that modes with $l+m$ even but $m$ odd have larger amplitudes in the unequal mass case than in the equal mass case; moreover, these modes are not in phase, due to their different quasinormal mode frequencies.
This leads to the slightly oscillatory behavior as the signal approaches zero at late times.
The $(l,m)=(3,0)$ mode has a similar behavior for unequal and equal masses, but the flux term undergoes a few more oscillations than the $(l,m)=(4,0)$ case on the left.

\subsection{Higher memory signals from a BBH like GW150914} \label{subsec:higherMemoryBBH}

Because in Sec.~\ref{subsec:accuracy} we showed the stationary-to-stationary approximation produces a reasonable notion of a nonoscillatory signal associated with offsets in the moments of the news, we now can use this prescription to compute the corresponding higher memory signals (i.e., the gravitational-wave strain). 
As a specific example, we consider source parameters similar to those inferred for GW150914~\cite{LIGOScientific:2016aoc}, the first GW event observed by the LIGO detectors.
Specifically, we assume that the dimensionless spin magnitudes vanish ($\chi_1=\chi_2=0$), the total source-frame mass is $M=65 \mathrm{M}_\odot$, the mass ratio is $q=1.25$, the luminosity distance is $d=410 \,\mathrm{Mpc}$, and the inclination angle is $\iota=\pi/6$. 
We generate the oscillatory GW modes using the NR surrogate \textsc{NRSur7dq4}~\cite{Varma:2019csw}, and we compute the associated higher memory signals.
We compute the strain contributions associated with the displacement, drift, and ballistic memory effects, and we analyze their time and frequency dependencies.

We will plot the GW polarizations $h_+$ and $h_\times$, which are defined from the multipole moments of the strain $h_{lm}$ by
\begin{equation} \label{eq:hPlusTimes}
    h_+ - i h_\times = \sum_{l,m} h_{lm} (_{-2}Y_{lm}) = \frac 1 r C_{AB} \bar m^A \bar m^B.
\end{equation}
The vectors $m^A$ and $\bar m^A$ are a complex dyad, which is described in more detail in Appendix~\ref{app:harmonics}.
We denote the polarizations of the oscillatory part of the signal by $h_{+,\times}^\mathrm{osc}$, which are related to the $m\neq 0$ spherical harmonic modes, $C_{lm}^\mathrm{osc,I}$.
We denote the displacement memory, computed in the stationary-to-stationary approximation by $h_+^\mathrm{disp}$, which is generated by the $m=0$ modes of $C_{AB}^\mathrm{0,E}$ (which includes just the flux contribution).
The spin and CM memory signals arise from the magnetic- and electric-parity parts of $C_{AB}^\mathrm{1}$, respectively, again in the $m=0$ modes.
These will be denoted by $h_\times^\mathrm{spin}$ and $h_+^\mathrm{CM}$.
Finally, the strain $C_{AB}^\mathrm{2,E}$ will have an associated plus polarization from the $(l,m)=(4,0)$ mode, which will be denoted by $h_+^\mathrm{bal}$.
We do not show the magnetic-parity part of $C_{AB}^\mathrm{2,B}$ in the $(l,m)=(3,0)$ mode, because we will not use it in our forecasts.

\begin{figure*}
  \centering
  \includegraphics[width=\columnwidth]{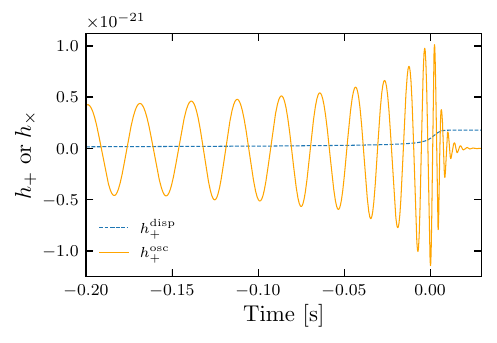}
  \includegraphics[width=\columnwidth]{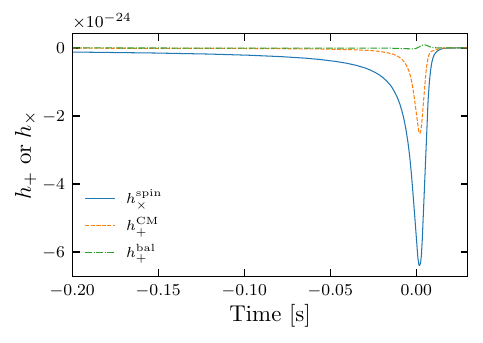}
  \includegraphics[width=\columnwidth]{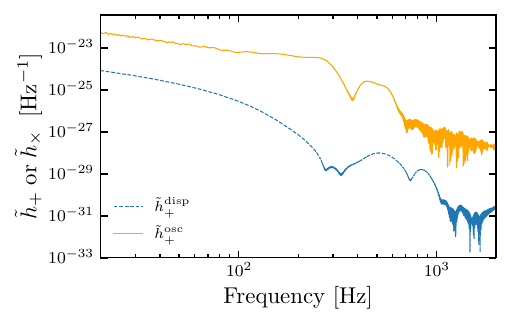}
  \includegraphics[width=\columnwidth]{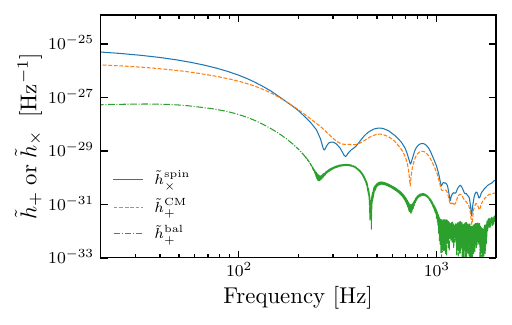}
  \caption[Time- and frequency-domain oscillatory strain, displacement memory, and higher memory signals for a BBH merger similar to GW150914]{\textbf{Time- and frequency-domain oscillatory strain, displacement memory, and higher memory signals for a BBH merger similar to GW150914.}
  \emph{Top row}: The waveforms in the time domain.
  Note that the memory signals accumulate primarily around the peak of the waveform at a time of zero, whereas the oscillatory signals have an amplitude with less variation over the times shown.
  \emph{Bottom row}: The Fourier transform of contribution of the signals above.
  At low frequencies, the oscillatory signal scales as $(Mf)^{-7/6}$, whereas the memory signal goes as $(Mf)^{-1}$.
  The spin and CM memory signals approach a constant at low frequencies, and the ballistic memory grows linearly with $Mf$ at frequencies lower than those depicted in the plot.
  \emph{Left column}: In both panels, the oscillatory modes of the waveform are solid orange curves, and the $l=2$, $m=0$ modes of the displacement memory signal are dashed blue curves.
  \emph{Right column}: In both panels, the spin-memory signal is the solid blue curve, the CM memory signal is the dashed orange curve, and the ballistic memory signal is the dash-dotted green curve.
  }
  \label{fig:memory signals}
\end{figure*}

The left column of Fig.~\ref{fig:memory signals} shows the plus polarization associated with the oscillatory modes ($m\neq 0$) of the strain in the \textsc{NRSur7dq4} surrogate model, and the displacement-memory signal [specifically, the plus polarization associated with the $(l,m)=(2,0)$ mode].
The memory signal is the dashed blue curve in both panels and the oscillatory signal is the solid orange curve.
The top panel shows the signals in the time domain and the bottom contains the frequency-domain signals.
In the time domain, the amplitude of the final offset in the displacement memory is roughly one order of magnitude smaller than the peak amplitude of the dominant oscillatory modes.

Because GW data analysis is often performed in the frequency domain, it is also useful to show the frequency-domain representation of these signals, which is displayed in the bottom-left panel of Fig.~\ref{fig:memory signals}. 
The dominant quadrupole mode of the signal scales with frequency like $(Mf)^{-7/6}$, which can be shown from evaluating the Fourier transform of the time-domain quadrupole waveform in the stationary phase approximation.
The memory signal scales like $(Mf)^{-1}$ in the low frequency limit, which is similar to the scaling of the Fourier transform of a step function.
At higher frequencies, both memory and oscillatory signals fall off more rapidly with frequency, which is consistent with the behavior of a smooth time-domain signal.

While the memory is a low-frequency phenomenon, the ratio of the oscillatory to memory signals actually grows (slowly) like $(Mf)^{-1/6}$ as the dimensionless frequency goes to zero.
The memory does not exceed the amplitude of the oscillatory signal at higher frequencies either, because both signals fall off more rapidly at high frequencies, and the memory signal undergoes this turnover to a faster fall off rate at a lower frequency than the oscillatory signal does.
Thus, the memory signal remains about two orders of magnitude smaller than the dominant oscillatory at all frequencies shown in the bottom-left of Fig.~\ref{fig:memory signals}.\footnote{The features in the frequency-domain signals at frequencies above $\sim 500$\,Hz are likely related to interpolation ``noise'' from the NR surrogate, which was used to compute the oscillatory and displacement memory signals.}

The higher memory signals are displayed in the right column of Fig.~\ref{fig:memory signals}. 
The top and bottom panels of the right column Fig.~\ref{fig:memory signals}, respectively, show the time-domain and frequency-domain signals associated with the $(l,m)=(2,0)$ mode of the CM memory, the $(l,m)=(3,0)$ mode of the spin memory, and the $(l,m)=(4,0)$ mode of the ballistic memory. 
For the CM and spin memories, these correspond to the largest, nonoscillatory multipoles of the signals. 
For the ballistic memory, however, we used the stationary-to-stationary approximation to compute the $l\geq 3$ multipole moments of the charge contribution to the ballistic memory.
However, this prescription does not constrain the $(l,m)=(2,0)$ charge contribution.
While the flux and pseudoflux contributions to the $(l,m)=(2,0)$ mode of the ballistic memory can be computed, we do not compute these in isolation, because we cannot verify that the charge contribution does not (partially) cancel these fluxes.
As a result, we focus on the next higher nontrivial, nonoscillatory electric-parity mode, which is the $(l,m)=(4,0)$.

This restriction to the $(l,m)=(4,0)$ of the ballistic memory could have a large impact on the forecasts in Sec.~\ref{sec:results}.
For example, the $(l,m)=(4,0)$ mode of the displacement memory signal is an order of magnitude smaller than the $(l,m)=(2,0)$ mode; the same is true of the charge contribution to the CM memory.
Whether this trend also applies to the ballistic memory is not immediately apparent.
Verifying it with a calculation beyond the stationary-to-stationary approximation would require a more detailed calculation of the charge contribution, as in~\cite{Blanchet:2020ngx,Blanchet:2023sbv}.
We leave such a calculation for future work.
However, this implies that our results using the $(l,m)=(4,0)$ charge contribution to the ballistic memory are likely underestimates of the effect from BBH mergers.

The top-right panel of Fig.~\ref{fig:memory signals} shows that in the time domain, the peak amplitude of the spin and CM signals is an order of magnitude (or more) smaller than the lasting offset in the displacement memory signal.
Their time dependencies look like a single pulse, because when integrated once in time, they produce an offset in the first moment of the news.
The $(l,m)=(4,0)$ mode of the ballistic memory is another order of magnitude smaller than both the spin and CM memory signals.
While its amplitude is small on the scale of the figure, its time dependence has the form of the derivative of the spin and CM memory signals (see the top-left panel of Fig.~\ref{fig:Q2_e_nospin_shear_contri} for the full time dependence).
This is also to be expected, as when it is integrated twice in time, it will produce an offset in the second moment of the news.

The bottom-right panel of Fig.~\ref{fig:memory signals} shows the corresponding signals in the frequency domain.
The spin and CM memory effects approach constants at low frequencies, because their time-domain signals have the morphology of the derivative of the displacement memory.
Thus, the Fourier derivative theorem implies that the signal will appear qualitatively like the frequency times the memory signal.
The ballistic memory has a time dependence that is like the derivative of the spin and CM memory effects, so its low-frequency behavior would be expected to scale linearly with frequency.
One can see a decreasing trend with frequency at the lowest frequencies shown in Fig.~\ref{fig:memory signals}, but it does not extend to low enough frequencies that the linearly increasing frequency dependence is manifest.
Finally, all three signals, spin, CM, and ballistic, turn over at higher frequencies and have a more rapid falloff, as the displacement-memory and oscillatory GW signals in the bottom-left panel of Fig.~\ref{fig:memory signals} do.

Because the CM and spin memory signs are of comparable magnitudes, and the spin memory signal was shown to be potentially detectable in next-generation ground-based detectors~\cite{Grant:2022bla,Goncharov:2023woe}, the detection prospects are expected to be comparable.
We will show more concrete results in Sec.~\ref{sec:results} after describing the approach we take to make our forecasts in the next section (Sec.~\ref{sec:forecastMethods}).

\section{Methods for forecasting the higher-memory detection prospects} \label{sec:forecastMethods}

To compute the forecasts for the detection prospects for the higher memory signals we closely follow the methods outlined in~\cite{Grant:2022bla} as implemented in the code~\cite{GWForecasts}.\footnote{The current public version of the code is that used in~\cite{Grant:2022bla}; an update to the code associated with this paper will appear soon.}
We give a brief review of the main elements of the forecast methods below and highlight a few relevant differences.
Further details about the methods can be found in~\cite{Grant:2022bla}.
The main elements of this method are using an effective signal-to-noise ratio (SNR) as a proxy for the total Bayes factor, limiting to events that satisfy a waveform degeneracy-breaking criteria in the higher GW multipole moments of the oscillatory GW signal, simulating synthetic populations of BBH mergers, and evaluating the effective SNR for these synthetic populations for appropriate waveform models and detector configurations.
These four constituents are described in the subsections below.

\subsection{Effective signal-to-noise ratio and Bayes factor}

Two of the main approaches to searching for evidence for the presence of GW memory in a population of BBH mergers have used Bayesian evidence ratios and hierarchical Bayesian inference (see, e.g.,~\cite{Lasky:2016knh,Hubner:2019sly,Hubner:2021amk,Cheung:2024zow,Rossello-Sastre:2026gah} and~\cite{Cheung:2024zow,Mitman:2026zfg}, respectively).
These approaches are applicable to the GW events measured by the LVK Collaboration, and they require posterior distributions for the 15 parameters that describe an inspiral of a BBH merger for every event in the population.

For making forecasts with ensembles of simulated populations of BBH mergers with CE, which involves simulating $O(10^2)$ populations (to account for variance at the population level), each with $O(10^4)$ events per year, running full Bayesian inference on each event to obtain the relevant posterior distributions is quite computationally expensive.
Instead, Ref.~\cite{Grant:2022bla} used an effective SNR introduced in~\cite{Lasky:2016knh} as a metric for the evidence in the BBH population, because~\cite{Grant:2022bla} showed that the natural logarithm of the expectation value of the Bayes factor reduced to the square of the SNR in the limit that the posterior distributions of the parameters are localized to exactly the true values of the parameters.
We review the primary features of this argument below.

The method starts with two hypotheses, $\mathscr{H}_\m$ and $\mathscr{H}_\nm$, where the signal model for the hypothesis $\mathscr{H}_\m$ is given by
\begin{equation}
    h_\s = h_{\nm} + h_{\m} ,
\end{equation}
and where $h_{\m}$ refers to the displacement memory or a higher memory signal (such as those illustrated in Sec.~\ref{sec:computingMemory}), and $h_{\nm}$ refers to just the $m\neq 0$ oscillatory modes of the GW strain that do not include any of the (higher) memory effects.
The waveform models are functions of a set of parameters, which will be denoted by $\theta$ (which represents however many parameters are needed to describe the model).
The basis of the method is Bayes' theorem, which is a statement about the equivalence of different ways of representing joint and conditional probabilities.
One description involves the likelihood $\mathscr{L}\boldsymbol(d|h(\theta)\boldsymbol)$ of the data $d$ given a signal model $h(\theta)$ associated with a hypothesis $\mathscr{H}$ and the prior probability distribution $\pi(\theta)$ for the parameters of the model $h(\theta)$.
The other makes use of a posterior probability distribution for the parameters conditioned on the data $p\boldsymbol(h(\theta)|d\boldsymbol)$ and the evidence for the data $Z(d)$.
Bayes' theorem states that these distributions are related by
\begin{equation}
    p\boldsymbol(h(\theta)|d\boldsymbol) Z(d) = \mathscr{L}\boldsymbol(d|h(\theta)\boldsymbol) \pi(\theta) .
\end{equation}
The evidence can be computed from integrating the right-hand side of Bayes' theorem over the entire parameter space, $\theta$, of the model:
\begin{equation}
    Z(d)=\int \ud \theta \mathscr{L}\boldsymbol(d|h(\theta)\boldsymbol) \pi(\theta) .
\end{equation}
Because the signal models $h_\s$ and $h_\nm$ depend on the same parameters $\theta$, the same prior $\pi(\theta)$ will be used for both hypotheses, and the evidence $Z(d)$ will differ in the two cases because of the influence of the signal model on the value of the likelihood function over the parameter space $\theta$.

The ratio of the evidences is defined as the Bayes factor, 
\begin{equation} \label{eq: Bayes fac}
    \mathscr{B}^\m _{\nm} \equiv \frac{Z_\m(d)}{Z_\nm(d)} = \frac{\displaystyle \int \ud \theta \mathscr{L}\boldsymbol(d|h_\m(\theta)\boldsymbol) \pi(\theta)}{\displaystyle \int \ud \theta \mathscr{L}\boldsymbol(d|h_\nm(\theta)\boldsymbol) \pi(\theta)} .
\end{equation}
Given that the priors for the hypotheses $\mathscr{H}_\m$ and $\mathscr{H}_\nm$ are equivalent in this case, Bayes' theorem can be used to eliminate the prior in the numerator of the the evidence ratio in Eq.~\eqref{eq: Bayes fac} so as to write it in terms of the integral of the likelihood ratio when averaged over the posterior of the $\mathscr{H}_\nm$ hypothesis:
\begin{equation} \label{eq: Bayes fac simpified}
    \mathscr{B}^\m _{\nm}= \int \ud \theta \frac{\mathscr{L}\boldsymbol(d|h_\m(\theta)\boldsymbol)}{\mathscr{L}\boldsymbol(d|h_\nm(\theta)\boldsymbol)} p\boldsymbol(h_\nm(\theta)|d\boldsymbol) .
\end{equation}
In this form, it is more evident that a Bayes factor greater than one corresponds to support for the hypothesis $\mathscr{H}_\m$, because the likelihood ratio of the two hypotheses favors $\mathscr{H}_\m$, when averaged over the posterior distribution of the $\mathscr{H}_\nm$ hypothesis.

As in~\cite{Grant:2022bla}, we assume that the data consists of a GW signal characterized by some parameters $\theta_0$ and additive Gaussian noise $n$ which is stationary, zero-mean, and characterized by a power spectral density $S_n(f)$.
The data is assumed to have both oscillatory and memory contributions, so that it can be written as
\begin{equation}
    d = h_\nm(\theta_0) + h_\m(\theta_0) + n .
\end{equation}
Under this assumption for the noise, the likelihood takes the form
\begin{equation}
    \mathscr{L}(d|h(\theta)) \propto \exp\left(-\frac{1}{2}\rho_{d-h}^2\right).
\end{equation}
The square of the SNR of a signal $\rho^2_h$ in the expression above is defined most naturally from the noise-weighted inner product of two signals: 
\begin{equation} \label{eq:InnerProduct}
    \langle a | b \rangle = 4\Re \int_{f_\mathrm{low}}^{f_\mathrm{high}} \ud f\; \frac{\tilde a(f) \bar{\tilde b}(f)}{S_n(f)} .
\end{equation}
With this notation, the SNR is 
\begin{equation} \label{eq:SNR}
    \rho_h^2 = \langle h | h \rangle .
\end{equation}

Thus far, no significant approximations were made in evaluating the evidence ratio $\mathscr{B}^\m _{\nm}$.
The main approximation in~\cite{Grant:2022bla} was to assume that the posterior distribution is a delta function at the true parameters:
\begin{equation} \label{eq:deltaPosterior}
    p(h_\nm(\theta)|d) \approx \delta(\theta-\theta_0) 
\end{equation}
(the $\approx$ symbol denotes this is an assumption about the posterior).
In this case, the evidence ratio in Eq.~\eqref{eq: Bayes fac simpified} reduces to the likelihood ratio of the models at the true parameters:
\begin{equation} \label{eq:BayesDelta}
     \mathscr{B}^\m _{\nm} \approx \exp\left( \frac 12 \rho^2_{h_\m(\theta_0)} + \langle h_\m(\theta_0)|n\rangle \right) .
\end{equation}
Although Eq.~\eqref{eq:BayesDelta} is an exact expression given the posterior in~\eqref{eq:deltaPosterior}, we still denote the relationship with an $\approx$ symbol given this strong assumption about the form of the posterior.
The inner product $\langle h_m(\theta_0)|n\rangle$ for different noise realizations $n$ is a zero-mean random variable with higher moments that depend on the SNR of the memory signal. 
When taking the expected value (denoted by $\mathbb E[\ldots]$) of the exponential of this variable, it was shown in~\cite{Grant:2022bla} that the natural logarithm is given by
\begin{equation} \label{eq: Bayes-SNR}
    \ln \mathbb E[\mathscr{B}_\nm^\m] \approx \rho^2_{h_\m} .
\end{equation}
Because each GW event is independent, the total Bayes factor in a population of BBH mergers is the product of the individual Bayes factors.
Thus, the logarithm of the expectation value of the total Bayes factor will be the sum of the squares of the memory SNR for each event, and it is natural to define an ``effective'' SNR squared for $N$ events (measured in $n$ detectors) as
\begin{equation} \label{eq:rhoEff}
    \rho_{\ef}^2 = \sum_{i=1}^{N} \sum_{j=1}^n \rho^2 _{h_{\m,ij}} .
\end{equation}
Here $\rho^2 _{h_{\m,ij}}$ is the square of the SNR of the memory signal in the $i^\text{th}$ event for the $j^\text{th}$ detector.
A common choice of a ``threshold'' effective SNR is $\rho_\ef\geq 3$, because it corresponds to a total evidence ratio in the population of order $\sim 10^4$.
In our forecasts in Sec.~\ref{sec:results}, we will display $\rho_\ef(T)$, which shows how the effective SNR accumulates as a function of time for a population of events that occur at times $T_i$.

This calculation assumed that the posterior had a single mode which was localized around the true values $\theta_0$.
However, there is a known degeneracy in the waveform when only the $(l,m)=(2,\pm 2)$ modes of the signal can be measured, which makes the posterior distributions multimodal.
We discuss this in more detail in the next subsection.

\subsection{Parameter degeneracy in the oscillatory waveform modes} \label{subsec:degeneracy}

The degeneracy in the waveform arises because the strain measured in the GW detectors is typically written as a sum of the polarizations $h_+$ and $h_\times$ in Eq.~\eqref{eq:hPlusTimes} times antenna functions $F_+$ and $F_\times$, which are functions of sky location (e.g., right ascension $\alpha$ and declination $\delta$) and a polarization angle $\psi$ of the gravitational waves (see, e.g.,~\cite{LIGOScientific:2026sit}).
Namely, GW detectors measure
\begin{equation} \label{eq:hF}
    h = F_+ h_+ + F_\times h_\times .
\end{equation}
Under a transformation of of the polarization angle $\psi \rightarrow \psi + \pi/2$, the antenna functions for an L-shaped interferometer change sign (i.e., $F_{+,\times} \rightarrow -F_{+,\times}$).
The strain $h$ changes sign, because the polarizations $h_+$ and $h_\times$ are not functions of $\psi$.

The polarizations $h_+$ and $h_\times$ are the sum of products of time-dependent spherical harmonic modes $h_{lm}$ and spin-weighted spherical harmonics $_{-2}Y_{lm}(\iota,\phi_\mathrm{ref})$.
Under a transformation $\phi_{\mathrm{ref}}\rightarrow \phi_{\mathrm{ref}}+\pi/2$, the spherical harmonics transform as $_{-2}Y_{lm}\rightarrow i^{m} {}_{-2}Y_{lm}$.
Thus, under the joint transformation
\begin{equation} \label{eq:phiPsi}
    \phi_{\mathrm{ref}}\rightarrow \phi_{\mathrm{ref}}+\pi/2, \quad \psi \rightarrow \psi + \pi/2,
\end{equation}
the full detector response $h$ in Eq.~\eqref{eq:hF} transforms differently for each multipolar component contributing to the full $h$.
Specifically, as first noted in~\cite{Lasky:2016knh}, the $m=\pm 2$ modes of the signal are invariant under this transformation, while the $m=0$ (including the $h_\m$ parts) change sign. 

If the majority of the SNR arises from the $(l,m) = (2,\pm 2)$ parts of $h_\nm$ then the posteriors will be bimodal about the true values of $\phi_\mathrm{ref}$ and $\psi$, and the true values plus $\pi/2$.
It was shown in~\cite{Grant:2022bla} that when both modes of the posterior distribution are equally favored, there is a cancellation that causes the logarithm of the expectation value of the Bayes factor to go as the SNR to the fourth power (note, however, that if one mode of the posterior is preferred, it still scales with SNR squared, but it is suppressed by a coefficient related to the fraction of the posterior in each mode).

How well the two modes of the posterior can be distinguished is a function of the SNR in the parts of the signal that change (in phase) under the transformation in Eq.~\eqref{eq:phiPsi}~\cite{Lasky:2016knh,Boersma:2020gxx,Grant:2021hga}.
We use the criteria in~\cite{Grant:2022bla} that when the part of the signal which is odd under the above transformation has SNR greater than two, then the degeneracy between the two modes of the posterior is strongly broken. 
In our forecasts, therefore, we follow~\cite{Grant:2022bla} and consider only signals for which the SNR squared of the odd part is greater than two when calculating the $\rho_\ef$ in Eq.~\eqref{eq:rhoEff}.

\subsection{Simulating BBH populations} \label{subsec:BBHpops}

A quasicircular BBH merger is characterized by 15 parameters (8 intrinsic and 7 extrinsic).
In this paper, we focus on nonspinning binaries, so that we do not need to model the six degrees of freedom in two spin vectors.
The intrinsic parameters that we use are the primary mass $m_1$ and the mass ratio $\varepsilon \equiv m_2/m_1$ 
The extrinsic parameters are the polar angle $\iota$ and azimuthal angle $\phi_\mathrm{ref}$ described in Sec.~\ref{subsec:degeneracy}; right ascension $\alpha$ and declination $\delta$; the polarization $\psi$ mentioned in Sec.~\ref{subsec:degeneracy}; the redshift $z$ (inferred from a cosmological model, because the luminosity distance $d_L$ is obtained from the GW observations); and a time elapsed between events $\Delta T$ (or equivalently a GPS time from a given reference GPS time).
We closely follow the procedure used in~\cite{Grant:2022bla} (as implemented in~\cite{GWForecasts}) to generate synthetic populations of BBH mergers that we use to produce our forecasts.

We generate populations with mass distributions given by the ``power law plus peak'' (PLPP) model used in GWTC-3~\cite{KAGRA:2021duu}, which is a joint distribution for the  primary mass $m_1$ and the mass ratio $\varepsilon$.
Although the population models have been updated following the release of the GWTC-4~\cite{LIGOScientific:2025pvj} and GWTC-5~\cite{LIGOScientific:2026ctl} catalogs, the updated models are still consistent with the GWTC-3 results; thus, the populations generated with the GWTC-3 models will give reliable forecasts, albeit with slightly larger variance than the updated models. 

For the extrinsic parameters, we exactly follow~\cite{Grant:2022bla}.
Specifically, we assume populations that are uniform in $\psi$ and uniform on the sphere for the sky position $(\alpha,\delta)$ and for the binary's orientation with respect to the detector $(\iota,\phi_\mathrm{ref})$.
For the redshift, we use the GWTC-3 model where the merger rate evolves with redshift as $(1+z)^\kappa$.
Although a maximum redshift of $z_\mathrm{max} \approx 2.3$ was used in GWTC-3, we will limit to events at redshifts of $z<1$ in our forecasts.
We continue to use the \textsc{Planck15} cosmological parameters~\cite{Planck2015} to compute the comoving volume in the redshift model (which we obtain from \textsc{astropy}~\cite{2013A&A...558A..33A,2018AJ....156..123A,2022ApJ...935..167A}).
For $\Delta T$, the time difference between mergers follows a Poisson process with a local rate of mergers $\mathcal R_0$, which is included in the evolving redshift model of GWTC-3.

For our forecasts, we generate 100 simulated populations each spanning 1 calendar year.
We generate each population by drawing the population parameters from the posterior distributions of the PLPP mass model and the redshift model.
These posterior samples are available from the public data release associated with GWTC-3~\cite{LIGO-P2000434}.
Rather than draw from the discrete samples provided by the LVK, we use \textsc{kombine}~\cite{kombine} to sample from a Gaussian kernel density estimate of the posterior.
With these populations, we can then draw the event parameters for each BBH system,  
which we describe next.

\subsection{Generating BBH events and computing SNRs}

The majority of the individual BBH parameters follow distributions with a simple analytical form, and we draw from those directly.
For the joint distribution of the primary mass and mass ratio, as well as the redshift distribution, we use MCMC sampling implemented through the \textsc{emcee} package~\cite{2013PASP..125..306F} to obtain individual event parameters.

To generate waveforms for each event, we transform the event parameters in Sec.~\ref{subsec:BBHpops} to those required by the waveform family.
These parameters are the redshifted total mass, the luminosity distance, and the GPS time $T$ of the event relative to some reference time.
As in~\cite{Grant:2022bla}, we use the \textsc{NRSur7dq4} surrogate~\cite{Varma:2019csw} for mass ratios up to 4 and \textsc{SEOBNRv4PHM}~\cite{Gadre:2022sed} for higher mass ratios. 
We use this approach because, as was noted in~\cite{Grant:2022bla}, the surrogate waveform is faster to evaluate, so we use EOB waveform only when the parameters are outside the domain of surrogate model. 
From these waveforms, we obtain the memory waveforms using the fluxes, pseudofluxes, and stationary-to-stationary parts of the charges, as described in Secs.~\ref{sec:multipole}--\ref{subsec:signals_nonprecessing}.

\begin{figure}
    \centering
    \includegraphics[width=\columnwidth]{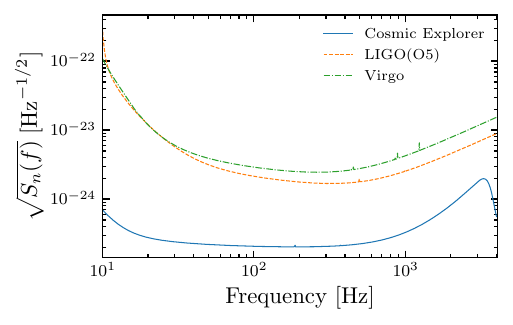}
    \caption[Amplitude spectral densities of Cosmic Explorer, LIGO, and Virgo]{\textbf{Amplitude spectral densities of Cosmic Explorer, LIGO, and Virgo}.
    The CE curve is the solid blue curve, whereas the LIGO and Virgo curves for O5 are the dashed orange and dashed-dotted green curves, respectively.
    Note that we only use CE in the forecasts in this paper; the LIGO and Virgo curves are shown for reference.
    }
    \label{fig:sensitivity_com}
\end{figure}

To compute the SNR, we need to specify a noise curve $S_n(f)$ to evaluate the noise-weighted innner product in Eq.~\eqref{eq:SNR}.
We will use a detector network of two 40\,km Cosmic Explorer detectors. 
We assume that the individual CE detectors have a duty cycle of 75\%, and we compute the SNR in our forecasts only for events that are measured by both CE detectors, as in~\cite{Grant:2022bla}.
For the locations of the CE detectors, we place one in the continental United States in Utah and one in Australia for the locations in~\cite{Hall:2019xmm} (though the precise locations do not have a significant impact on our forecasts because the sky localization of individual events does not have a large impact on our effective SNR calculations).
The noise amplitude spectral density for the CE detectors is available at~\cite{CEnoiseCurve}.
We plot the CE noise curve in Fig.~\ref{fig:sensitivity_com} as a solid blue curve.
The projected fifth observing run (O5) LIGO and Virgo noise curves are shown in orange (dashed) and green (dash-dotted), respectively, in Fig.~\ref{fig:sensitivity_com}.
The LIGO and Virgo spectral densities are given just for comparison; they are not being used in these forecasts.

\section{Forecasts for the higher memory effects} \label{sec:results}

\begin{figure}
\centering
\includegraphics[width=\columnwidth]{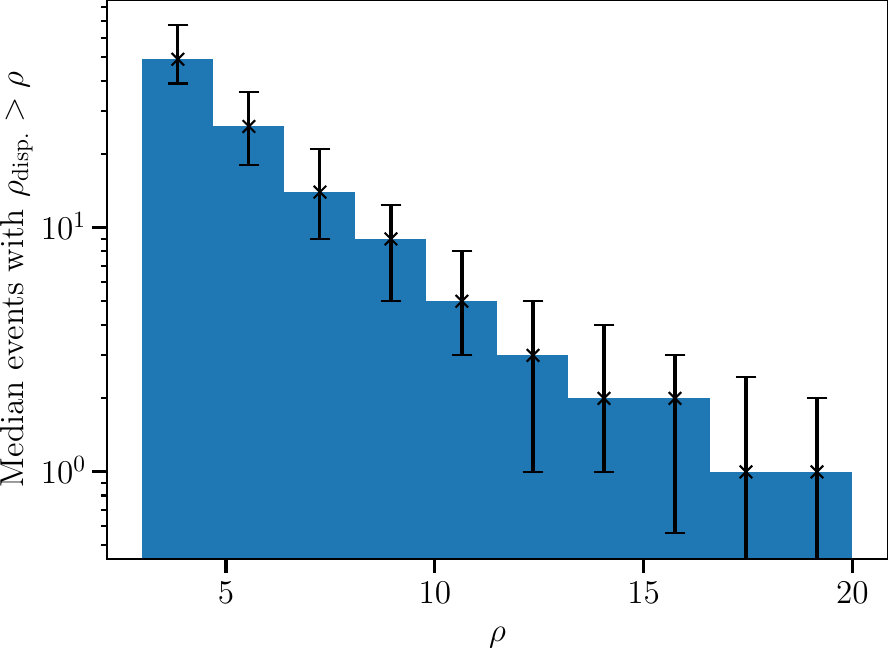}
\caption[Histogram of events with memory SNR greater than a given threshold]{\textbf{Histogram of events with memory SNR greater than a given threshold}.
This figure updates Fig.~7 of~\cite{Grant:2022bla} and gives the number of events in one year with a SNR for the displacement memory greater than the SNR $\rho$ plotted on the abscissa.
The height of each bin indicates the median value for the 100 populations used, and the ranges are the corresponding 68\% symmetric credible regions.
The text of Sec.~\ref{sec:results} explains in more detail the differences between this figure and the corresponding one in~\cite{Grant:2022bla}.}
\label{fig: Displacement memory}
\end{figure}

\begin{figure}
\centering
\includegraphics[width=\columnwidth]{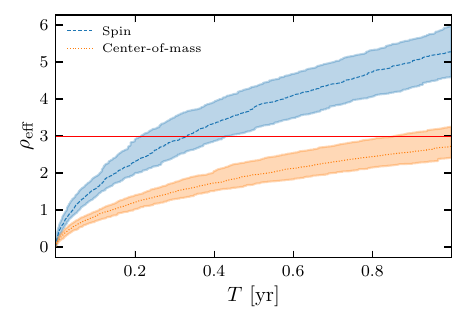}
\includegraphics[width=\columnwidth]{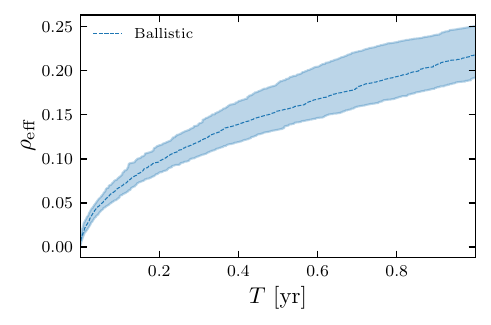}
\caption[Effective SNR for the spin, CM, and ballistic memory signals versus detector time $T$]{\textbf{Effective SNR for the spin, CM, and ballistic memory signals versus detector time $T$}.
In both panels, the curves in each plot show the median value for 100 simulated populations and the shaded regions are the corresponding 68\% symmetric credible regions.
\emph{Top panel}:
The spin-memory effective SNR, which was computed from the flux contribution to the $(l,m)=(3,0)$ waveform mode, is shown in blue (with the median being the dashed curve).
The orange region is the CM-memory effective SNR, which was computed from both charge and flux contributions to the $(l,m)=(2,0)$ (with the median being the dotted curve).
The red line at $\rho_\ef=3$ is shown to indicate a threshold for strong evidence for the presence of memory in the BBH population.
\emph{Bottom panel}: 
Here the effective SNR from the flux contribution to the $(l,m)=(4,0)$ mode of the ballistic memory is being plotted.
This effective SNR serves as a lower limit for the detection prospects of the ballistic memory, because it does not include the lower multipole $(l,m)=(2,0)$ contributions, as discussed in Sec.~\ref{subsec:higherMemoryBBH}.
}
\label{fig:higher memories}
\end{figure}

The main results of the forecasts are given in Figs.~\ref{fig: Displacement memory} and~\ref{fig:higher memories}.
We discuss and interpret these results next.

In Fig.~\ref{fig: Displacement memory}, we revise the forecasts for observing a memory signal with a SNR greater than a given threshold with two CE detectors.
It differs from the similar figure (Fig.~7) in~\cite{Grant:2022bla} in several important ways.
First, it corrects for the error that arose involving the CE noise curve in the forecasts in~\cite{Grant:2022bla}.
Second, it uses only the nonoscillatory memory signal (as discussed in Sec.~\ref{subsec:nonosc}).
Third, it uses a nonspinning BBH population rather than the more general precessing spin one used in~\cite{Grant:2022bla}.
The largest difference from~\cite{Grant:2022bla} comes from correcting the error involving the noise curve, which significantly increases [by an $O(10)$ factor] the number of events with a displacement memory SNR above any given threshold.
It is worth noting, however, that although there are tens of events (median) that have an SNR of at least five, this is still a small fraction of the $O(10^4)$ BBH mergers that can be detected by CE in a year.

In the top panel of Fig.~\ref{fig:higher memories}, we show the effective SNR as a function of detector time $T$ for the spin and CM memory signals (respectively, the blue and orange curves and shaded regions).
The spin-memory effective SNR was computed from just the $(l,m)=(3,0)$ mode of the flux term and the CM memory from the $(l,m)=(2,0)$ mode of the charge and flux terms (as discussed in Sec.~\ref{sec:computingMemory}).
The median values (for the 100 realizations of the populations) are the solid curves, and the shaded regions are the 68\% symmetric credible regions about the median.
For both the spin and CM memory signals, although the SNR for each individual event is low, the large number of events which CE will be able to measure still allows for a more significant effective SNR for the population to be reached.

For a year-long observing run of two 40\,km CE detectors at their design sensitivities, there are enough events that the spin memory clearly crosses the threshold of an effective SNR of three, which indicates strong evidence for the spin memory effect in the population of BBH mergers.
Comparing with the results in~\cite{Grant:2022bla}, the time for the median $\rho_\ef$ to reach a value of three is shorter by an order of magnitude.
As with the displacement memory, the correction to the error related to the noise curve outweighs the decrease in the SNR that arises from including only the nonoscillatory part of the spin-memory signal.
While not directly comparable, the results here are more closely in line with those forecast for Einstein Telescope in~\cite{Goncharov:2023woe}.

For the CM memory, the upper limit of the confidence interval crosses this effective SNR value in one year, though the median does not quite reach it.
Given that CE is expected to run for longer than one year, it will also cross the threshold during CE's operation time.
This is consistent with the fact that the typical amplitude of the CM memory signal is a factor of two smaller than the spin memory one (see the top-right panel of Fig.~\ref{fig:memory signals}).
Given that the effective SNR grows with the square root of the number of events, four times as many events would be required and it would take four times as long to reach the same threshold.

The bottom panel of Fig.~\ref{fig:higher memories} shows the effective SNR for the $(l,m)=(4,0)$ flux contribution to the ballistic memory.
The solid curve and shaded region represents the median and 68\% symmetric credible region for the different populations (as in the top panel).
This effective SNR is small, and it will not reach the threshold of three, even in a five-year-long observing run of CE.
As we had previously discussed in Sec.~\ref{subsec:higherMemoryBBH}, the $(l,m)=(4,0)$ contribution to the ballistic memory, which can be computed in the stationary-to-stationary approximation, neglects a potentially larger signal in the $(l,m)=(2,0)$ mode that cannot be obtained in this approximation.
For this reason, the results here should be interpreted as a lower limit for the effect, rather than a definitive indication of the effect being weaker than the resolving power of the CE detectors.

\section{Conclusions} \label{sec:conclusions}

In this paper, we investigated the detection prospects of higher memory effects from binary-black-hole mergers with the Cosmic Explorer detector network.
We updated the forecasts for the displacement and spin memory, which had been previously considered in~\cite{Grant:2022bla} (see also~\cite{Goncharov:2023woe} for results for the Einstein Telescope detector).
We found that there will likely be tens of individual BBH events where the SNR of the displacement memory will exceed five.
For the spin memory, after less than one year of operation of two CE detectors at their design sensitivities, there will be strong evidence for the spin memory signal being present in the population of BBH mergers.
We also showed that there would be strong evidence for the center-of-mass memory signal in a population of BBH mergers, for the same detector configuration, in about one year.
Finally, we placed lower limits on the evidence for the ballistic memory signal.

In addition to these observational forecasts, this paper provided several theoretical and computational results that helped enable the results of the forecasts which we just summarized.
First, we focused in this paper on nonoscillatory waveform multipoles (those with the spherical-harmonic index $m=0$), which allowed us to more straightforwardly determine the parts of a GW signal that produce an offset in the strain or the higher moments of the news tensor.
Even restricting to these nonoscillatory waveform multipoles, there were other oscillatory components of the GW signal (related to quasinormal modes during the ringdown phase of the waveform) that did not contribute substantially to offsets in the moments of the news.
For the displacement and spin memory signals, the flux contributions in the $m=0$ multipoles do not have these residual oscillatory components, as had been previously identified elsewhere~\cite{Mitman:2020pbt,Grant:2023jhd}; the oscillations were restricted to the charge part of the signals, which were smaller than the flux parts.
We could identify a nonoscillatory signal by restricting to the flux contribution in this case.
For the CM memory signal, the flux contributions to the nonoscillatory modes are comparable to the charge contributions.
This required us to develop a prescription for computing the charge contribution, which separated the oscillatory and the slowly growing parts of the signal.

To compute the charge contribution, we introduced a notion of a stationary-to-stationary part of the time-dependent charge contribution to the memory signal.
It was based on taking the constraints on Bondi metric functions and charges in a stationary region of spacetime and applying them to the nonstationary regions as well.
These stationary-to-stationary parts of the charges give rise to time-dependent signals that capture the offsets in the charge contributions to the moments of the news, but which have minimal oscillations unrelated to the offset which accumulates.
This generalized a similar approximation in~\cite{Nichols:2018qac} (restricted to the CM memory in the PN approximation) to apply to higher memory effects more generally.
Comparing this approximation to full NR CCE simulations of BBH systems with zero (or aligned) spins showed that it provided a reasonable notion of a nonoscillatory charge contribution to a memory signal which reproduced the expected moment of the news.
It was used as part of the forecasts for the CM memory signal.

When the stationary-to-stationary approximation was applied to compute the charge contribution to the ballistic memory signal in the $(l,m)=(4,0)$ waveform multipoles, the charge part was found to be significantly smaller than the flux part.
We thus made the forecasts using the fluxes, which showed that CE would be unlikely to find evidence for this portion of the ballistic memory signal.
However, the stationary-to-stationary approximation does not allow for the quadrupolar part of the charge contribution to the memory signal to be computed, which is likely the largest component.
The forecasts for the ballistic memory, therefore, were not fully conclusive.

These limitations on the forecasts for the ballistic memory point to an obvious future direction: namely, to generalize the stationary-to-stationary approximation to allow for the charge contribution to the lowest multipole of the ballistic memory to be computed.
The expressions for the Bondi metric functions in terms of the PN multipoles in~\cite{Blanchet:2020ngx,Blanchet:2023pce} are a natural starting point for this calculation, though they are currently restricted to the linear approximation for all multipoles and just the quadrupole-quadrupole interactions for the nonlinear terms.

Another future direction would be to generalize the forecasts here for nonspinning binaries to more general precessing binaries.
Such a generalization would require a notion of extracting a nonoscillatory signal from a signal that has precession-induced modulations.
This could be obtained from an appropriate smoothing operation or through restricting to $m=0$ modes in a co-precessing frame before transforming to the asymptotically inertial frame.

Finally, there are improvements to the forecast method that could be made.
For example, one could use a Gaussian approximation to the modes of the posterior (rather than the delta-function approximation used here).
This would allow the impact of larger variance in individual measurements to be assessed.
It could also allow for a forecast based on hierarchical population-based inference to be performed as in~\cite{Mitman:2026zfg}.

\acknowledgments

D.A.N.\ and S.S.\ acknowledge support from the NSF grant No.\ PHY-2309021 and from the NSF CAREER Award PHY-2439893.
A.M.G. acknowledges support from the Royal Society under
Grant No. RF\textbackslash ERE\textbackslash 221005, as well as ERC Consolidator/UKRI Frontier Research Grant GWModels (selected by the ERC and funded by UKRI [grant number EP/Y008251/1]). 
The authors acknowledge Research Computing at The University of Virginia for providing computational resources and technical support that have contributed to the results reported within this publication.
They also thank Keefe Mitman for his input on fixing the BMS frame in NR simulations.

\appendix

\section{Additional results on the charge contribution to the higher memory signals} \label{app:moreCharges}

\begin{figure}
  \centering
  \includegraphics[width=\columnwidth]{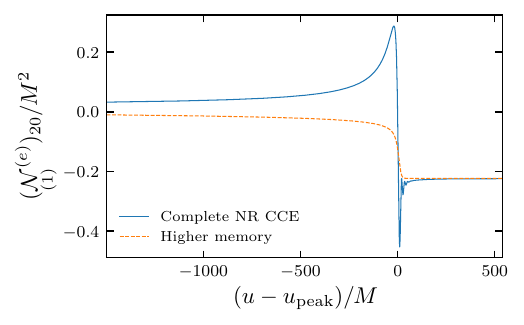}
  \caption[Charge contribution to the first moment of the news for an equal-mass, aligned-spin BBH merger]{\textbf{Charge contribution to the first moment of the news for an equal-mass, aligned-spin BBH merger.}
  This figure shows the same quantities as the left column of Fig.~\ref{fig:Q1_e_nospin}.
  Here the NR data came from the simulation SXS:BBH\_ExtCCE:0002, and the charge contribution from the simulation is shown as the solid blue curve.
  The dashed orange curve is the higher memory signal computed from the stationary-to-stationary approximation.
  }
  \label{fig:Q1_q1_aligned_spin}
\end{figure}

In Sec.~\ref{subsec:accuracy}, we commented that adding aligned spins has a small impact on the charge contribution to the first moment of the news in the stationary-to-stationary approximation.
We illustrate this in Fig.~\ref{fig:Q1_q1_aligned_spin}, which shows the similarity with the $q=1$ results in the top-left panel of Fig.~\ref{fig:Q1_e_nospin}.
The relevant NR CCE data came from the simulation SXS:BBH\_ExtCCE:0002, which is an equal-mass simulation where the individual black-hole spins are aligned with the orbital angular momentum and have dimensionless spins $\chi_1 = \chi_2 = 0.2$.

\begin{figure*}
  \centering
  \includegraphics[width=\columnwidth]{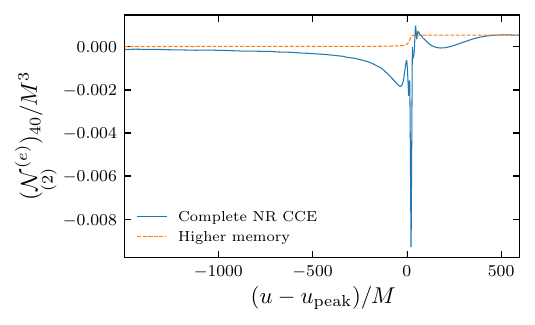}
  \includegraphics[width=\columnwidth]{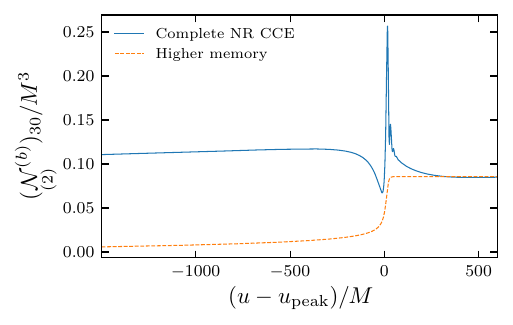}
  \includegraphics[width=\columnwidth]{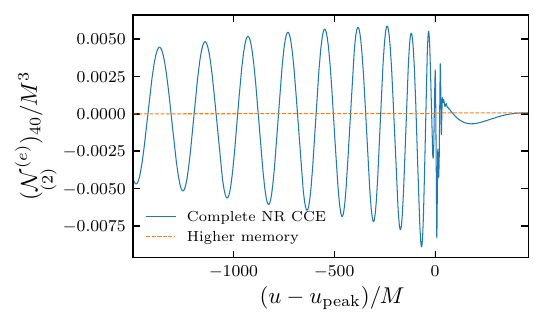}
  \includegraphics[width=\columnwidth]{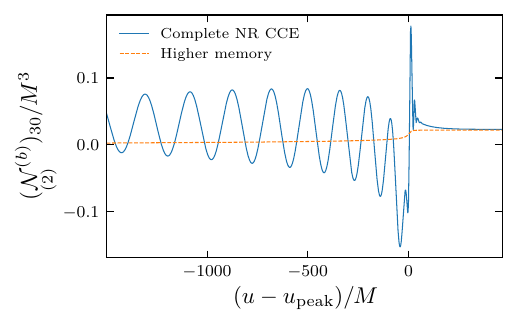}
  \caption[Charge contribution to the second moment of the news for two BBH mergers]{\textbf{Charge contribution to the second moment of the news for two BBH mergers.}
  \emph{Top row}: Results for an equal-mass ($q=1$), nonspinning BBH.
  \emph{Bottom row}: Results for an unequal-mass ($q=4$), nonspinning BBH. 
  \emph{Left column}: The $(l,m)=(4,0)$ electric-parity part of the second moment of the news.
  As in the left column of Fig.~\ref{fig:Q1_e_nospin}, the solid blue curves are computed from NR CCE data and the dashed orange curves show the stationary-to-stationary part.
  The stationary-to-stationary part captures the small accumulation of the offset in the second moment hidden below larger oscillatory features that do not produce such an offset.
  However, it takes more time to converge to the final value than it did for the first moment of the news.
  The $q=4$ waveform has oscillatory features during the inspiral that also do not have a simple physical explanation.
  \emph{Right column}: Similar to the left, except for the $(l,m)=(3,0)$ magnetic-parity part of the second moment of the news.
  Here the convergence to the final value is more prompt, but the signals during inspiral have more significant differences.
  }
  \label{fig:Q2_nospin}
\end{figure*}

We also mentioned in Sec.~\ref{subsec:accuracy} that we would present results for the second moment of the news and the associated memory signal for the same two BBH binaries discussed in Sec.~\ref{subsec:accuracy} and Figs.~\ref{fig:Q1_e_nospin} and~\ref{fig:Q2_e_nospin_shear_contri}.
These results are shown in Fig.~\ref{fig:Q2_nospin}.
The computation of the NR result is similar to that with the first moment of the news and uses the procedure described in Secs.~\ref{sec:stationary} and~\ref{subsec:signals_nonprecessing}.
The solid blue curves in Fig.~\ref{fig:Q2_nospin} are computed from the same NR CCE data, and the orange dashed curves are the stationary-to-stationary part.
The left column shows the $(l,m)=(4,0)$ mode of the charge contribution to the second moment of the news, $(\E{\ord{2}{\mathcal N}})_{40}$, whereas the right column shows the $(l,m)=(3,0)$ mode of the charge contribution.

Although BMS frame fixing was able to remove smaller oscillatory features in the NR data in Fig.~\ref{fig:Q1_e_nospin} and in the top panels of Fig.~\ref{fig:Q2_nospin}, it did not do so in the bottom panel of Fig.~\ref{fig:Q2_nospin}.
We are unsure of the origin of this oscillatory behavior, but it is possible that it is related to a small, residual misalignment of the orbital angular momentum of the binary with the polar axis of the NR simulation's coordinates. 
Nevertheless, we can see that the stationary-to-stationary approximation to the memory signal does recover the offset in the second moment of news from the charge contribution, for the $(l,m)=(4,0)$ mode.
However, it takes a much longer time for the second moment of the news to converge to the final value than it did for the first moment of the news (cf.~Fig.~\ref{fig:Q1_e_nospin}).

For the $(l,m)=(3,0)$ modes, the late-time behavior of the NR CCE signals do not have the late-time drifts present in the  $(l,m)=(4,0)$ modes.
However, at early times, the stationary-to-stationary part and the NR CCE calculation disagree more substantially.
We do not have a clear explanation for this discrepancy.
Thus, we cannot clearly determine if the disagreement is a limitation of the stationary-to-stationary approximation or of the NR CCE data used to compute the NR results.

\begin{figure}
  \centering
  \includegraphics[width=\columnwidth]{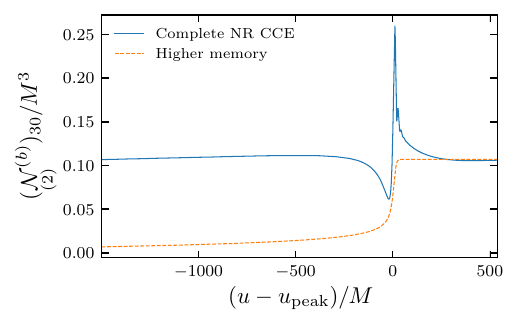}
  \includegraphics[width=\columnwidth]{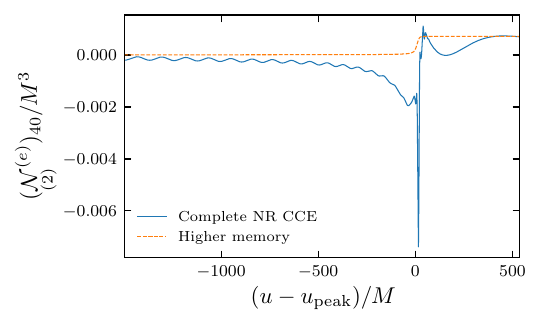}
  \caption[Charge contribution to the second moment of the news for an equal-mass, aligned-spin BBH merger]{\textbf{Charge contribution to the second moment of the news for an equal-mass, aligned-spin BBH merger.}
  These results show the equivalent results to those depicted in the two panels of the top row of Fig.~\ref{fig:Q2_nospin}, though the spinning simulation used in Fig.~\ref{fig:Q1_q1_aligned_spin} is used in this figure.
  Again the solid blue curves are the NR data and the dashed orange curve uses the stationary-to-stationary approximation.
  Including spins makes small changes to the charge contribution to the second moment of the news.
  }
  \label{fig:Q2_q1_aligned_spin}
\end{figure}

Finally, in Fig.~\ref{fig:Q2_q1_aligned_spin}, we show additional results for the aligned-spin simulation used in Fig.~\ref{fig:Q1_q1_aligned_spin}.
The top and bottom panels in Fig.~\ref{fig:Q2_q1_aligned_spin} should be compared with the top-right and top-left panels in Fig.~\ref{fig:Q2_nospin}, respectively, which contain the equivalent multipole moments of the charge contribution to the second moment of the news.
The addition of small aligned spins does not qualitatively change the results, although there are some small quantitative differences (for example, the additional small oscillations during the inspiral in the bottom panel of Fig.~\ref{fig:Q2_q1_aligned_spin}).

\section{Tensors and spin-weighted scalars on the 2-sphere} \label{app:harmonics}

This appendix gives the definitions of our spin-weighted spherical harmonics and rank-$s$ tensor harmonics.
It relates the two notions of these harmonics, and gives expressions for writing products of these harmonics in terms of sums of harmonics and Gaunt coefficients.

\subsection{Dyad and spin-weighted tensors}

Many of the calculations that we perform are expressed conveniently in terms of 
a complex dyad $m^A$, $\bar m^A$, which is defined such that
\begin{equation} \label{eq:dyadDef}
    m^A \bar m_A = 1, \qquad m^A m_A = 0.
\end{equation}
These equations imply that the 2-sphere metric $q_{AB}$ can be expressed in terms of the dyad by
\begin{equation}
    q_{AB} = 2 m_{(A} \bar m_{B)}.
\end{equation}
Moreover, we define this dyad such that
\begin{subequations}\label{eqn:eps_2}
\begin{align} 
    &\epsilon_{AB} = 2 im_{[A} \bar m_{B]} = 2 \sin \theta (\ud \theta)_{[A} (\ud \phi)_{B]},\\
    &\epsilon_{AB}m^A = i\,m_B,\;\;\epsilon_{AB}\bar m^A = -i\,\bar m_B,
\end{align}
\end{subequations}
where $\epsilon_{AB}$ is the usual 2D volume form on the 2-sphere.
In spherical coordinates, these two equations fix (up to an overall sign) that\footnote{\label{fn:signs}In this paper, we will be using this particular convention for the sign of $m_A$.
Note that it disagrees with the sign often used in the Newman-Penrose formalism~\cite{Newman:1961qr,Newman:1966ub},
but it has the advantage that it is the restriction to the sphere of the usual $m_a$ that is used at null infinity [see Eq.~\eqref{eq: SpECTRE tetrads}], as well as the Kinnersley tetrad~\cite{Kinnersley:1969zza} in the Kerr spacetime.
To have the spin-weighted spherical harmonics, as functions of spherical coordinates, agree with those in the literature (e.g.,~\cite{Newman:1966ub}), this dyad choice requires that the signs in Eqs.~\eqref{eq:sYlmDef} and~\eqref{eqn:eth_ladder} do not.}
\begin{equation}
    m_A = \frac{1}{\sqrt 2} \left[(\ud \theta)_A + i \sin \theta (\ud \phi)_A\right]. \label{eq: def dyad}
\end{equation}

Using this dyad, tensorial equations can be contracted with the dyad to become tensor equations of a lower rank; in doing so, however, it is useful to keep track of some aspects of this dyad.
This can be achieved by assigning a spin weight to a tensor.
Specifically, if a tensorial quantity $S_{A_1 \cdots A_p}$ transforms as
\begin{equation}
    S_{A_1 \cdots A_p} \to e^{is \theta} S_{A_1 \cdots A_p}
\end{equation}
under the transformation $m^A \to e^{i\theta} m^A$, then $S_{A_1 \cdots A_p}$ is said to have spin weight $s$.
Spin weight is most commonly discussed in the context of spin-weighted scalars, but it is useful to also introduce spin-weighted tensors: for example, the members of the dyad $m^A$ and $\bar m^A$ have spin weights $1$ and $-1$, respectively.

\subsection{Tensor and spin-weighted harmonics}

For scalar functions on the 2-sphere, the spherical harmonics form a convenient basis; for tensorial or spin-weighted quantities, it is more natural to use tensor or spin-weighted spherical harmonics instead.
The tensor harmonics can be defined in terms of the covariant derivative on the sphere.
First, note that the Hodge decomposition implies that any vector on the sphere can be written as the sum of a gradient and the dual of a gradient.
A similar type of argument applies to \emph{any} symmetric, trace-free tensor field.
Thus, we define the ``electric'' and ``magnetic'' tensor harmonics by
\begin{subequations} \label{eq:TlmAsDef}
    \begin{align}
        (T^{\rm E}_{lm})_{A_1 \cdots A_s} &\equiv 2^{(s - 1)/2} \mathcal A_{ls} \STF \mathscr D_{A_1} \cdots \mathscr D_{A_s} Y_{lm}, \\
        (T^{\rm B}_{lm})_{A_1 \cdots A_s} &\equiv (^* T^E_{lm})_{A_1 \cdots A_s},
    \end{align}
\end{subequations}
where the coefficient
\begin{equation}
    \mathcal A_{ls} \equiv \sqrt{\frac{(l - s)!}{(l + s)!}} 
\end{equation}
was introduced (and is related to normalizing the tensor harmonics).
These harmonics form a basis for symmetric, trace-free tensor fields on the sphere.

Because tensors can also be written in terms of spin-weighted scalars and the dyad, the existence of the tensor-harmonic basis suggests that there is also a basis for the spin-weighted scalars; these are the spin-weighted spherical harmonics.
Like the tensor harmonics, they can be defined by taking appropriate derivatives of the spherical harmonics; however, to take into account spin weight, they should also depend on $m^A$ in the appropriate way.
For a tensor $S_{A_1 \cdots A_p}$ of spin weight $s$, the tensors
\begin{subequations} \label{eq:ethSAs}
    \begin{align}
        \eth S_{A_1 \cdots A_p} &\equiv [m^C \mathscr D_C - s \bar m^B (m^C \mathscr D_C m_B)] S_{A_1 \cdots A_p}, \\
        \bar \eth S_{A_1 \cdots A_p} &\equiv [\bar m^C \mathscr D_C - s m^B (\bar m^C \mathscr D_C \bar m_B)] S_{A_1 \cdots A_p},
    \end{align}
\end{subequations}
have spin weight $s + 1$ and $s - 1$, respectively.
The operators $\eth$ and $\bar \eth$ are the spin-weight raising and lowering operators.
Moreover, one can show that
\begin{equation}
    \eth m^A = 0, \qquad \bar \eth m^A = 0 .
\end{equation}
The operators $\eth$ and $\bar \eth$ both follow the Leibniz rule, which allows us to commute these operators through the dyad when they act on a tensor defined in terms of spin-weighted scalars and the dyad.

Our definition of the spin-weighted spherical harmonics is
\begin{equation} \label{eq:sYlmDef}
    {}_s Y_{lm} \equiv 2^{|s|/2} \mathcal A_{l|s|} \begin{cases}
        (-\eth)^s Y_{lm} & s \geq 0 \\
        \bar \eth^{-s} Y_{lm} & s \leq 0
    \end{cases}
\end{equation}
(see Footnote~\ref{fn:signs}).
The definitions of the spin-weighted spherical harmonics and the tensor harmonics can be related.
By contracting Eq.~\eqref{eq:TlmAsDef} with $m^A$ and $\bar m^A$ on all indices, iteratively using Eqs.~\eqref{eq:ethSAs}--\eqref{eq:sYlmDef}, applying Eq.~\eqref{eqn:eps_2} in the magnetic case, and reconstructing the tensor harmonics from these scalars and the dyad, we obtain the relationship between the spin-weighted and tensor harmonics:
\begin{subequations} \label{eq:Tlm2sYlm}
    \begin{align}
        (T^{\rm E}_{lm})_{A_1 \cdots A_s} &= \frac{1}{\sqrt 2} \Big[{}_{-s} Y_{lm} m_{A_1} \cdots m_{A_s} \nonumber \\
        &\hspace{3.2em}+ (-1)^s {}_s Y_{lm} \bar m_{A_1} \cdots \bar m_{A_s}\Big], \\
        (T^{\rm B}_{lm})_{A_1 \cdots A_s} &= -\frac{i}{\sqrt 2} \Big[{}_{-s} Y_{lm} m_{A_1} \cdots m_{A_s} \nonumber \\
        &\hspace{4.2em}- (-1)^s {}_s Y_{lm} \bar m_{A_1} \cdots \bar m_{A_s}\Big].
    \end{align}
\end{subequations}
These two equations can be combined into a simpler form after making a few definitions.
First, let the index $\rm I$ be either $\rm E$ or $\rm B$, as in Sec.~\ref{subsec:multipoles}.
Next, construct a complex dyad on the space of $\rm E$ and $\rm B$ indices given by
\begin{equation}
    M^{\rm I} = \frac{1}{\sqrt 2} \left(\delta^{\rm EI} - i \delta^{\rm BI}\right),
\end{equation}
where $\delta^{\rm IJ}$ is the usual Kronecker delta.
It then follows that
\begin{equation} \label{eqn:T_from_Y}
    \begin{split}
        (T^{\rm I}_{lm})_{A_1 \cdots A_s} &= {}_{-s} Y_{lm} M^{\rm I} m_{A_1} \cdots m_{A_s} \\
        &\hspace{1.1em}+ (-1)^s {}_s Y_{lm} \bar M^{\rm I} \bar m_{A_1} \cdots \bar m_{A_s}.
    \end{split}
\end{equation}
This dyad is similar to the spacetime dyad $m^A$, because it satisfies the conditions
\begin{subequations}
\begin{gather}
    \sum_{\rm I} M^{\rm I} M^{\rm I} = 0, \qquad \sum_{\rm I} M^{\rm I} \bar M^{\rm I} = 1, \\
    \delta^{\rm IJ} = M^{\rm I} \bar M^{\rm J} + \bar M^{\rm I} M^{\rm J}.
\end{gather}
\end{subequations}
Given a tensor $S_{A_1 \cdots A_s}$, it can be written in terms of either of the harmonic bases as
\begin{equation} \label{eq:StensorSpin}
    \begin{split}
        S_{A_1 \cdots A_s} &= \sum_{{\rm I}, l, m} S^{\rm I}_{lm} (T^{\rm I}_{lm})_{A_1 \cdots A_s} \\
        &= \sum_{l, m} \Big[\;_{-s} S_{lm} \;_{-s} Y_{lm} m_{A_1} \cdots m_{A_s} \\
        &\hspace{3.2em}+ \;_s S_{lm} \;_s Y_{lm} \bar m_{A_1} \cdots \bar m_{A_s}\Big] .
    \end{split}
\end{equation}
The equivalence of the two equalities in Eq.~\eqref{eq:StensorSpin} and the properties of the dyad imply that we can express the tensorial and spin-weighted coefficients in terms of one another:
\begin{subequations}
    \begin{align}
        \;_{-s} S_{lm} &= \sum_{\rm I} M^{\rm I} S^{\rm I}_{lm}, \\
        \;_s S_{lm} &= (-1)^s \sum_{\rm I} \bar M^{\rm I} S^{\rm I}_{lm}, \\
        S^{\rm I}_{lm} &= \bar M^{\rm I} \;_{-s} S_{lm} + (-1)^s M^{\rm I} \;_s S_{lm}. \label{eqn:tensor_project}
    \end{align}
\end{subequations}

We now summarize a few additional operations that can be performed on the spin-weighted and tensor harmonics.
The first two are raising and lowering the respective spin-weight and rank of the harmonics.
For the spin-weighted harmonics, it follows directly from the definition, together with commutation relations for $\eth$ and $\bar \eth$, that
\begin{subequations} \label{eqn:eth_ladder}
    \begin{align}
        \eth \;_s Y_{lm} &= -\frac{1}{\sqrt 2} \mathcal B_{ls} \;_{s + 1} Y_{lm}, \\
        \bar \eth \;_s Y_{lm} &= \frac{1}{\sqrt 2} \mathcal B_{l(s - 1)} \;_{s - 1} Y_{lm},
    \end{align}
\end{subequations}
where
\begin{equation}
    \mathcal B_{ls} \equiv \mathcal A_{ls}/\mathcal A_{l(s + 1)} = \sqrt{(l - s)(l + s + 1)}.
\end{equation}
Moreover, from the definition of the tensor harmonics, raising is similarly given by
\begin{subequations}
\begin{equation} \label{eqn:tensor_raise}
    \STF \mathscr D_{A_1} (T^{\rm I}_{lm})_{A_2 \cdots A_{s + 1}} = \frac{1}{\sqrt 2} \mathcal B_{ls} (T^{\rm I}_{lm})_{A_1 \cdots A_{s + 1}},
\end{equation}
for $s\geq 1$.
The scalar case $s=0$ is given by
\begin{equation} \label{eqn:scalar_raise}
    \mathscr D_A Y_{lm} = \mathcal B_{l0} (T^\mathrm{E}_{lm})_A.
\end{equation}
\end{subequations}
For lowering, a calculation making use of the definition of the tensor harmonics in terms of spin-weighted harmonics implies, for $s > 1$, that
\begin{subequations}
\begin{equation} \label{eqn:tensor_lower}
    D^B (T^{\rm I}_{lm})_{B A_1 \cdots A_{s - 1}} = -\frac{1}{\sqrt 2} \mathcal B_{l(s - 1)} (T^{\rm I}_{lm})_{A_1 \cdots A_{s - 1}} .
\end{equation}
For the case where $s = 1$, we find that
\begin{equation} \label{eqn:vector_lower}
    \mathscr D^A (T^{\rm E}_{lm})_A = \epsilon^{AB} \mathscr D_A (T^{\rm B}_{lm})_B = -\mathcal B_{l0} Y_{lm},
\end{equation}
while
\begin{equation} \label{eqn:vector_lower_zero}
    \mathscr D^A (T^{\rm B}_{lm})_A = \epsilon^{AB} \mathscr D_A (T^{\rm E}_{lm})_B = 0.
\end{equation}
\end{subequations}
The raising and lowering results are used in Sec.~\ref{sec:multipole} to compute the multipolar expansions of the 2-sphere covariant derivatives of the Bondi metric functions.
Finally, we give the action of the spherical Laplacian $\mathscr D^2$ on the tensor harmonics.
Since
\begin{equation}
    \mathscr D^2 Y_{lm} = -l(l + 1) Y_{lm} = -\mathcal B_{l0}^2 Y_{lm},
\end{equation}
it follows that
\begin{equation} \label{eqn:tensor_D2}
    \begin{split}
        \mathscr D^2 (T^{\mathrm I}_{lm})_{A_1 \cdots A_s} &= [-l(l + 1) + s^2] (T^{\mathrm I}_{lm})_{A_1 \cdots A_s} \\
        &= (-\mathcal B_{ls}^2 - s) (T^{\mathrm I}_{lm})_{A_1 \cdots A_s}.
    \end{split}
\end{equation}

\subsection{Products of spin-weighted or of tensor harmonics}

We now consider multiplication of harmonics.
This is simplest for the spin-weighted spherical harmonics: for $s = s' + s''$, and $m = m' + m''$, we have that
\begin{equation} \label{eqn:Y_expansion}
    \;_{s'} Y_{l'm'} \;_{s''} Y_{l''m''} = \sum_l \mathscr C^{lm,s's''}_{l'm'l''m''} \;_s Y_{lm},
\end{equation}
where the Gaunt coefficient $\mathscr C^{lm,s's''}_{l'm'l''m''}$ is defined in Eqs.~\eqref{eq:Cdef3Ylm} and~\eqref{eq:int3Ylm}.
By the properties of the Wigner $3$-$j$ symbols (specifically the ``time-reversal'' property), it follows that these coefficients satisfy
\begin{equation} \label{eqn:C_symmetry}
    \mathscr C^{lm,s's''}_{l'm'l''m''} = (-1)^{l + l' + l''} \mathscr C^{lm,(-s')(-s'')}_{l'm'l''m''}.
\end{equation}
This property will be used below to write the tensor harmonic expressions more compactly.

For the tensor harmonics, there are multiple cases that need to be considered: (i) contraction of two tensor harmonics of equal rank, (ii) contraction of two tensor harmonics of unequal rank, (iii) multiplication of a tensor harmonic by a scalar harmonic, and (iv) the STF part of the outer product of two tensor harmonics.
To derive each of these four cases, we use Eq.~\eqref{eqn:T_from_Y} first to write the tensor harmonics in terms of the spin-weighted harmonics, next Eq.~\eqref{eqn:Y_expansion} to rewrite the result in terms of a sum of individual spin-weighted harmonics, and finally Eq.~\eqref{eqn:tensor_project} to map this expression onto the tensor-harmonic basis.
We can then use Eq.~\eqref{eqn:C_symmetry} to simplify the expression, so that the summand involves just one spin-weighted Gaunt coefficient.
This gives a final result that is written in terms of the dyad $M^{\rm I}$.
To convert the result to a more useful form, we note that the $\eta^{\rm II'I''}_{ll'l''}$ coefficient in Eq.~\eqref{eq:etaCoeff} can be expressed in terms of the dyad $M^\mathrm{I}$ by
\begin{equation}
    \eta^{\rm II'I''}_{ll'l''} = 2 \sqrt 2 \left[\bar M^{\rm I} M^{\rm I'} \bar M^{\rm I''} + (-1)^{l + l' + l''} M^{\rm I} \bar M^{\rm I'} M^{\rm I''}\right] .
\end{equation}
Given the definition of $M^\mathrm{I}$, note that
\begin{equation}
    \eta^{\rm EI'I''}_{ll'l''} = 2 \left[M^{\rm I'} \bar M^{\rm I''} + (-1)^{l + l' + l''} \bar M^{\rm I'} M^{\rm I''}\right].
\end{equation}
All of the dyad-dependent pieces can be compactly written in terms of $\eta^{\rm II'I''}_{ll'l''}$.
Thus, the four cases of products of tensor harmonics can be summarized 
as follows:
\begin{widetext}
\begin{subequations} \label{eq:TlmProducts}
    \begin{align}
        (T^{\rm I'}_{l'm'})_{A_1 \cdots A_{s'}} (T^{\rm I''}_{l''m''})^{A_1 \cdots A_{s'}} &= \frac{(-1)^{s'}}{2} \sum_l \eta^{\rm EI'I''}_{ll'l''} \mathscr C^{lm,(-s')s'}_{l'm'l''m''} Y_{lm}, \\
        (T^{\rm I'}_{l'm'})_{A_1 \cdots A_{s' - s''} B_1 \cdots B_{s''}} (T^{\rm I''}_{l''m''})^{B_1 \cdots B_{s''}} &= \frac{(-1)^{s''}}{2 \sqrt 2} \sum_{I, l} \eta^{\rm II'I''}_{ll'l''} \mathscr C^{lm,(-s')s''}_{l'm'l''m''} (T^{\rm I}_{lm})_{A_1 \cdots A_{s' - s''}}, \\
        Y_{l'm'} (T^{\rm I''}_{l''m''})_{A_1 \cdots A_{s''}} &= \frac{1}{2} \sum_{I, l} \eta^{\rm EII''}_{ll'l''} \mathscr C^{lm,0s''}_{l'm'l''m''} (T^{\rm I}_{lm})_{A_1 \cdots A_{s''}}, \\
        \STF[(T^{\rm I'}_{l'm'})_{A_1 \cdots A_{s'}} (T^{\rm I''}_{l''m''})_{A_1 \cdots A_{s''}}] &= \frac{1}{2 \sqrt 2} \sum_{I, l} \eta^{\rm I'II''}_{ll'l''} \mathscr C^{lm,s's''}_{l'm'l''m''} (T^{\rm I}_{lm})_{A_1 \cdots A_{s' + s''}}.
    \end{align}
\end{subequations}
\end{widetext}
Recall that the Gaunt coefficients are nonzero when $m = m' + m''$.

\section{Relating Bondi metric functions and Newman-Penrose scalars} \label{app:Bondi2NP}

The NR CCE simulations provide the leading-order in $1/r$ parts of the Newman-Penrose scalars rather than the Bondi metric functions used in this paper.
The two sets of quantities can be related (this has been done in~\cite{Grant:2023jhd}, for example), and we review some key aspects of this correspondence in this appendix.
Specifically, we relate the Bondi metric functions described in Sec.~\ref{sec:review} to the Newman-Penrose scalars from the SXS CCE simulations using the conventions of the \textsc{scri} package.
The primary uses of the expressions below are for computing the different higher memory signals from NR waveforms, which were used for the comparisons in Figs.~\ref{fig:Q1_e_nospin}, \ref{fig:Q2_e_nospin_shear_contri}, and~\ref{fig:Q1_q1_aligned_spin}--\ref{fig:Q2_q1_aligned_spin}.

The Newman-Penrose formalism relies upon a null tetrad $\{\tl^a, \tn^a, \tm^a, \bar{\tm}^a\}$, which is normalized so that the only nonvanishing inner products are
\begin{equation}
    \tl^a\tn_a=-1,\;\; \tm^a \bar{\tm}_a=1
\end{equation} 
The tetrad used by the implementation of CCE in Bondi coordinates in the \textsc{SpECTRE} code~\cite{Kidder:2016hev} takes the following form~\cite{Moxon:2020gha,Grant:2023jhd}:
\begin{subequations}\label{eq: SpECTRE tetrads}
    \begin{align}
        \tl^a \equiv {} & \frac1{\sqrt{2}} (\partial_r)^a , \label{eq: l}\\
        \tn^a \equiv {} & \sqrt{2} \exp^{-2\beta/r} \Bigg[(\partial_u)^a+\frac1{r^2}\mathcal{U}^A(\partial_A)^a\nonumber\\
        &-\frac{1}{2}\left(1-\frac{2V}{r}(\partial_r)^a\right)\Bigg] , \label{eq: n}\\
        \tm^a\equiv {} & \frac{1}{\sqrt{2} r}(\partial_A)^a m_B\left( \sqrt{1+\gamma}h^{AB}- \frac{1}{r\sqrt{1+\gamma}}\mathcal{C}^{AB} \right) , \label{eq: m}
    \end{align}
\end{subequations}
where $m_A$ is the dyad defined in~\eqref{eq: def dyad}.

The output of the CCE data includes the shear $\sigma$ and the leading-order pieces of the Weyl scalars, $\Psi_i$. 
The Newman-Penrose shear $\sigma$ can be obtained from the Bondi shear tensor $C_{AB}$ by contracting with the dyad:
\begin{equation}
    \sigma=\frac1{2}{m}^A {m}^B C_{AB} . \label{eq: def shear}
\end{equation}
The shear $\sigma$ is related to the leading-order (in $1/r$) piece of the gravitational strain~\eqref{eq:hPlusTimes} by
\begin{equation}
    \sigma=\frac{1}{2}\underset{r\rightarrow\infty}{\mathrm{lim}} r\bar h \label{eq: shear strain rel}
\end{equation}

The five (complex) Weyl scalars are defined by
\begin{subequations}\label{eq: def Weyl scalars}
    \begin{align}
        \Psi_0&= C_{abcd}\tl^a\tm^b \tl^c \tm^d , \label{eq: psi0}\\
        \Psi_1&= C_{abcd}\tl^a\tn^b \tl^c \tm^d , \label{eq: psi1}\\
        \Psi_2&= C_{abcd}\tl^a\tm^b \bar{\tm}^c \tn^d , \label{eq: psi2}\\
        \Psi_3&= C_{abcd}\tl^a\tn^b \bar{\tm} ^c \tn^d , \label{eq: psi3}\\
        \Psi_4&= C_{abcd}\tn^a \bar{\tm}^b \tn ^c \bar{\tm}^d . \label{eq: psi4}
    \end{align}
\end{subequations}

The asymptotic Bondi data returned by the \textsc{scri} package contains just the leading-order part of the Weyl scalars defined in Eq.~\eqref{eq: def Weyl scalars}, but it uses a slightly different null tetrad from the one introduced in Eq.~\eqref{eq: SpECTRE tetrads} (the conventions for the tetrad used by \textsc{scri} are described in~\cite{Boyle:2015nqa}\footnote{Note that Eq.~(12b) of~\cite{Boyle:2015nqa}, which makes use of the complex stereographic coordinate $\zeta = e^{i\phi} \cot(\theta/2)$, suggests that the complex dyad differs from the dyad in Eq.~\eqref{eq: def dyad} by a spin transformation $m_A \rightarrow e^{-i\phi} m_A$. 
The tetrad implemented in~\textsc{scri} does not seem to differ by this spin transformation, however.}).
The two conventions are related by a simple rescaling.
Denoting by $\psi_i$ the leading-order in (in $1/r$) part of the Weyl scalars returned by \textsc{scri}, their relation to the leading-order Weyl scalars $\Psi_i$ was given in~\cite{Grant:2023jhd}:
\begin{equation}
    \psi_i\equiv \underset{r\rightarrow\infty}{\mathrm{lim}} r^{5-i}(-\sqrt{2})^{2-i}\Psi_i
\end{equation}
These asymptotic Weyl scalars are related to the Bondi aspects and the leading-order piece of shear $\sigma$ by (see~\cite{Moxon:2020gha}) 
\begin{subequations}\label{eq: Weyl- Bondi data relation}
    \begin{align}
        \psi_0 & =-3\big(\ord{0}{\mathcal{E}} _{AB}\;m^Am^B -{\sigma}\left|\sigma\right|^2\big), \label{eq: psi0 E rel}\\
        \psi_1 & =  m^A N_A, \label{eq: psi1 N rel}\\
        \psi_2 & = -\mass - (\mathrm{Im}[\eth^2 \bar{\sigma}]+ {\sigma}\dot{\bar{\sigma}}), \label{eq: psi2 m rel}
    \end{align}
\end{subequations}
The other two asymptotic Weyl scalars depend on just the shear,
\begin{subequations}
    \begin{align}
        \psi_3 & = \eth \dot{\bar{\sigma}}, \label{eq: psi3 shear rel} \\
        \psi_4 & =  -\ddot{\bar{\sigma}}, \label{eq: psi4 shear rel}
    \end{align}
\end{subequations}
Equation~\eqref{eq: Weyl- Bondi data relation} allows us to write the charges and fluxes associated with different memory effects in terms of the NR CCE data.

To express the fluxes and pseudofluxes in Sec.~\ref{subsec:chargeFlux} in terms of the Bondi shear and its angular derivatives, we rewrite the tensorial expressions in terms of the Newman-Penrose scalars and the complex dyad $m_A$. 
The calculations make use of Eqs.~\eqref{eq:dyadDef}--\eqref{eqn:eps_2}.
The expressions are conveniently written in terms of the operator $\eth$ in Eq.~\eqref{eq:ethSAs} as follows:
\begin{subequations}
\begin{align}
    \E{\mathcal{F}_0}& = -\dot\sigma\dot{\bar{\sigma}}, \\
    \E{\mathcal F_1}-i \B{\mathcal F_1} & =  [\bar{\eth}(\dot{\bar{\sigma}}\eth\sigma +3\sigma\eth\dot{\bar{\sigma}}-{\bar{\sigma}}\eth\dot\sigma -3\dot\sigma\eth{\bar{\sigma}})], \\
    \E{\mathcal{F}_2} -i \B{\mathcal{F}_2}& = 4[\bar{\eth}^2(\sigma(\sigma\dot{\bar\sigma}+\bar\sigma\dot\sigma)], \\
    \E{\mathcal{G}_2} -i\B{\mathcal{G}_2}& =  2\big[\bar\eth^2(m\sigma+i\sigma\mathrm{Im}[\eth^2\bar\sigma]\nonumber, \\
    &\qquad\qquad+\frac{1}{6}\eth(\bar\sigma\eth\sigma+3\sigma\eth\bar\sigma))\big].
\end{align} 
\end{subequations}
Note, however, that the $\eth$ refers to the definition in Appendix~\ref{app:harmonics}; the $\eth$ in the \textsc{scri} package has a relative minus sign (see the discussion in Footnote~\ref{fn:signs} about sign conventions).

A similar calculation can be performed for the charges.
The details of the calculations are similar to those for the fluxes, and the results are
\begin{subequations}\label{eq:Charge CCE}
\begin{align}
    \E{Q_1} -i\B{Q_1}  & =  2\big[\bar\eth \big\{ \psi_1+\frac 12(\bar\sigma\eth\sigma+3\sigma\eth\bar\sigma)\big\}\big], \\
    \E{Q_2}-i\B{Q_2} & =  2\Big[\bar\eth^2\Big(-\frac13\psi_0 +\sigma| \sigma|^2\Big)\Big].
\end{align}
\end{subequations}
The results can also be obtained in the stationary-to-stationary approximation:
\begin{subequations}\label{eq:STS Charge CCE}
\begin{align}
    \E{Q_{1,\mathrm S}} -i \B{Q_{1,\mathrm S}}& = 2\big[\bar\eth\big\{3\eth^{-1} (\sigma\psi_2)_{\mathrm S}+\frac 12(\bar\sigma\eth\sigma+3\sigma\eth\bar\sigma)_\mathrm{S}\big\}\big], \\
    \E{Q_{2,\mathrm S}}- i\B{Q_{2,\mathrm S}} & = -\frac{16}3[\bar\eth^2(\eth\bar\eth+\bar\eth\eth+2)^{-1}\bar\eth(\psi_1\sigma)_{\mathrm S}]+ O(\sigma_\mathrm{S}^3).
\end{align}
\end{subequations}
The notation $\eth^{-1}$ means to compute the inverse of the $\eth$ operator, which is most straightforwardly done by expanding the relevant expressions on which $\eth^{-1}$ acts in terms of appropriate spherical harmonics.
Here the subscript $\mathrm S$ is used to indicate that this is the stationary-to-stationary part of the charge (as discussed in Sec.~\ref{sec:stationary}).

\bibliography{refs}

\end{document}